\documentclass[useAMS,usenatbib]{mn2e}
   \usepackage{graphicx}
   \usepackage{amsmath}
   \usepackage{color}
   \usepackage{epstopdf}
   \usepackage{pdfsync}
   \usepackage{indentfirst}
   \usepackage{gensymb}
   \usepackage{float}
   \usepackage{ulem}
\font\fiverm=cmr5             \font\sevenrm=cmr7

          \font\sixrm=cmr6       

\def\dover#1#2{\hbox{${{\displaystyle#1 \vphantom{(} }\over{
   \displaystyle #2 \vphantom{(} }}$}}

{\catcode`\@=11                                                  
\gdef\SchlangeUnter#1#2{\lower2pt\vbox{\baselineskip 0pt\lineskip0pt    
\ialign{$\m@th#1\hfil##\hfil$\crcr#2\crcr\sim\crcr}}}}           
\def\gtrsim{\mathrel{\mathpalette\SchlangeUnter>}}               
\def\lesssim{\mathrel{\mathpalette\SchlangeUnter<}}

\def\sigt{\sigma_{\hbox{\fiverm T}}}                                
\def\taut{\tau_{\hbox{\fiverm T}}}                                

\def\omegaB{\omega_{\hbox{\sixrm B}}}

\def\teq#1{$\, #1\,$}                         
\def\mns{M_{\hbox{\sixrm NS}}}
\def\LEdd{L_{\hbox{\sevenrm Edd}}}

\def\betavec{\overrightarrow{\beta}}

\def\wcyc{\omega_{\hbox{\fiverm B}}}
\def\Bvec{\boldsymbol{B}}
\def\Bvechat{\hat{\boldsymbol{B}}}
\def\Evec{\boldsymbol{E}}
\def\calEvec{\boldsymbol{\cal E}}
\def\calEvechat{\hat{\boldsymbol{\cal E}}}

\def\calEthet{{\cal E}_\theta}
\def\calEphi{{\cal E}_\phi}

\def\kvec{\boldsymbol{k}}
\def\kvechat{\hat{\boldsymbol{k}}}
\def\rvec{\boldsymbol{r}}
\def\vvec{\boldsymbol{v}}
\def\avec{\boldsymbol{a}}

\def\alphavec{\boldsymbol{\alpha}}

\def\alphavecre{\boldsymbol{\alpha}_{\hbox{\fiverm R}}}
\def\alphavecim{\boldsymbol{\alpha}_{\hbox{\fiverm I}}}

\def\thetaB{\theta_{\hbox{\sixrm B}}}
\def\betavec{\boldsymbol{\beta}}
\def\betavechat{\hat{\boldsymbol{\beta}}}
\def\SigmaB{\Sigma_{\hbox{\sixrm B}}}
\def\DeltaB{\Delta_{\hbox{\sixrm B}}}

\def\IP{I_{\hbox{\fiverm P}}}
\def\QP{Q_{\hbox{\fiverm P}}}
\def\UP{U_{\hbox{\fiverm P}}}
\def\VP{V_{\hbox{\fiverm P}}}

\title[Polarized Radiation Transfer in Neutron Star Surface Layers]
{Polarized Radiation Transfer\\ in Neutron Star Surface Layers}
\author[J.~A. Barchas, K. Hu \& M.~G. Baring]{
Joseph A. Barchas,$^{1,2}$\thanks{E-mail: joseph.barchas@hccs.edu (JAB); kh38@rice.edu (KH); baring@rice.edu (MGB)\vskip -80pt} 
Kun Hu$^{2}$ and Matthew G. Baring$^{2}$\\
$^{1}$Natural Sciences, Southwest Campus, Houston Community College, 5601 W. Loop S., Houston, Texas 77081, USA \\
$^{2}$Department of Physics and Astronomy - MS 108, Rice University,
6100 Main Street, Houston, Texas 77251-1892, USA
\vspace{-15pt}}

\begin{document}

\date{Accepted November 9, 2020; Received October 19, 2020; in original form August 31, 2020.}

\newcommand{\vol}[2]{$\,$\rm #1\rm , #2}                 

\pagerange{\pageref{firstpage}--\pageref{lastpage}} \pubyear{2020}

\maketitle

\label{firstpage}
\begin{abstract} 
The study of polarized radiation transfer in the highly-magnetized surface
locales of neutron stars is of great interest to the understanding of accreting
X-ray pulsars, rotation-powered pulsars and magnetars.  This paper explores scattering
transport in the classical magnetic Thomson domain that is of broad
applicability to these neutron star classes.  The development of a Monte Carlo
simulation for the polarized radiative transfer is detailed: it employs an
electric field vector formalism to enable a breadth of utility in relating
linear, circular and elliptical polarizations.  The simulation can be applied to
any neutron star surface locale, and is adaptable to accretion column and
magnetospheric problems.  Validation of the code for both intensity and Stokes
parameter determination is illustrated in a variety of ways.  Representative
results for emergent polarization signals from surface layers are presented for
both polar and equatorial magnetic locales, exhibiting contrasting signatures
between the two regions.   There is also a strong dependence of these characteristics on the ratio of
the frequency \teq{\omega} of a photon to the cyclotron frequency \teq{\wcyc
=eB/mc}. Polarization signatures for high opacity domains are presented,
highlighting compact analytic approximations for the Stokes parameters and
anisotropy relative to the local field direction for an extended range of frequencies.
These are very useful in defining injection conditions deep in the simulation
slab geometries, expediting the generation of emission signals from highly
opaque stellar atmospheres. The results are interpreted throughout using the
polarization characteristics of the magnetic Thomson differential cross section.
\end{abstract}

\vspace{-15pt}

\section{Introduction}

Observations of thermal emission from both isolated and accreting neutron stars have provided a
richesse of information about their surfaces and interiors. Radii of isolated neutron stars can
be estimated or constrained by applying the Stefan-Boltzmann law to their quasi-thermal spectra
\citep[][]{Potekhin14}. Soft X-ray pulse profiles of younger and middle-aged neutron stars have
been used to estimate geometric parameters like the sizes and locales of hot spots and observer
viewing angles to the spin axes: see \cite{Gotthelf10} for an example of central compact object,
and \cite{Younes20} for a magnetar. For much older neutron stars, recycled pulsars (PSRs), pulse
profiles of atmospheric X rays from millisecond PSR J0030+0451 by  the Neutron Star Interior
Composition Explorer ({\sl NICER}) have enabled precise mass and radius measurements
\citep{Riley19,Miller19}. Spectral features such as absorption lines help to estimate the
gravitational redshifts \citep[][]{Hambaryan11} and thereby also inform mass-to-radius ratios.
Comparisons between surface temperatures from cooling neutron stars and theoretical cooling
curves yield insights into the thermal heating and neutrino transport in neutron star interiors,
deriving constraints on the equation of state \citep{Yakovlev04}. All these probes can leverage
a sophisticated understanding of neutron star atmospheres, which serves as the principal
objective of this paper.

Of great interest is the persistent soft X-ray emission of magnetars, neutron stars that possess
super-strong magnetic fields. Magnetars' surface fields (\teq{B_p\sim 10^{13}-10^{15}}G) and
relative young ages are inferred directly from their long spin periods (\teq{P \sim 2-12}s) and
large spin-down rates (\teq{\dot{P}\sim 10^{-14}-10^{-10}\text{s s}^{-1}}), presuming that their
rotational spin down is due to magnetic dipole torques \citep[e.g., see][]{Kouveliotou98}.
Magnetars have historically been divided into two observational groups: Soft-Gamma Repeaters
(SGRs) and Anomalous X-ray Pulsars (AXPs). However, the observed quiescent emission from SGRs
and the discovery of SGR-like bursts in several AXPs diluted the difference between the two
groups, suggesting a ``unification paradigm'' where SGRs and AXPs belong to a single class.
On the theoretical side, \cite{DT92} and \cite{TD96} postulated the fireball scenario where the flare
emission is triggered by sub-surface magnetic activity.  This focuses on magnetic stresses in
the crust and structural rearrangements and partial decay of the sub-surface fields \citep{TD95, TD01}. 

The persistent soft X-ray signals from magnetars are very bright, with typical luminosities of
\teq{L_X\sim10^{{35}}\text{erg s}^{-1}}, often exceeding their rotational energy loss rates. The
spectra of this emission can be approximately fit using blackbodies of temperature
\teq{\sim0.3-0.6}keV, connected to a soft power-law tail with photon index \teq{\lesssim2-4}
\citep[e.g.,][]{Mereghetti08,Vigano13}. These temperatures are higher than those of
typical isolated neutron stars -- see \cite{Becker09} for a review of X-ray pulsar emission. The
thermal component likely comes from the stellar surface, providing information about the
physical properties of its atmosphere, like chemical composition, ionization equilibrium, and
the state of matter. Pulse profiles of magnetars constrain the locale of the emission region,
and thereby help improve the understanding of sub-surface thermal heat transport, and informing
paradigms of particle bombardment of the surface \citep{Beloborodov16}. For some magnetars, the
best fit is obtained using a two-blackbody model, which is possibly a signature of temperature
gradients across the stellar surface \citep{Gotthelf05}.

Models for magnetic neutron star surface emissions have been constructed by \cite{Shibanov92},
\cite{Pavlov94}, \cite{Zavlin1996AandA} and \cite{Zane00}, considering fully-ionized hydrogen or
helium atmospheres with moderate magnetic fields \teq{\sim10^{13}}G. Partially-ionized
atmosphere models have been explored by \cite{Ho03ApJ}, \cite{Potekhin04}, \cite{Ho08} and
\cite{Suleimanov09}, using sophisticated opacity information and updated equations of state.
Atmosphere models addressing the magnetar field domain were treated by \cite{HoLai-2001-MNRAS},
\cite{Ozel-2001-ApJ}, \cite{HoLai-2003-MNRAS}, \cite{Ozel-2003-ApJ}, \cite{Adelsberg06} and
\cite{Taverna20}. Since the fields of magnetars generally exceeds the critical field \teq{B_{\rm
cr}=4.414 \times 10^{13}}Gauss, detailed consideration of the polarization of the magnetized
quantum vacuum is necessary \citep{LaiHo-2003-ApJ}. In addition, thermal emission
from condensed surfaces at relatively low temperatures was studied by \cite{Turolla04} and
\cite{Medin07}. For magnetic fields $\lesssim10^{12}$G, this magnetic condensation is expected
to be minimal, and the atmospheres are likely fully ionized  \citep{Medin06,Medin07}.

A common feature of these previous studies is that they use scattering and free-free opacity to
mediate the photon transport in terms of two orthogonal polarization modes. These opacities help
to support the atmosphere hydrostatically and shape the calculated spectra, which deviate
substantially from a pure Planck form. 
Furthermore, X-ray emission from the surfaces of magnetars is expected to be polarized, because the strong
magnetic fields introduce anisotropy to the plasma medium, the quantum vacuum, and the
scattering process, as is highlighted here.  These interesting physics elements can be
probed using polarimetric measurements of magnetars in the soft X-ray band, which are expected
to be provided by future missions like IXPE \footnote{https://ixpe.msfc.nasa.gov/}
\citep{Weisskopf16} and eXTP \citep{Zhang16}. Such polarimetry introduces an extra dimension to
diagnostics of source physical properties that complement spectroscopy, and should permit the
detection of the QED vacuum effect from magnetar atmospheres \citep[see][]{Taverna20}.

In this paper, we detail the construction of a new Monte Carlo simulation of the magnetic
Thomson scattering of polarized X-rays in neutron star surface layers. In contrast to previous
studies, we simulate the transport via a tracking of wave electric field vectors. This
encapsulates full polarization information, linear and circular and their interplay throughout.
The code does not presently treat self-consistent hydrostatic atmospheric structure 
and the free-free opacity that dominates \citep{HoLai-2001-MNRAS} at photon energies 
below around 1--2 keV and deep in the atmosphere.  Thus, the simulation is readily applied to 
outer atmospheres and is adaptable to the more tenuous environments of magnetar magnetospheres.
The simulation is developed in the special case of zero
dispersion in order to enable its validation and a detailed understanding of emergent
polarization signatures; it is routinely extendable to treat elliptical eigenmodes of
propagation in dispersive media. Our Monte Carlo code generates angular distributions of
intensity and Stokes parameters for arbitrary field orientations and a representative range of
photon frequencies. 
The code is validated by direct comparison of results with those from prior investigations.

Although our present focus is on surface layers of magnetars and neutron stars of lesser
magnetizations, the Monte Carlo technique can be easily extended to treat Comptonization in hot
plasmas, which is important in the context of other phenomena like accreting X-ray pulsars and
magnetar bursts and giant flares.  The atmosphere results presented here can serve as a general guide to the expectations for
photospheric outer envelopes of optically-thick magnetospheric bursts in magnetars 
\citep{Taverna17}. Detailed simulations of the
radiation transport in the photospheres of both magnetar fireballs and accretion column of X-ray
pulsars are of interest of future hard X-ray polarimeters, and will be the subject of future
extensions of the code.

The paper begins with a review of elements of the classical electrodynamical formulation 
of magnetic Thomson scattering and the description of polarization in Sec.~\ref{sec:transfer_theory}.
In Sec.~\ref{sec:MC_technique} we describe the Monte Carlo technique underpinning the 
{\sl MAGTHOMSCATT} code, including sampling and binning.  Intensity and polarization 
results for slab surface volumes in are detailed in Sec.~\ref{sec:validation} for two special locales 
on a neutron star surface, the magnetic pole and the equator.  Comparisons with previous 
works are forged therein as a means of code validation.  Polarization results at high opacity 
are given in Sec.~\ref{sec:high_opacity_pol}, together with empirical functions for anisotropic 
photon distribution and Stokes parameters.  The resultant understanding of this polarization 
information in such high opacity domains facilitates the accurate modeling of complete 
atmospheres with simulations possessing modest computational demands.  Some contextual 
discussions for neutron star applications are offered in Sec.~\ref{sec:discussion}.

\section{Polarized Radiation Transfer in Strong Magnetic Fields}
 \label{sec:transfer_theory}
 
The technical approach adopted in the simulation we detail in this paper models photon transport 
and magnetic scattering using an electric field vector formalism. This distinguishes it from 
previous studies that have tracked Stokes parameter information \citep[][]{Whitney-ApJS91,Whitney-ApJ91} 
or linearly polarized states in the resonant cyclotron approximation \citep[][]{Fernandez07,Nobili-2008-MNRAS,Fernandez11}. 
This electric vector approach is more fundamental, more general, and is elegant in how it isolates key characteristics of 
the scattering transport, identifying the critical interplay between linear and circular polarizations
that is missing in most works.  
Employment of the tracking of electric (polarization) vectors also affords greater versatility 
for future extensions, like adding up emission from extended regions on the stellar surface 
or perhaps from magnetospheric locales, general relativistic polarization transport and 
dispersive propagation out to an observer at infinity.  Such versatility is not permitted by pure
Stokes parameter transport approaches such as in \cite{Whitney-ApJS91,Whitney-ApJ91} .

The structure of the radiative transfer simulation described in Sec.~\ref{sec:MC_technique}
is to inject polarized light waves at the base of an upper atmosphere where free-free opacity 
is low, propagate them as 
discrete photons, scattering them as classical electromagnetic waves in the magnetic Thomson 
domain, and eventually allow them to escape at the top of the atmospheric slab.  
The simulation will assume an effectively cold plasma, which well approximates the 
outer surface layers of a neutron star.  Given the focus on the simulation 
development and validation, dispersive influences due to warm plasma and 
the quantum magnetic vacuum will be neglected, their treatment being 
deferred to future stages of our program.

The intensity and polarization of a classical, transverse electromagnetic wave can
be described with a {\sl complex} electric field 3-vector \teq{\Evec} that is orthogonal 
to the direction of propagation \teq{\kvec}:
\begin{equation}
   \Evec \; = \; \Evec_0 \exp\Bigl\{ i \bigl[\kvec\cdot\rvec - \omega t -\phi (t) \bigr] \Bigr\} 
   \; =\; \calEvec (\rvec, \, t) \, e^{-i\omega t} \; .
 \label{eq:Evec_genform}
\end{equation}
The magnetic field component \teq{\Bvec = \kvechat \times \Evec}
of this wave is then automatically captured and trivially determined. 
The polarization vector \teq{\calEvec (\rvec, \, t)} incorporates the spatial dependence, 
which usually factors out of measures when time averages of the field are taken; 
we will ignore it hereafter.  

\subsection{Magnetic Thomson Scattering}
 \label{sec:magThom}

The classical electromagnetic theory of electron scattering in the absence of external 
fields is detailed in texts such as \cite{Jackson75,LandL1975,Rybicki79}.  Such Thomson
scattering formalism is routinely adapted to treat the case of gyrational motion of 
electrons in a magnetic field, and a seminal formulation was presented in \cite{Canuto71PRD}
in the context of plasma dispersion.  The classical formulation is distilled here to
identify elements germane to the simulation algorithms of Sec.~\ref{sec:MC_technique}, 
and the presentation of results in Secs.~\ref{sec:validation} and~\ref{sec:high_opacity_pol}.

The incident electromagnetic wave has a direction of propagation given by the unit vector \teq{\kvechat_i},
and an electric field \teq{\Evec(t) = \calEvec_i e^{-i\omega t}}, with \teq{\calEvec_i\cdot\kvechat_i=0}.
The oscillating wave field drives the acceleration of an electron subject to the 
influence of a magnetic field \teq{\Bvec \equiv B\Bvechat},  
motion described by the Newton-Lorentz equation:
\begin{equation}
   m_e\dover{d\vvec (t)}{dt} \; =\; -e \Evec(t) - \dover{e}{c} \, \vvec (t) \times \Bvec \quad .
 \label{eq:Newton_Lorentz}
\end{equation}
Since \teq{\Evec (t) \propto e^{-i\omega t}}, it is readily seen that the time 
dependence of the acceleration contains the same exponential factor:
\teq{\avec \equiv d\vvec /dt \propto e^{-i\omega t}} with velocity
\teq{\vvec = \avec/(-i\omega) \propto e^{-i\omega t}} also.  
By factoring out the time dependence, the induced acceleration then satisfies
\begin{equation}
   \avec \, e^{i\omega t}\; =\; - \dover{e}{m_e} \, \dover{\alphavec \, \vert \calEvec_i \vert }{\omega^2 - \wcyc^2} \quad ,
 \label{eq:scatt_accel}
\end{equation}
where
\begin{equation}
   \alphavec \; =\; \omega ^2 \calEvechat_i
      -i \omega \wcyc \,\calEvechat_i \times \Bvechat
      - \wcyc^2 ( \calEvechat_i \cdot \Bvechat ) \Bvechat \quad ,
 \label{eq:alpha_vec_def}
\end{equation}
and \teq{\wcyc = eB/m_ec} is the electron cyclotron frequency.
In general, \teq{\kvechat_i \times \Bvechat \neq \mathbf{0}} so that 
\teq{\calEvec_i \cdot \Bvechat \neq \mathbf{0}}.  
We have introduced the scaled incident polarization 
vector \teq{\calEvechat_i = \calEvec_i /\vert \calEvec_i \vert} to simplify 
ensuing expressions for the differential and total cross section. 
The motion is clearly oscillatory at the driving frequency \teq{\omega},
generally with different amplitudes in each of the three dimensions, 
leading to elliptical polarization for the scattered photon.
The \teq{\calEvechat_i \times \Bvechat} term is the driver for circular polarization, 
and its influence is maximized near the cyclotron frequency.
%
%

The accelerating, non-relativistic charge radiates a scattered wave.
The dipole radiation formula can be employed 
for an accelerating electron to obtain the electric field after scattering (\citealt{LandL1975}):
\begin{equation}
   \Evec_f (\rvec,t)\; \equiv\; \dover{r_0}{R}\, \calEvec_f e^{-i\omega t}
   \; =\; - \dover{e}{Rc^2}\, \kvechat_f \times \Bigl( \kvechat_f \times \avec \Bigr)\quad ,
 \label{eq:calEf_def}
\end{equation}
where \teq{\kvechat_f} is the direction of propagation of the scattered wave. The frequency 
of the outgoing wave is just that of the incoming one, the signature of Thomson scattering. 
Here, \teq{R} is the distance from the oscillating/radiating dipole to a point of observation,
and \teq{r_0=e^2/m_ec^2} is the classical electron radius.  
Inserting the acceleration from Eq.~(\ref{eq:scatt_accel}), 
\begin{equation}
   \calEvechat_f \;\equiv\; \dover{\calEvec_f}{ \vert \calEvec_i \vert} \; =\; 
    \dover{ \kvechat_f \times \bigl( \kvechat_f \times \alphavec \bigr)}{\omega^2 - \wcyc^2} \quad .
 \label{eq:calEf_eval}
\end{equation}
Thus the transversality condition \teq{\calEvec_f \cdot \kvechat_f =0} follows.  
The dipole radiation formalism then delivers the differential cross section for magnetic 
Thomson scattering via the ratio \teq{R^2\,\vert\Evec_f \vert^2 / \vert\Evec_i \vert^2} 
of final to initial wave Poynting fluxes: 
\begin{equation}
   \dover{d\sigma}{d\Omega_f} 
   \;=\; r_0^2 \, \calEvechat_f \cdot \calEvechat_f^*    
   \; =\; r_0^2 \, \dover{ \bigl( \kvechat_f \times \alphavec \bigr) \cdot  
       \bigl( \kvechat_f \times \alphavec^{\ast} \bigr)}{ (\omega^2 - \wcyc^2 )^2}\; .
 \label{eq:dsig_magThom}
\end{equation}
This distillation has employed Eqs.~(\ref{eq:alpha_vec_def}) and~(\ref{eq:calEf_eval}).
Using a polarization tensor, Eq.~(\ref{eq:dsig_magThom}) can be modified to treat dispersive cases, 
for example in accounting for the dielectric response of a plasma
\citep{Canuto71PRD,Ventura79_PRD}.

Employing a standard vector identity for the numerator of Eq.~(\ref{eq:dsig_magThom}),
it is routinely integrated over solid angles \teq{d\Omega_f} to yield the total 
scattering cross section:
\begin{equation}
   \sigma \; =\; \dover{8\pi}{3} \, r_0^2 \, \dover{ \alphavec \cdot  \alphavec^{\ast} }{ (\omega^2 - \wcyc^2 )^2}  \quad .
 \label{eq:sig_mag_Thom_form}
\end{equation}
Details of the derivation are posited in Appendix A.  Further, expanding 
\teq{\alphavec \cdot  \alphavec^{\ast}} using Eq.~(\ref{eq:alpha_vec_def}) then 
quickly gives
\begin{eqnarray}
   \sigma & = & \dover{\sigt}{ (\omega^2 - \wcyc^2)^2}
      \biggl[ \omega^4 + \wcyc^2 (\wcyc^2 - 2\omega^2) \Bigl\vert \calEvechat_i \cdot \Bvechat \Bigr\vert^2 
       \nonumber\\[-5.5pt]
 \label{eq:sig_magThom}\\[-5.5pt]
   &&  \qquad  + \omega^2 \wcyc^2 \Bigl\vert \calEvechat_i \times \Bvechat \Bigr\vert^2
         + 2i\, \omega^3 \wcyc \, \Bvechat \cdot \bigl( \calEvechat_i \times \calEvechat_i^* \bigr)
         \biggr] \quad ,\nonumber
\end{eqnarray}
%
%
where \teq{\sigt=8\pi r_0^2/3} is the familiar Thomson cross section 
in the absence of an external field. The term proportional to \teq{i\omega^3\wcyc} 
is actually real: this character can be established by adding 
\teq{\calEvechat_i \times \calEvechat_i^*} to its complex conjugate 
to demonstrate that this vector is always purely imaginary. 
Observe that because of transversality, \teq{\calEvechat_i \times \calEvechat_i^*} 
is parallel to \teq{\kvechat_i}, a result that is quickly established using the 
expansion of the vector triple product \teq{\kvechat_i \times (
\calEvechat_i \times \calEvechat_i^*) \equiv \mathbf{0}}.
Eq.~(\ref{eq:sig_magThom}) can be evaluated for any 
polarization configuration for the incident photon, as is done in Appendix~B.  
These vector forms for the differential and the total 
cross sections appear not to have been derived before, and 
complement other expositions in the literature
\citep[e.g.,][]{Hamada1974PASJ,Boerner1979PlPh}.  

Total cross sections for the linearly-polarized states \teq{\perp} and \teq{\parallel}
are presented in Fig.~\ref{fig:csect_mag} in Appendix~B, illustrating the prominence 
of the resonance at the cyclotron frequency, \teq{\omega = \wcyc}.  
In the classical picture, radiation reaction consumes some of the kinetic 
energy of the driven electron, leading to an extra term in the dynamical 
equation in Eq.~(\ref{eq:Newton_Lorentz}) \citep[e.g.,][]{LandL1975}.  This yields damped 
waves with solutions approximated by \teq{\omega \to \omega + i\Gamma /2} for 
cyclotron radiative width \teq{\Gamma}, 
so that a Lorentz profile proportional to \teq{1/[ (\omega -\wcyc)^2 + \Gamma^2/4]} 
replaces the divergent factor \teq{1/(\omega - \wcyc )^2}.  
The same type of Breit-Wigner modification arises in a 
quantum description \citep[e.g.][]{HD91}, where the width is now due to the finite 
cyclotron decay lifetime of the intermediate virtual electron state.
Accordingly, the divergence is truncated, yielding finite values for the cross section that are of the order of 
\teq{(\wcyc/\Gamma )^2 \sigt} when the magnetic field is 
highly-subcritical, \teq{B\ll B_{\rm cr}}; details can be found in the papers by \cite{bwg11} and \cite{Gonthier14}.

\subsection{Parameterizing Polarization}
 \label{sec:polarization} 

The electric field vector approach to handling transport and scattering of individual photons is
elegant and powerful.  Yet, for the purposes 
of polarization accounting for large numbers of photons emergent from the simulation geometry, 
and for comparison with other studies, it is expedient to also adopt
the convenient parameterization identified by \cite{Stokes1851}
encapsulated in the Stokes vector \teq{\boldsymbol{S} \equiv (I,Q,U,V)}. 
The Stokes \teq{I} parameter describes the intensity of the radiation. 
The Stokes \teq{Q} and \teq{U} parameters are related to the degree and angle of linear polarization.
The Stokes parameter \teq{V} captures information concerning circular polarization.  
In this paper, we will use the convention that the right-hand rule applies to {\it increasing} phases
to specify the sense of \teq{\calEvec} rotation for a circularly polarized wave.
With this convention, \teq{V/I=1 (-1)} corresponds to fully right- (left-) handed circularly polarized radiation. 

A principal measure featuring in the simulation output and graphical illustrations of Sections~\ref{sec:validation}
and~\ref{sec:high_opacity_pol} is the {\sl degree of polarization} \teq{\Pi}: 
\begin{equation}
   \Pi\; =\; \sqrt{(Q/I)^2+(U/I)^2+(V/I)^2}
   \; ,\quad
   0\; \leq\; \Pi\; \leq\; 1\; .
 \label{eq:poldeg_def}
\end{equation}
One can similarly define the linear polarization degree \teq{\Pi_{\rm lin} = \Pi (V\to 0)} 
by setting \teq{V=0} in Eq.~(\ref{eq:poldeg_def}).
In the Monte Carlo radiation transfer simulation that is described in
Section~\ref{sec:MC_technique}, each light wave is a monochromatic photon 
with 100\% polarization, i.e. \teq{\Pi = 1}, 
regardless of whether the light is linearly, circularly or elliptically polarized.  
Accordingly, the construction can comfortably accommodate the elliptical eigenmodes that 
arise in dispersive propagation in plasma or the magnetized quantum vacuum. 

The Stokes parameters for a photon can naturally be expressed in terms of its 
electric field vector information for general propagation directions using 
spherical polar coordinates rather than employing a Cartesian basis.  The propagation
direction \teq{\kvechat} is the radial direction, and the spherical polar angles give  
\begin{equation}
   \calEvec \; =\; {\cal E}_\theta\hat{\theta} + {\cal E}_\phi\hat{\phi}
   \;\equiv\; \bigl\vert \calEvec \bigr\vert \bigl( \hat{\cal E}_\theta\hat{\theta} + \hat{\cal E}_\phi\hat{\phi}  \bigr) 
 \label{eq:Evec_polar}
\end{equation}
for the polarization vector, with \teq{\hat{\cal E}_\theta = {\cal E}_\theta / \vert \calEvec \vert }
and \teq{\hat{\cal E}_\phi = {\cal E}_\phi / \vert \calEvec \vert }.  Using the \teq{\phi =0} 
plane for reference, with a correspond convention that \teq{U=0}, the Stokes parameter definition in this basis is 
\begin{equation}
   \boldsymbol{S} \; \equiv\; 
   \left( \begin{array}{c}
   I \\[2pt]
   Q \\[2pt]
   U \\[2pt]
   V
   \end{array} \right) 
   \; =\; 
   \left( \begin{array}{c}
   \langle \calEthet \calEthet^* \rangle + \langle \calEphi \calEphi^* \rangle \\[2pt]
   \langle \calEthet \calEthet^* \rangle - \langle \calEphi \calEphi^* \rangle \\[2pt]
   \langle \calEthet \calEphi^* \rangle + \langle \calEthet^* \calEphi \rangle \\[2pt]
   i \, \langle  \calEthet \calEphi^* -  \calEthet^* \calEphi \rangle
   \end{array} \right) \quad .
 \label{eq:Stokes_polar_def}
\end{equation}
The brackets \teq{\langle \dots \rangle} signify time averages of the products of wave 
field components.  
This coordinate choice is naturally suited to a fixed observer direction, with 
\teq{(\theta, \, \phi)} constituting zenith polar angles relative to the atmospheric slab, 
to be described in Sec.~\ref{sec:MC_technique}.
One can also a form a reduced Stokes parameter 3-vector
\teq{\hat{\boldsymbol{S}} = (\hat{Q}, \, \hat{U}, \, \hat{V}) \equiv (Q/I, \, U/I, \, V/I)}, 
using ratios of the polarization quantities of interest.   These will be employed 
at the recording stage when waves/photons exit the atmospheric slabs.
 
The total cross section in Eq.~(\ref{eq:sig_magThom}) can be expressed using 
the Stokes parameters.  This is best done using the normalized 
forms \teq{\hat{I}_i=1, \hat{Q}_i} and \teq{\hat{V}_i} for the incident photon, 
noting that in our coordinate description, \teq{\hat{U}_i=0} can be chosen without 
loss of generality.  If \teq{\theta_i = \arccos \mu_i} 
is the angle of the initial photon to the magnetic field direction \teq{\Bvechat}, then 
using the polar coordinate forms for the Stokes parameters in 
Eq.~(\ref{eq:Stokes_polar_def}),
and the electric field vector form in Eq.~(\ref{eq:Evec_polar}), one quickly 
derives the result
\begin{eqnarray}
   \sigma  & = & \sigt \biggl\{ \SigmaB (\omega ) \,\hat{I}_i
       + \dover{1}{2} \Bigl[ 1 - \SigmaB (\omega ) \Bigr]  \bigl( \hat{I}_i + \hat{Q}_i \bigr) \, \bigl( 1- \mu_i^2\bigr)  \nonumber\\[-5.5pt]
 \label{eq:sigma_tot_Stokes}\\[-5.5pt]
   && \hspace{100pt} + \DeltaB (\omega )\, \hat{V}_i \, \mu_i \biggr\} \quad , \nonumber
\end{eqnarray}
where the frequency dependence is encapsulated in two 
simple functions \teq{\SigmaB (\omega )} and \teq{\DeltaB (\omega )}
defined in Appendix B in Eqs.~(\ref{eq:SigmaB_def}) and~(\ref{eq:DeltaB_def}), 
respectively.  This clearly identifies the contributions of linear and circular 
polarizations to the scattering, to be elaborated upon in due course.
Eq.~(\ref{eq:sigma_tot_Stokes}) concurs with Eq.~(4) of \cite{Whitney-ApJS91} 
and Eq.~(2.26) of \cite{Barchas17}, both of which 
were derived from the polarization phase matrix analysis of \cite{Chou86}.

\newpage

\section{Monte Carlo Simulation}
 \label{sec:MC_technique}

\begin{figure}
\vspace*{-0pt}
\centerline{\includegraphics[width=9.0cm]{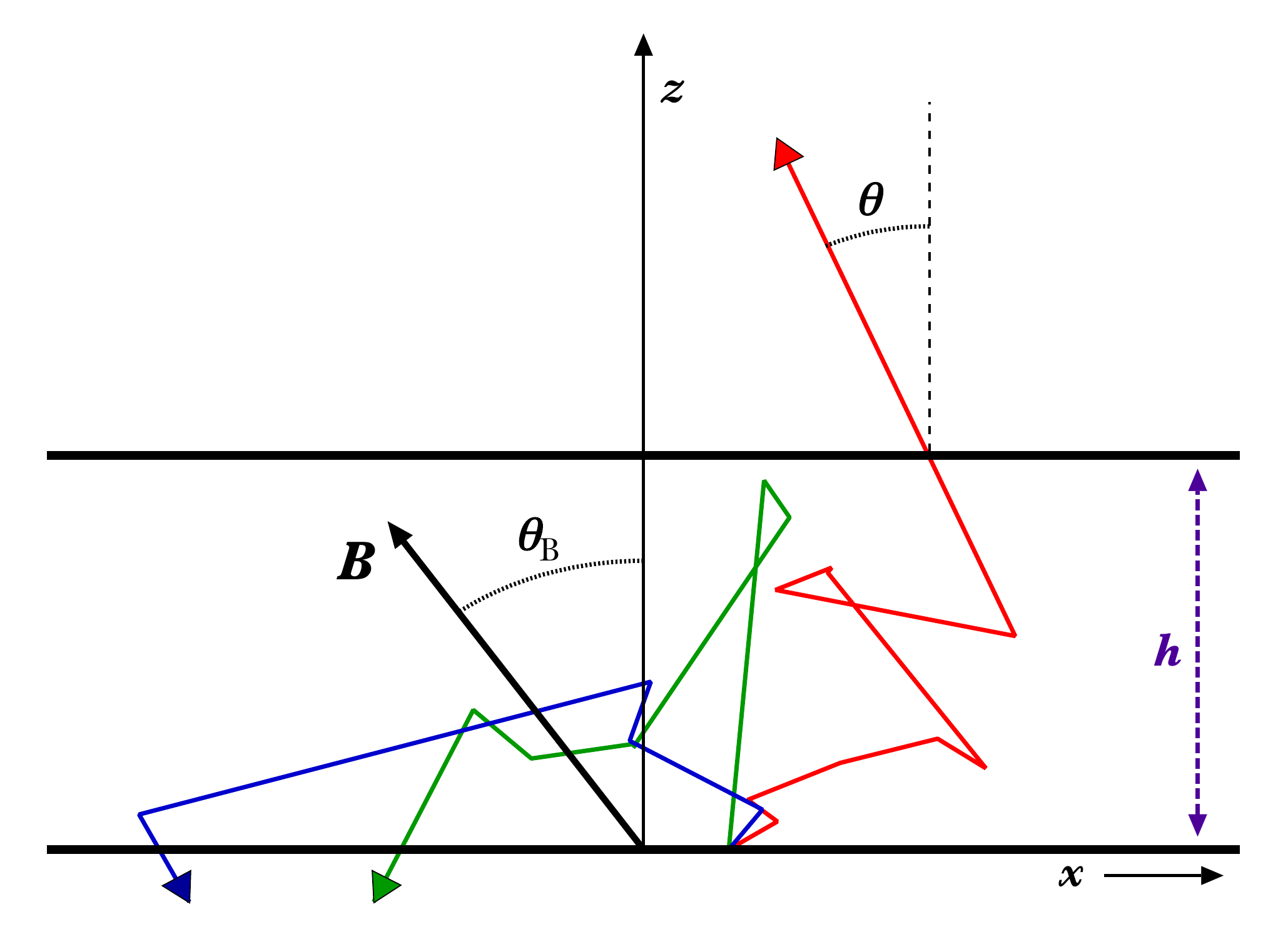}}
\vspace*{-5pt}
\caption{Simulation geometry for transfer of photons through an atmospheric slab 
of height \teq{h} with a
normal direction along the \teq{z} axis (zenith) and magnetic field in the \teq{x-z} plane,
at an angle \teq{\thetaB} to the local zenith.
The red trajectory (projected on the \teq{x-z} plane) corresponds to a photon
that scatters 10 times before exiting the top of the slab, reaching the observer.
The blue and green trajectories are for photons that scatter and eventually exit 
the bottom of the slab and are therefore unobservable; they represent a sizable majority 
of photons (see Section~\ref{sec:validation}) that are not included in the Stokes parameter data throughout.
\label{fig:slab_geom}}
\vspace{-20pt}
\end{figure}

This Section outlines the structure of the Monte Carlo simulation that has been developed 
to treat scattering transfer of polarized electromagnetic waves/photons in 
neutron star atmospheres.  
This paper will focus on magnetic Thomson transport, though the technique can easily 
capture electron-photon energy exchange in the full Compton process.  We will also 
restrict considerations here to locally uniform thin slabs.  The familiar
Thomson optical depth \teq{\taut} serves as the key parameter controlling emergent 
intensities, anisotropies and Stokes polarization parameters.  Note that the results 
presented in Section~\ref{sec:validation} are applicable also 
to stratified atmospheric slabs with the same values of \teq{\taut}, as long as 
the influence of temperature gradients on \teq{e-\gamma} energy exchange can be neglected.
A schematic diagram of the slab geometry and representative photon transfer is given in
Fig.~\ref{fig:slab_geom}.  The injection of photons in the simulation can be anywhere within the slab, 
though for our results in Section~\ref{sec:validation}, it will occur at 
the base of the slab, i.e. at \teq{z=0} in the diagram in Fig.~\ref{fig:slab_geom}
and closest to the center of the star.  

%
%

The Monte Carlo technique is computationally efficient: for intensive simulations in high 
opacity domains, the C++ code described herein normally takes several hours or less to 
run on a desktop computer with a multi-core CPU.  As the algorithm operates on a 
photon-by-photon basis, the code is parallelized.  The Monte Carlo method has been applied by 
\cite{Whitney-thesis89,Whitney-ApJS91,Whitney-ApJ91} in the context of white dwarf atmospheres,
for magnetar atmospheres by \cite{Bulik-1997-MNRAS,Niemiec-2006-ApJ}, 
and for magnetar coronae by \cite{Fernandez07,Nobili-2008-MNRAS,Zane-2011-AdSpR,Taverna17}.
It serves as a complementary approach to integro-differential equation 
radiation transfer methods \citep{Chandra60} that are common in the magnetar literature
\citep[e.g.,][]{Ozel-2001-ApJ,HoLai-2001-MNRAS}.  Flat 
spacetime is presumed when tracking polarization (\teq{\calEvec}) and propagation
(\teq{\boldsymbol{k}}) vectors, since the general relativistic influences can be captured 
with single redshift and field distortion parameters that apply uniformly throughout 
the thin slabs at particular surface locales.

\subsection{Photon Injection and Scattering}
 \label{sec:intensity}

The initial injection of photons at the bottom of an atmospheric layer will 
often (but not always) be distributed isotropically in intensity, i.e. \teq{I=}(const.). 
Assuming {\it flux isotropy}, the differential number of photons in an interval 
\teq{(\theta,\, \theta+d\theta)} in polar angle and \teq{(\phi, \,\phi+d\phi)} 
in azimuthal angle can be expressed as 
\begin{equation}
   \frac{dN}{d\theta d\phi} \;=\; f(\theta,\phi)
   \; =\; \dover{{\cal N}_i}{\pi} \, \cos\theta\sin\theta \quad .
 \label{eq:inject_dist_def}
\end{equation}
Here the total number of photons \teq{{\cal N}_i} 
passing through the surface where the injection occurs is obtained 
via an integration over all angles:
\begin{equation}
   \int_0^{\pi/2}\int_0^{2\pi} f(\theta,\phi) \, d\theta d\phi \; =\; {\cal N}_i \quad .
 \label{eq:Ndef}
\end{equation}
The integration over \teq{\theta} spans the interval \teq{[0, \pi /2]} because we 
consider only the radiation emerging from the hemisphere below the injection surface. 

To generate the initial propagation direction \teq{\kvechat_0} of photons at the 
base of the slab, the assumption of flux-weighted isotropy in Eq.~(\ref{eq:inject_dist_def})
will be imposed in most of the illustrative results of Sec.~\ref{sec:validation}.
Thus, for a particular choice of polar coordinates specified relative to the \teq{z} axis, 
\begin{equation}
   \kvechat_0 \; =\; \Bigl( \sin\theta_0\cos\phi_0, \, \sin\theta_0\sin\phi_0 ,\, \cos\theta_0 \Bigr)\quad .
 \label{eq:kvechat_zero_def}
\end{equation}
For this restrictive case of isotropy, the functional \teq{f(\theta,\phi)} applies, which is azimuthal 
symmetric.  Therefore it is analytically invertible in both polar and azimuthal angles, 
yielding familiar forms expressed as functions of the two pertinent 
random variates, \teq{\xi_{\theta}} and \teq{\xi_{\phi}}:
\begin{equation}
   \theta_0 \; =\; \dover{1}{2} \arccos \bigl( 2\xi_\theta-1\bigr)
   \quad ,\quad
   \phi_0 \; =\; 2\pi\xi_\phi \quad .
 \label{eq:isotropic_flux}
\end{equation}
This defines a uniform distribution weighted by the factor \teq{\cos\theta_0} that constitutes
the angle-dependent flux of photons through the base of the slab. The factor of \teq{1/2} in the
\teq{\theta_0} formula is introduced to render it appropriate for injection in the upward
hemisphere only.  Therefore, two random number choices are required to specify the direction of
the injected photon.  In some of our high
opacity simulation runs addressed in Section~\ref{sec:high_opacity_pol}, the assumption of
isotropy is relaxed via a routine adaptation of the injection algorithm. 

The choice of polarizations for injected photons is not unique, and as will become apparent, the
emergent polarization signatures for moderate opacity slabs will be dependent on the injection
choice.  In such circumstances, the scattering simulation will not yet have reached a truly
Markovian domain.   The only true zero
polarization injection is to randomly select all components \teq{\calEvechat_{\theta} =
\hat{\theta} \exp\{ i\varphi_{\theta} \} } and \teq{\calEvechat_{\phi} = \hat{\phi}
\exp\{ i\varphi_{\phi} \} } both in vector direction (\teq{\hat{\theta} , \hat{\phi} })
and in complex phase (\teq{\varphi_{\theta} , \varphi_{\phi} }), simultaneously isotropizing
\teq{\kvechat_0}. This then statistically generates zero for all the averages defining the
Stokes parameters in Eq.~(\ref{eq:Stokes_polar_def}), resulting in polarization isotropy 
on the Poincar\'e sphere.  

The distance \teq{s} a photon propagates between scatterings or before its first collision is
determined probabilistically using the total cross section \teq{\sigma}, which is a function of
polarization \teq{\calEvechat_i} and propagation vector \teq{\kvec_i}: see
Eq.~(\ref{eq:sig_magThom}). If the photon was initially at position \teq{\rvec_i}, then the
position at which it scatters is \teq{\rvec_f=\rvec_i+s \kvechat_i}. For a {\it uniform}
electron number density \teq{n_e}, the propagation distance is sampled according to Poisson
statistics: 
\begin{equation}
   \xi_s\; =\; \exp\left(-n_e\sigma s\right)
   \quad \Leftrightarrow\quad
   s \; =\; - \dover{\log \xi}{n_e\sigma} \quad .
 \label{eq:scatt_length_stat}
\end{equation}
Accordingly, \teq{\xi_s} is the random variate on the interval \teq{[0,1]} that depends directly
on the {\bf optical depth} \teq{\tau = n_e \sigma s}, yielding a mean free path \teq{\lambda =
1/(n_e\sigma )} for scattering. This form was actually used to compute the
trajectories illustrated in Fig.~\ref{fig:slab_geom}. For non-uniform electron densities, this
algorithm is easily adapted by working in optical depth space so that \teq{n_e s} is replaced by
an integral of \teq{n_e} over pathlength.

Once a scattering is determined to occur, the differential cross section must also be sampled
probabilistically in order to determine the new propagation direction \teq{\kvec_f} and
polarization \teq{\calEvec_f} after scattering.  This is accomplished by applying the 
{\it accept-reject method} to the suitably-normalized polarization-dependent differential 
cross section: 
\begin{eqnarray}
   p \bigl( \xi_{\theta f} ,\, \xi_{\phi f} \bigr) & \equiv & 
    \dover{8\pi}{3\sigma} \, \dover{d\sigma}{d\Omega_f} 
   \; =\; \dover{\sigt}{\sigma} \, \dover{ \calEvec_f \cdot \calEvec_f^*}{\vert \calEvec_i \vert^2}  \nonumber\\[-5.5pt]
 \label{eq:dsig_mag_acc_rej}\\[-5.5pt]
   & \equiv &  \dover{ \bigl( \kvechat_f \times \alphavec \bigr) \cdot  
       \bigl( \kvechat_f \times \alphavec^{\ast} \bigr)}{ \alphavec \cdot  \alphavec^{\ast} } \nonumber
\end{eqnarray}
Here \teq{\calEvec_f} is calculated using Eq.~(\ref{eq:calEf_eval}), or 
equivalently, \teq{\alphavec} is formed using Eq.~(\ref{eq:alpha_vec_def}).  
Observe that both \teq{\sigma} and \teq{d\sigma/d\Omega} are dependent 
on the initial polarization state and propagation direction.   The form on 
the second line of Eq.~(\ref{eq:dsig_mag_acc_rej}) is obtained simply 
from the ratio of Eqs.~(\ref{eq:dsig_magThom}) and~(\ref{eq:sig_mag_Thom_form}),
and so expressing its numerator using the Binet-Cauchy identity from vector analysis, 
it is quickly established that \teq{0 \leq p \bigl( \xi_{\theta f} ,\, \xi_{\phi f} \bigr) \leq 1} for all possible 
\teq{ \kvechat_f}.  Accordingly, \teq{p \bigl( \xi_{\theta f},\, \xi_{\phi f} \bigr) } is well posed 
for the accept-reject method on the unit cube, and it represents a scaled two-dimensional probability 
distribution in the variables \teq{\theta_f} and \teq{\phi_f}.  We therefore employ 
two uniform and independent random variates \teq{\xi_{\theta f}} and \teq{\xi_{\phi f}}, 
identified with the values of \teq{(1-\cos\theta_f)/2} and \teq{\phi_f/2\pi}, respectively:
\begin{equation}
   \theta_f \; =\; \arccos\left(2\xi_{\theta f}-1\right)
   \quad ,\quad
   \phi_f \; =\; 2\pi\xi_{\phi f} \quad .
 \label{eq:thetaf_phif_variates}
\end{equation}
The normalization condition for \teq{p} can be written
\begin{equation}
    \int_0^1 d\xi_{\theta f} \int_0^1 d\xi_{\phi f} \; p \bigl( \xi_{\theta f} ,\, \xi_{\phi f} \bigr) \; =\; \dover{2}{3}\quad ,
 \label{eq:2Dprob_dist_norm}
\end{equation}
since \teq{d\xi_{\theta f}\, d\xi_{\phi f} = d\Omega_f/4\pi}, and this gives an 
indication of good numerical efficiency of this protocol, given that
the normalization is not much inferior to unity.
The accept-reject method then selects an additional random variate \teq{\xi_p} 
to represent the probability function \teq{p}.  Thus the three random numbers
\teq{\xi_{\theta f},\, \xi_{\phi f}, \, \xi_p} identify a point \teq{P} within a 
rectangular prism volume that is bifurcated by the \teq{p( \xi_{\theta f} ,\, \xi_{\phi f})}
surface.  If \teq{P} lies below the surface, then the values of \teq{\xi_{\theta f}}
and \teq{ \xi_{\phi f}} are accepted, and both \teq{\kvec_f} and \teq{\calEvec_f} 
are then determined.  If, however, the point \teq{P} lies above the surface, 
then the selection is rejected, and the Monte Carlo decision process is 
initiated anew.  In this way, on average, the differential cross section 
for scattering is sampled representatively by the volume beneath the surface.

\section{Illustrative Results and Code Validation for Moderate Opacities}
 \label{sec:validation}

This Section outlines some basic results from the polarized radiation transport simulation, and
provides a level of validation via comparison with prior numerical analyses of the magnetic
Thomson problem.  Throughout, the magnetic field will be assumed uniform, though of different
orientations with respect to the slab normal: see  Fig.~\ref{fig:slab_geom}.  The photons are 
monochromatic at select frequencies, and their injection is both isotropic and an unpolarized 
mix of linear polarization states.  The zenith angle \teq{\theta_z} of the observer direction is a
suitable choice for the polar angle of the anisotropic emergent radiation.  The thickness of the
slab will be specified via an optical depth parameter. Photons can exit either through the top
of the slab at \teq{z>h}, escaping to an observer at infinity with their Stokes parameter
information being recorded, or through the bottom of the slab at \teq{z<0}, to be absorbed deep
in the atmosphere.  The loss of photons via absorption at \teq{z<0} turns out to be quite
significant at high optical depths, thereby limiting the computational efficiency of the
simulation.

Results will be presented for two magnetic field orientations, along the local zenith and
parallel to the slab surface.  These cases bracket the range of possibilities on the neutron
star surface, and one anticipates that as a neutron star rotates, the emergent polarization and
intensity signals will be a pulsing mix of results from these examples and those pertaining to
interstitial field orientations. The coordinate reference system will be chosen so that the
Stokes \teq{U} parameter is effectively zero, to within photon count statistics, with Stokes
\teq{Q} and \teq{V} parameters being the depicted polarization measures.  This simplification
can be maintained for arbitrary magnetic colatitudes by orienting one Cartesian coordinate axis
to coincide with a particular magnetic longitudinal plane. However, when extending to summing
over different longitudes, \teq{U} is no longer zero in general.

It is of interest to explore how the emergent Stokes measures vary with thickness/depth of the
slab, and to assess at what thickness the scattering/diffusion is saturated, for which
dependence on the photon injection angular and polarization distributions is minimal.  There is
no unique optical depth measure, since the scattering cross section depends on the photon
energy.  Here we will be guided by the choice of \cite{Whitney-ApJS91}, centered on two
preferred directions, namely parallel \teq{\parallel} and perpendicular \teq{\perp} to the
slab's field direction. Note that these labels are below applied to optical depths and must be
distinguished from the linear photon polarization state labels. For unpolarized (up) radiation,
the scattering cross section in Eq.~(\ref{eq:sig_perp_para_sum}) in Appendix~B yields optical
depths parallel to and orthogonal to the field \teq{\Bvec} of
\begin{eqnarray}
   \tau_{\parallel} & \!\! = \!\! & n_e h\, \sigma_{\rm up}(\theta_i =0^{\circ})
   \; =\; \taut \dover{\omega^2 (\omega^2+\wcyc^2)}{(\omega^2 - \wcyc^2)^2} \quad ,\nonumber\\[-5.5pt]
 \label{eq:tau_par_per_def}\\[-5.5pt]
   \tau_{\perp} & \!\! = \!\! & n_e  h\, \sigma_{\rm up}(\theta_i = 90^{\circ} )
   \; =\; \dover{\taut}{2} \left[1+  \dover{\omega^2 (\omega^2+\wcyc^2)}{(\omega^2 - \wcyc^2)^2} \right] \;\; .\nonumber
\end{eqnarray}
These can be applied to any field orientation.  Since the cross section is strongly frequency
dependent, fixing the value of \teq{\tau_{\parallel}} or \teq{\tau_{\perp}} provides a useful
path to comparing results for different frequencies.  Accordingly, this parameter appears in the
ensuing plots of this Section, as opposed to the familiar Thomson value \teq{\taut = n_eh
\sigt}. Furthermore, it enables direct comparison with results from the Monte Carlo simulation
of magnetic Thomson transport in \cite{Whitney-ApJS91,Whitney-ApJ91}.

While the framing of the discussion is centered on neutron star atmospheres, the results are
also potentially applicable to dynamic hard X-ray bursts in magnetar magnetospheres, where the
slabs can be considered to be localized portions of the outer layers of a Compton-thick cloud
contained in closed field line regions.  A caveat to this extension is that there are locales in
the magnetosphere where the photon energy and the cyclotron energy exceed 20 keV and the
magnetic Thomson restriction must be relaxed in favor of a full QED treatment. A second caveat
is that the cold electron gas approximation must also be relinquished in order to accurately
model magnetar bursts, thereby defining an extension that will be explored in future work.

\subsection{The Polar Case}
 \label{sec:mag_pole}

\begin{figure*}
 \begin{minipage}{17.5cm}
\vspace*{-10pt}
\centerline{\hskip 0pt \includegraphics[width=17.5truecm]{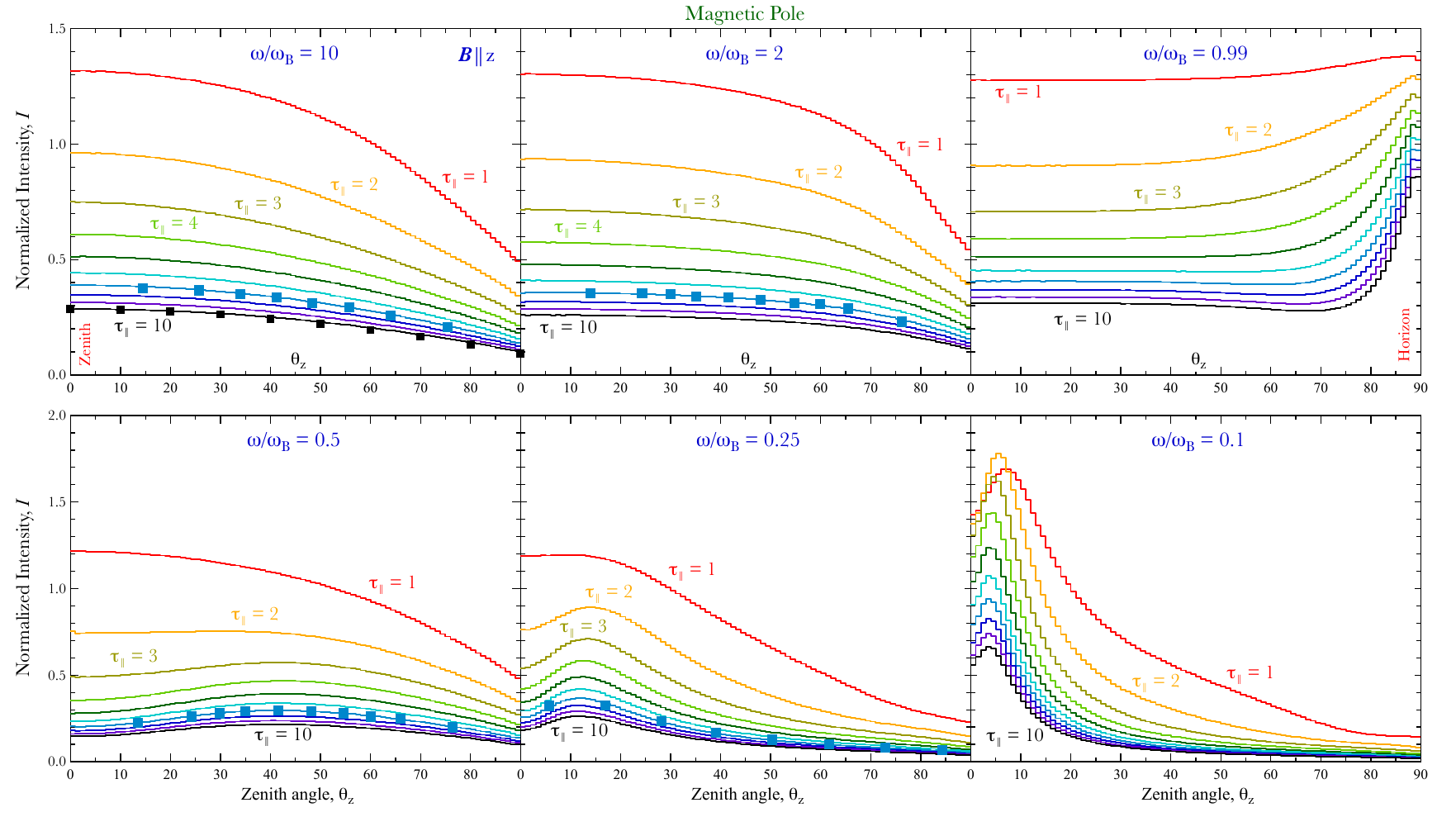}}
\vspace*{-5pt}
\caption{Angular distributions of intensity, as functions of the observer's zenith angle \teq{\theta_z}, 
for the case of \teq{\Bvec} being in the zenith direction along the slab normal, appropriate 
to the magnetic pole.  Each panel depicts results for one of six photon frequencies \teq{\omega},
specified in terms of the electron cyclotron frequency \teq{\wcyc = eB/m_ec}. The colored histograms are 
for different optical depth parameters \teq{\tau_{\parallel}}, as defined in Eq.~(\ref{eq:tau_par_per_def}),
ranging from \teq{\tau_{\parallel}=1} (red) to \teq{\tau_{\parallel}=10} (black), in unit increments;
larger \teq{\tau_{\parallel}} incurring higher loss rates below the atmospheric slab into the stellar interior.
The injection at the base of the slab was isotropic, with a net polarization of zero 
as a superposition of the \teq{\perp} and \teq{\parallel} linear polarization states with their 
particular Stokes parameters.  The distributions are integrated over azimuthal angles about the 
zenith, so that they represent intensities captured in entire conical sectors of angular width
\teq{\Delta \theta_z = 1^{\circ}}.  Simulation runs were for an injection of  \teq{{\cal N}_i=10^9} photons in
all cases, except for \teq{\omega /\wcyc = 0.1}, where  \teq{{\cal N}_i=10^8}. 
Aqua squares for the \teq{\omega/\wcyc = 0.25, 0.5, 2, 10} examples 
represent data from simulations of Whitney (1991a); see text.  The black squares for 
\teq{\omega /\wcyc = 10} are results 
adapted from Sunyaev \& Titarchuk (1985)  for non-magnetic Thomson transport, discussed 
in connection with Fig.~\ref{fig:B_parallel_x_polarize} below.
 \label{fig:B_parallel_z_intense}}
\end{minipage}
\vspace*{-4pt}
\end{figure*}

Fig.~\ref{fig:B_parallel_z_intense} displays the angular profiles of the emergent intensity
distribution at the magnetic pole, where \teq{\theta_z=0^{\circ}} is along the field, and the
horizon direction \teq{\theta_z=90^{\circ}} is perpendicular to \teq{\Bvec}. The distributions
(linear scale) are integrated over azimuthal angles about the zenith, and assigned to bins of
width \teq{1^{\circ}} in \teq{\theta_z}.  The normalization is determined by dividing by the
total number of photons injected, which was \teq{{\cal N}_i=10^9}, and the escape probability
from the slab is addressed in Fig.~\ref{fig:B_parallel_z_escape}. Profiles are displayed for
different slab thickness \teq{h} using the optical depth parameter \teq{\tau_{\parallel}} as a
proxy.  The angle-integrated intensity generally declines monotonically with increasing
\teq{\tau_{\parallel}} as the ability of photons to escape out of the top of the slab drops, and
more photons cross the slab base into the stellar interior.

The distributions in Fig.~\ref{fig:B_parallel_z_intense} were generated using a flux-weighted
isotropic injection of photons at the slab base, with equal numbers of linear polarizations
\teq{\perp} and \teq{\parallel} chosen.  The emergent distributions are sensitive to the
injection choice.  For example, the thesis results in Figs. 4.2 of \cite{Barchas17} illustrate
how even with isotropic injection, the angular profiles depend on the specification of an
unpolarized injection at the base, with isotropy of photon electric field vectors on the
Poincar\'e sphere yielding different intensity (and polarization) distributions from the
\teq{\perp/\parallel} mode parity choice adopted here, and also a plasma eigenmode parity.  The
differences are substantial for thin slabs with \teq{\tau_{\parallel}=1,2,3}.  Yet, the
\teq{\tau_{\parallel} \gtrsim 7} examples correspond to circumstances where the radiative
transfer has generally saturated in generating angular and polarization distributions. The
shapes of the intensity profiles are then approximately independent of \teq{\tau_{\parallel}}
and are only slightly sensitive to the injection choice; the exception to this is at low
frequencies \teq{\omega/\wcyc \lesssim 0.1} and around the resonant frequency.  We will expand
upon this injection nuance in Section~\ref{sec:high_opacity_pol}.

The six panels of Fig.~\ref{fig:B_parallel_z_intense} sample a representative array of photon
frequencies/energies straddling the cyclotron frequency, thereby identifying a diversity of
character for the angular profiles.  In this sequence, the Thomson optical depth spans a wide
range of values, with the \teq{\omega = 0.99\,\wcyc} example representing the resonant cyclotron
case.  While not depicted here, in all cases, the full angular distributions exhibited no
dependence on the longitudinal angle (azimuth) within statistical precision. For
\teq{\omega/\wcyc\gg 1}, when the character approaches that of an unmagnetized plasma, the
distributions possess a modest anisotropy in polar angle.  In this domain,
Eq.~(\ref{eq:tau_par_per_def}) indicates \teq{\tau_{\parallel}\approx \tau_{\perp}\approx
\taut}, yet the differential cross section contains anisotropy that permeates into the angular
profiles realized in the radiative transfer.   Remembering that these are intensity
representations, they include a \teq{1/\cos\theta_z} flux weighting factor.   Accordingly, 
the true light density anisotropy \teq{n_{\gamma}(\theta_z, \, \phi_z)} at the slab surface is 
proportional to \teq{I\cos\theta_z} and thus is skewed more markedly towards the field direction.

Below the cyclotron frequency, the anisotropy is more profound, evincing a strong collimation
aligned with the magnetic field; note that the reduction in solid angle about \teq{\Bvec}
imposes a modest decline very near \teq{\theta_z=0}. This collimation is the essence of the 
``pencil beam'' distributions familiar in historic studies of accreting X-ray pulsars
\citep[e.g.,][]{Gnedin74_AA,MB81_ApJ,Burnard1991ApJ}. For these \teq{\tau_{\parallel} \leq 10}
examples, the origin of this collimation is mostly due to a high percentage of \teq{\parallel} 
photons injected almost along \teq{\Bvec} emerging unscattered because of a markedly 
reduced cross section for these directions. This simulation bias is eliminated when the code is run in much
higher opacity domains, and the beaming becomes much more muted, as will become apparent in
Sec.~\ref{sec:high_opacity_pol}. For the low frequency, modest opacity cases, the angle
\teq{\theta_{\rm p}} of the peak of the profile appears to correlate with frequency as
\teq{\theta_{\rm p}\sim \omega/\wcyc}, a correlation that persists to lower frequencies than are
exhibited here in supplemental runs performed for \teq{0.025 \gtrsim \omega/\wcyc \gtrsim 0.01}.
This coupling is naturally expected from the interplay between the incident photon angle and
frequency present in the cross section for the \teq{\parallel} polarization (ordinary) mode, an
interplay illustrated in Fig.~\ref{fig:csect_mag}.

The intensity profile right in the cyclotron resonance is in striking contrast to those at other
frequencies.  Eq.~(\ref{eq:dsig_perp_para}) shows that relatively proximate to the cyclotron
resonance, spanning \teq{0.25 \lesssim \omega /\wcyc \lesssim 4}, scatterings into
\teq{\parallel} modes preferentially occur in the direction of \teq{\Bvec}.  The \teq{\perp}
mode does not have this bias.  Combined, they enhance the emergent intensity closer to the field
direction i.e. for smaller zenith angles.  The exception is right in the resonance, where the
non-resonant contribution to \teq{\parallel\to\parallel} scatterings is generally negligible. 
The scatterings are then azimuthally symmetric, with a modest bias 
\teq{\langle \sigma_{\perp} \rangle > \langle \sigma_{\parallel}\rangle} 
in the angle-averaged cross sections; see Eq.~(\ref{eq:sig_perp_para_sum}) or
Fig.~\ref{fig:csect_mag}.  This makes for a generally broad distribution of photon zenith angles
in density, that maps over to an intensity profile \teq{I(\theta_z)} that is moderately peaked
near the horizon, \teq{\theta_z = 90^{\circ}} due to the flux weighting factor.  This peaking is
diminished by non-resonant \teq{\parallel\to\parallel} conversions that help drive the escape of
light close to the field direction at other frequencies.

As a validation of the code, we can compare some of our angular distributions with those
obtained in \cite{Whitney-ApJS91}, who employed a different prescription for the magnetic
Thomson cross section from the electric field vector formalism adopted here.  Nevertheless, our
code should reproduce Whitney's anisotropies, and it does.  Only limited direct comparison is
possible, specifically exhibited here for frequencies \teq{\omega/\wcyc = 0.25, 0.5, 2, 10} and
\teq{\tau_{\parallel}=7}. Data are actually taken from figures in Whitney's thesis
\citep{Whitney-thesis89}, being binned rather broadly (\teq{\Delta\theta_z \sim 10^{\circ}}),
limited by the computational statistics available at the time. Whitney's intensity distributions
for the magnetic pole are normalized in a particular fashion, and so are adjusted to
approximately match the normalization generated here.  Therefore, angular profile shapes are the
available diagnostic, and the comparisons in Fig.~\ref{fig:B_parallel_z_intense} indicate strong
agreement. Simulation results were also compared for other select \teq{\omega /\wcyc} ratios
adopted by \cite{Whitney-thesis89,Whitney-ApJS91}, with comparable agreement, yet are not
illustrated here. \cite{Whitney-thesis89} also produced an array of results for low optical
depths \teq{\tau_{\parallel}=0.1} that more directly capture the information imprinted by the
scattering cross section.  Angular profiles including polar plot diagrams were generated for
\teq{\tau_{\parallel}=0.1} cases using our code, and replicated results in
\cite{Whitney-thesis89,Whitney-ApJS91} with good precision and without exception; again, these
are not depicted here.

\begin{figure*}
 \begin{minipage}{17.8cm}
\vspace*{-10pt}
\centerline{\hskip 0pt \includegraphics[width=17.8truecm]{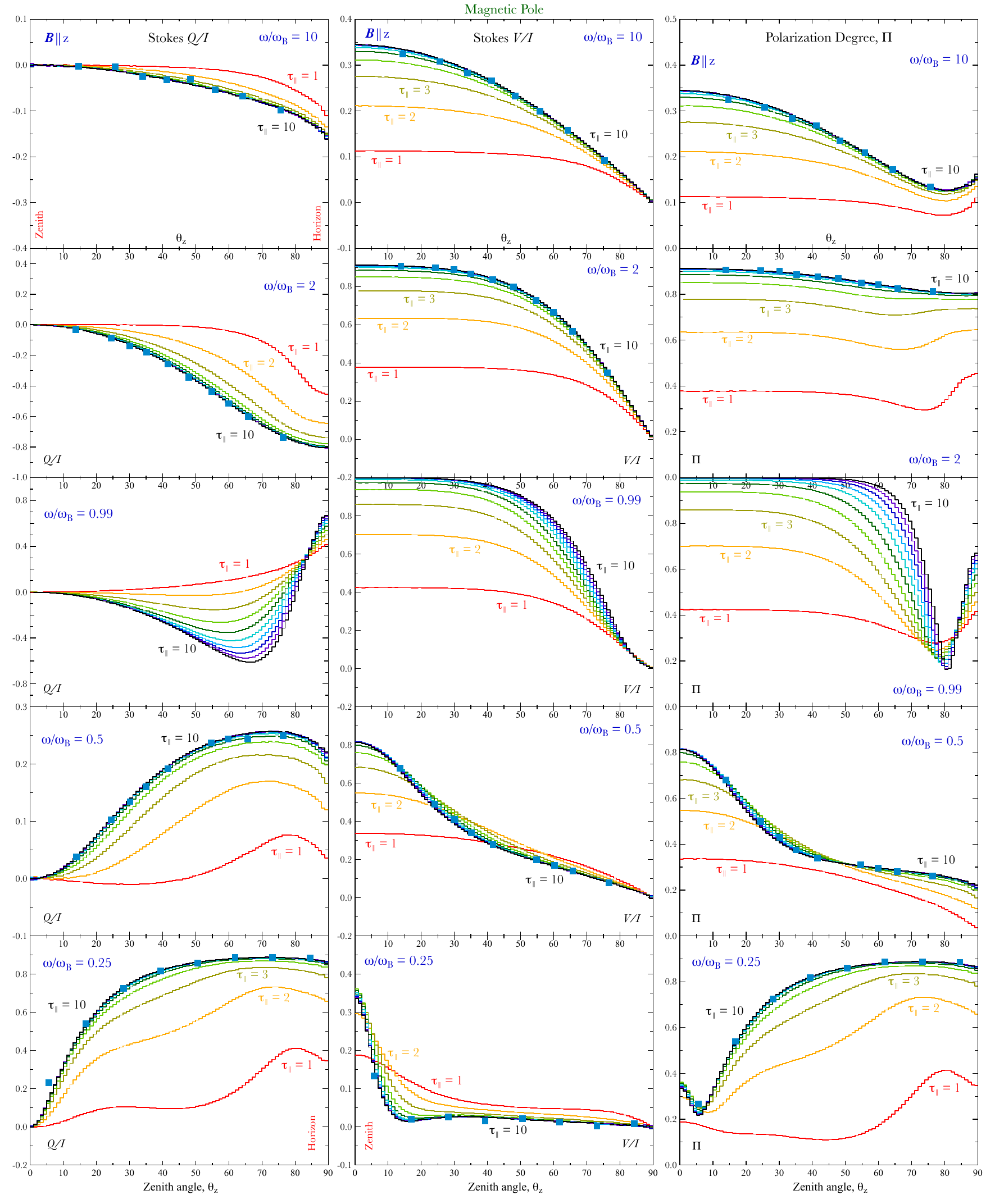}}
\vspace*{-5pt}
\caption{A companion plot for Fig.~\ref{fig:B_parallel_z_intense} generated using the same 
simulation runs.   Angular distributions 
of Stokes parameters \teq{Q, V} and the polarization degree \teq{\Pi}, 
as functions of the observer's zenith angle \teq{\theta_z}, again
for the ``polar'' case of \teq{\Bvec} being in the zenith direction along the slab normal.  
Each row depicts results for one of five photon frequencies \teq{\omega},
specified in terms of the electron cyclotron frequency \teq{\wcyc = eB/m_ec}. The colored histograms are 
for different optical depth parameters \teq{\tau_{\parallel}}, coded as in Fig.~\ref{fig:B_parallel_z_intense}.
The polarization angular profiles approximately saturate at large \teq{\tau_{\parallel}}.
The injection at the base of the slab was isotropic and unpolarized as for Fig.~\ref{fig:B_parallel_z_intense}.
The distributions are again integrated over azimuthal angles about the 
zenith, applying to conical sectors of angular width
\teq{\Delta \theta_z = 1^{\circ}}.  Simulation runs were for an injection of  \teq{{\cal N}_i=10^9} photons. 
Aqua squares represent data from simulations of Whitney (1991); see text.  
 \label{fig:B_parallel_z_polarize}}
\end{minipage}
\end{figure*}

The polarization properties corresponding to five of these intensity panels are exhibited in
Fig.~\ref{fig:B_parallel_z_polarize}, namely for \teq{\omega/\wcyc = 0.25, 0.5, 0.99, 2, 10},
again with impressive statistics because the injection was for \teq{{\cal N}_i=10^9} photons. 
These signatures are principally the Stokes Q (linear polarization) and V (circular
polarization) parameters, both normalized to the emergent intensity \teq{I}, as functions of
observer zenith angle \teq{\theta_z}.  At the right is the corresponding degree of polarization
\teq{\Pi} as defined in Eq.~(\ref{eq:poldeg_def}), and this trio displays a wealth of
observational signatures. Statistically, \teq{U} is essentially zero in the simulations for all
\teq{\theta_z} bins other than the noisy \teq{0^{\circ}} and \teq{180^{\circ}} bins, which are
subject to small number statistics.  Accordingly, only two of the three polarization quantities
are independent. Approximate convergence of the polarization measures to fixed angular profiles
is generally only realized for \teq{\tau_{\parallel}\gtrsim 7} in the examples shown.  We note
again that runs at \teq{\omega /\wcyc \lesssim 0.2} do not saturate at \teq{\tau_{\parallel}=10}
due to the large disparity in cross sections between the two linear polarization states. This
occurs only for much larger values of \teq{\tau_{\parallel}}, for which run times increase
dramatically.  Note also that the results do not saturate for \teq{\omega /\wcyc = 0.99} at high
zenith angles, in this case due to the complicated interplay of the circular polarization
characteristics.

The coordinate choice combined with a viewing perspective in the general direction of
\teq{\Bvec} dictate generally positive values for \teq{V/I}. This follows from the general
preponderance of circular polarization mode conversions \teq{-\to +} near to the field
direction, inferred from Eq.~(\ref{eq:mode_convert_circular}), combined with \teq{V>0} for
\teq{+} helicity in these directions, deducible from the circular polarization eigenvectors in
Eq.~(\ref{eq:plus_minus_polarvec}). A prominent feature of the plots is that substantial
circular polarization emerges for \teq{\omega/\wcyc \gtrsim 0.5} when \teq{\tau_{\parallel}=10},
except when the viewing angle is orthogonal to the field (horizon). In particular, \teq{\vert
V/I\vert} is maximized when looking along the field. This is expected since circular
polarization states are the eigenmodes of propagation along the field in a magnetized plasma
\citep[e.g.,][]{Canuto71PRD}.  An accompanying signature is that the linear polarization
\teq{\Pi_{\rm lin}\approx \vert Q/I\vert} is zero along \teq{\Bvec} in general, and strong
perpendicular to the field when either \teq{\omega/\wcyc \gtrsim 0.5} or \teq{\omega/\wcyc
\lesssim 0.25}.   These properties are direct consequences of the scattering cross sections,
which evince strong circular polarization and small linear polarization along \teq{\Bvec}, with
this bias reversed transverse to the field: see Eqs.~(\ref{eq:sig_perp_para_sum})
and~(\ref{eq:sig_pm}) in Appendix B.

In terms of the variation with frequency,  \teq{\vert V/I\vert} is generally quite large in the
window \teq{0.5 \lesssim \omega/\wcyc \lesssim 2}, straddling the cyclotron resonance.  The
origin of this is the comparative strength of the gyrational motion of an electron induced by
the incoming wave, so that the curl term in Eq.~(\ref{eq:alpha_vec_def}) contributes
significantly to the overall value of \teq{\alphavec}, i.e., the acceleration vector.  A
consequence of this is a relative balance between scatterings into the \teq{\perp} and
\teq{\parallel} states, a balance maximized at \teq{\omega /\wcyc = 1/\sqrt{3}} that seeds
\teq{\vert Q/I\vert} dropping to values below 0.3; see Fig.~\ref{fig:csect_mag}. At high
frequencies \teq{\omega/\wcyc \gtrsim 10}, the scattering is essentially non-magnetic and both
the circular and linear polarization measures are smaller, on average. At small frequencies
\teq{\omega/\wcyc \lesssim 0.25}, the large disparity in opacities between these two linear
polarization states yields a rapid rise in emergent \teq{\vert Q/I\vert} as \teq{\omega}
declines. This is accompanied by a drop in the circular polarization that is retarded somewhat
when viewing along \teq{\Bvec}, as exemplified in the \teq{\omega/\wcyc = 0.25} illustration.
Also notable is that the sign of \teq{Q/I} changes from negative to positive as the frequency
drops through the cyclotron resonance, character persisting for all lower frequencies. 
This is a consequence of the frequency dependence of the balance between scatterings of
\teq{\perp} and \teq{\parallel} states, which drives a change in the relative apportionment of
polar \teq{\hat{\cal E}_{\theta}} and toroidal \teq{\hat{\cal E}_{\phi}} field components.

The polarization degree column of panels on the right of Fig.~\ref{fig:B_parallel_z_polarize}
reveals a rich behavior with both zenith angle and frequency that can afford significant
diagnostic potential for X-ray polarimeters, which nominally are not expected to detect circular
polarization, instead measuring \teq{Q, U} and \teq{\Pi}. Extremely strong \teq{\Pi} emerges
around the cyclotron frequency, near 100\% for \teq{\theta_z < 50^{\circ}}, and polarization
degrees generally above 50\% for sub-cyclotronic photon frequencies around \teq{\omega \sim
\wcyc/2}.  In contrast, in the ``non-magnetic domain'' of \teq{\omega /\wcyc =10}, lower values
of \teq{0.2 \lesssim \Pi \lesssim 0.4} result. The strong evolution of linear and total
polarization degree in the frequency range straddling the cyclotron resonance signals the
potential for these signatures to be powerful diagnostics on the emission environs in neutron
stars of much lower magnetization than magnetars.

The last feature of Fig.~\ref{fig:B_parallel_z_polarize} to note is the comparison of \teq{Q/I},
\teq{V/I} and \teq{\Pi} with the results displayed in \cite{Whitney-thesis89}.  As with
intensity, this was possible for \teq{\tau_{\parallel}=7} and \teq{\omega/\wcyc = 0.25, 0.5, 2,
10}, again revealing a satisfying agreement for all three polarization quantities. As noted
above, the binning of Whitney's data is coarse in zenith angle, dictated by statistics limited
by the contemporaneous computational capability. Given that
\cite{Whitney-thesis89,Whitney-ApJS91} modeled the magnetic Thomson transfer in a manner
somewhat different from the electric field vector protocol here, using Jones matrix cross
section evaluations, the favorable comparison of three Stokes quantities \teq{I, Q, V} serves as
an important code validation for the magnetic polar case.
We note that insightful comparison with the Monte Carlo models of 
\cite{Fernandez07,Nobili-2008-MNRAS,Fernandez11}
that treat scattering of linearly-polarized photons only in the cyclotron resonance 
is not possible.

\begin{figure*}
 \begin{minipage}{17.8cm}
\vspace*{-10pt}
\centerline{\hskip 0pt \includegraphics[width=17.8truecm]{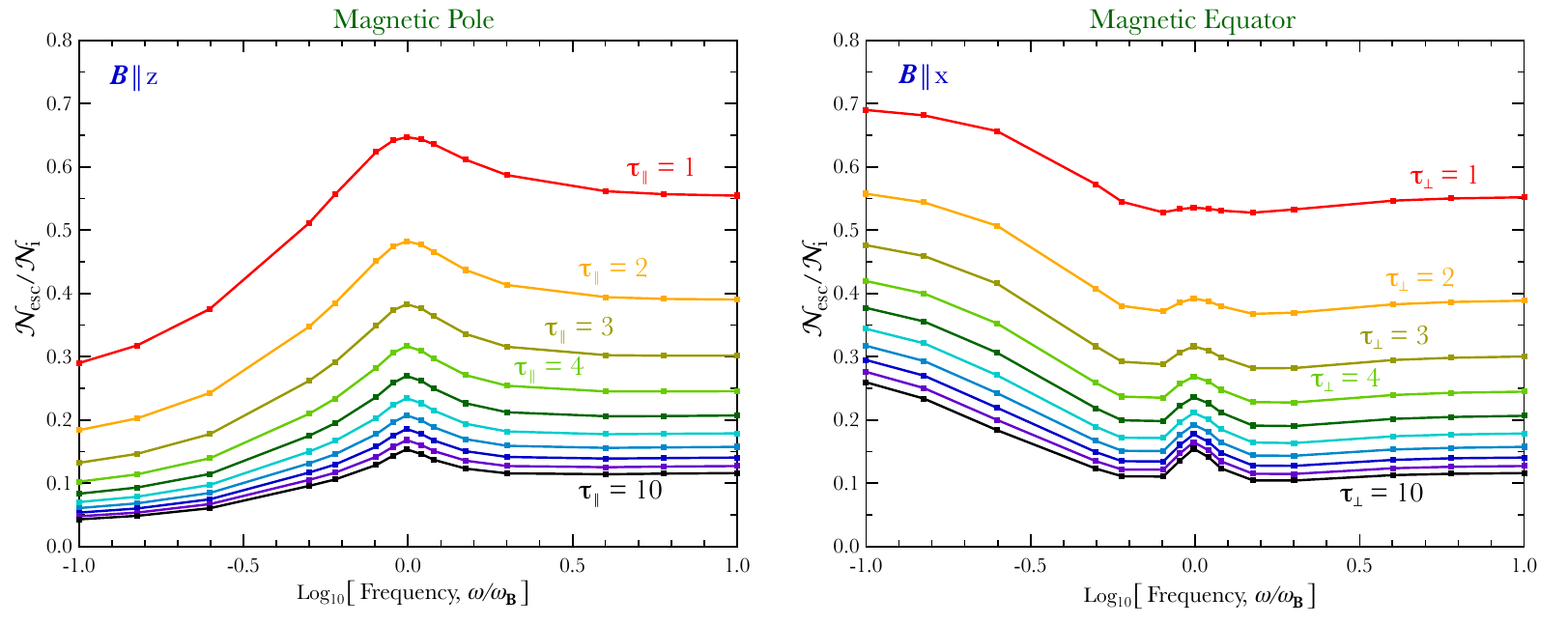}}
\vspace*{-5pt}
\caption{Outwards escape probabilities, \teq{{\cal N}_{\rm esc} / {\cal N}_i}, the ratios of the number 
of photons \teq{{\cal N}_{\rm esc}} emerging at the top of the slab to the 
number \teq{{\cal N}_i} injected at its base.  These are formed via integration over the 
angular distributions in Fig.~\ref{fig:B_parallel_z_intense}.
On the left is the magnetic polar case where the field is normal to the slab and in the 
zenith direction, and on the right is the magnetic equator case, for which \teq{\Bvec}
is oriented parallel to the slab boundaries (horizon); these are respectively
\teq{\thetaB=0} and \teq{\thetaB=\pi /2} in Fig.~\ref{fig:slab_geom}.
The dots at discrete frequencies represent data for runs with \teq{{\cal N}_i=10^9} photons, 
and they are joined by straight lines to guide the eye.  The expected monotonic decline of the 
escape probabilities with \teq{\tau_{\parallel}} and \teq{\tau_{\perp}} is obvious.  The frequency 
dependence for the two cases is discussed in the text.
 \label{fig:B_parallel_z_escape}}
\end{minipage}
\end{figure*}

\subsection{Slab Escape Probabilities}
 \label{sec:slab_escape} 

The construction of these simulation runs guarantees a monotonic decline with
\teq{\tau_{\parallel}} of the total numbers of photons escaping through the top of the slab. 
This naturally arises due to an increase in the cumulative probability that photons will diffuse
deep into the interior of the atmosphere as net opacity increases, emerging from the bottom of
the slab where they are injected. If one divides the intensities by the solid-angle-integrated
net intensity, one arrives at the ratio of the number of photons \teq{{\cal N}_{\rm esc}}
emerging from the top of the slab to the number \teq{{\cal N}_i} injected at its base.  This
serves as a measure of efficiency of the simulation, i.e. the capturing of useful polarization
data.  This ratio \teq{{\cal N}_{\rm esc} / {\cal N}_i}, the outwards escape probability, is
plotted in Fig.~\ref{fig:B_parallel_z_escape} as a function of the scaled photon frequency
\teq{\omega/\wcyc}, for the set of \teq{\tau_{\parallel}} values presented in the magnetic polar
simulations in Figs.~\ref{fig:B_parallel_z_intense} and~\ref{fig:B_parallel_z_polarize}. This
information is also presented there for the magnetic equator case that will be addressed in
detail shortly.

The dots in this figure present results for runs with \teq{{\cal N}_i=10^9} photons.  A
ubiquitous monotonic decline with \teq{\tau_{\parallel}} is apparent.  For both polar and
equatorial cases, the escape probability \teq{{\cal N}_{\rm esc} / {\cal N}_i} varies with
photon energy/frequency just as the differential cross section does. For the magnetic polar case
on the left, the escape ratio peaks at the cyclotron frequency and is particularly low when
\teq{\omega \ll \wcyc}.  The peaking is a consequence of the scattering cross section for the
two linear polarization modes being similar at the cyclotron frequency, influenced also to some
extent by the actual angular dependence of the differential cross section \teq{d\sigma
/d\Omega}.  The number of scatterings per photon is still modest since scaling the runs by
\teq{\tau_{\parallel}} essentially moderates the enhanced opacity at the cyclotron resonance. At
highly-sub-cyclotronic frequencies, the scattering conversions \teq{\perp\to \parallel}, while
comparatively rare, do actually arise in the runs since fixing \teq{\tau_{\parallel}} as defined
in Eq.~(\ref{eq:tau_par_per_def}) effectively guarantees them.  Subsequently,
\teq{\parallel\to\parallel} scatterings enhance and dominate the opacity, thereby biassing the
escape of photons towards the lower boundary of the slab and reducing the \teq{{\cal N}_{\rm
esc} / {\cal N}_i} ratio.

For the magnetic equatorial case on the right of Fig.~\ref{fig:B_parallel_z_escape}, for which
we use \teq{\tau_{\perp}} as defined in Eq.~(\ref{eq:tau_par_per_def}) to tag the optical depth,
the general dependence with frequency is quite different from that for the polar runs.
Noticeably, the escape probability/efficiency increases in \teq{\omega /\wcyc \ll 1} domains. 
This too is a consequence of the action of prolific \teq{\parallel\to\parallel} scatterings
subsequent to \teq{\perp\to \parallel} mode conversions.  However, while the polar case samples
escape somewhat aligned to the field direction, the equatorial case preferentially selects the
complementary directions, those that are highly oblique to \teq{\Bvec}.  Accordingly it captures
the signals contributed by diffusion away from the field direction, and for \teq{\omega /\wcyc
\ll 1} this constitutes considerably larger numbers of photons that underpin the inhibition of
the \teq{{\cal N}_{\rm esc} / {\cal N}_i} ratio when \teq{\Bvec} is along the zenith, i.e. for
the polar case at left.

We remark that the detailed values of \teq{{\cal N}_{\rm esc} / {\cal N}_i} will depend on the
injection conditions, so that different values will be realized if the injection at the slab
base is either anisotropic or polarized.  Nonetheless, one may anticipate that the general shape
of the escape probability curves will be somewhat similar to those depicted in
Fig.~\ref{fig:B_parallel_z_escape}.

\subsection{Equatorial Atmospheric Zones}
 \label{sec:mag_equat}

\begin{figure*}
 \begin{minipage}{17.8cm}
\vspace*{-10pt}
\centerline{\hskip 0pt \includegraphics[width=17.8truecm]{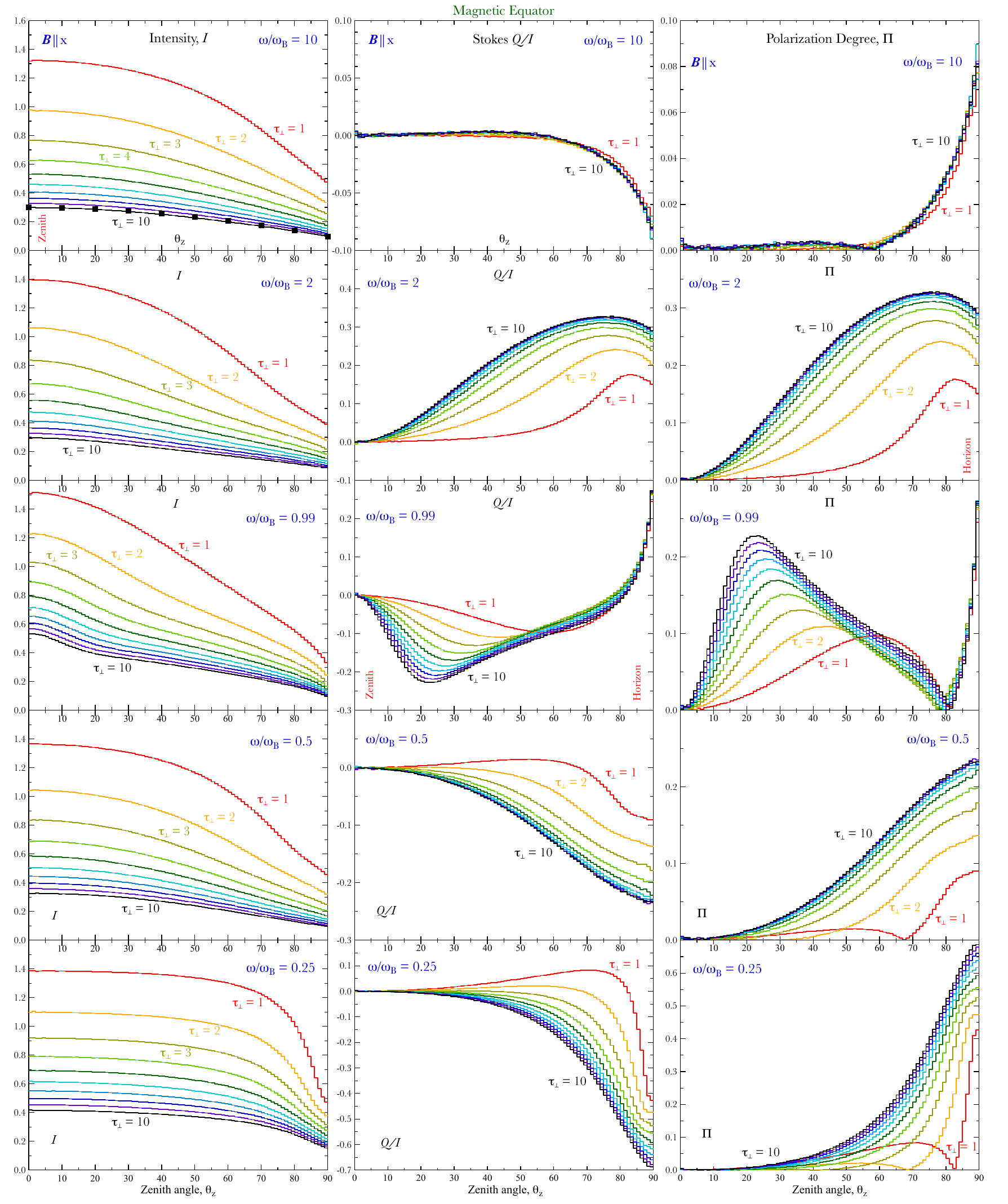}}
\vspace*{-5pt}
\caption{Angular distributions for intensity \teq{I}, Stokes parameter \teq{Q} and the 
polarization degree \teq{\Pi}, as functions of the observer's zenith angle \teq{\theta_z}, 
now for the ``equatorial'' case of \teq{\Bvec} being parallel to the slab surface.  
Each row depicts results for one of five photon frequencies \teq{\omega},
specified in terms of the electron cyclotron frequency \teq{\wcyc = eB/m_ec}. The 
colored histograms are for different optical depth parameters \teq{\tau_{\perp}}, as 
defined in Eq.~(\ref{eq:tau_par_per_def}).  The angular profiles approximately saturate 
at large \teq{\tau_{\perp}}.  The injection at the base of the slab was isotropic and 
unpolarized as for the polar case.  The distributions are again integrated over azimuthal 
angles about the zenith, applying to conical sectors of angular width \teq{\Delta \theta_z = 1^{\circ}}.  
Simulation runs were for an injection of  \teq{{\cal N}_i=10^9} photons.  The black 
squares for \teq{\omega /\wcyc = 10} are intensity results (suitably scaled) from 
Sunyaev \& Titarchuk (1985)  for non-magnetic Thomson transport.
 \label{fig:B_parallel_x_polarize}}
\end{minipage}
\end{figure*}

The emission anisotropy and polarization characteristics depend strongly on the orientation of
the magnetic field to the slab normal.  To illustrate this and provide a contrast to the polar
case in Sec.~\ref{sec:mag_pole}, an ``equatorial'' example with the field aligned parallel to
the planar slab boundaries (\teq{\thetaB=\pi /2}) is depicted in
Fig.~\ref{fig:B_parallel_x_polarize} \citep[note that an intermediate \teq{\thetaB=\pi /4} case
is explored in][]{Barchas17}. The simulations generating these distributions for
\teq{\thetaB=\pi /2} also produced the escape probability results displayed on the right of
Fig.~\ref{fig:B_parallel_z_escape}.  Since the distributions are integrated over the azimuthal
angle \teq{\phi_z} about the zenith direction, the circular polarization \teq{V} in this set-up
is statistically equal to zero, and therefore is not displayed in the figure; non-zero \teq{V}
appears when particular \teq{\phi_z} values are selected.  Since \teq{U} is effectively zero due
to the choice of coordinates, the Stokes parameter \teq{Q} (center column) and polarization
degree \teq{\Pi \approx \vert Q/I\vert} (right) in Fig.~\ref{fig:B_parallel_x_polarize} provide
essentially redundant representations of the same simulation output information for each
\teq{\omega/\wcyc} row.

The intensity profiles in the left column monotonically decline with increasing \teq{\theta_z}
for all photon frequencies \teq{\omega} and optical depths \teq{\tau_{\perp}}.  While this
replicates the general behavior for \teq{\omega/\wcyc > 1} evident in
Fig.~\ref{fig:B_parallel_z_intense}, it differs vastly with the magnetic polar case at
sub-cyclotronic frequencies \teq{\omega /\wcyc <1}.  The reason for this disparity in character
is simply that for this equatorial example, viewing angles generally do not coincide with the
field direction and so the zenith azimuth \teq{\phi_z} integration acts to convolve information
from the differential cross section at mostly large angles relative to the field direction
\teq{\Bvec}.  Geometrically, one can isolate photon directions almost parallel to the field by
selecting particular \teq{\phi_z} azimuths with \teq{\theta_z\sim 90^{\circ}}, and then the
intensity profile does possess more variation with \teq{\omega /\wcyc}, though generally the
fluxes  of photons emergent through the upper surface of the slab are then low since scattering
into high \teq{\theta_z} directions is improbable, as is evident in the Figure's intensity
panels. Thus, the intensity information in Fig.~\ref{fig:B_parallel_x_polarize} can be described
as essentially ``non-magnetic'' in character.

The top panel in Fig.~\ref{fig:B_parallel_x_polarize} illustrates that the linear polarization
\teq{Q/I} is very small for zenith angles less than around \teq{55^{\circ}}, and then increases
somewhat  as viewing angles more or less align with \teq{\Bvec}, but nevertheless remain small:
\teq{\Pi_{\rm max} < 10}\%.  This \teq{\omega \gg \wcyc} example is approximately a non-magnetic
regime.  An extensive analysis of polarization signatures from slabs for classical unmagnetized
Thomson scattering was presented by \cite{ST85}, hereafter ST85, for the context of accretion
disks near black holes.  Fig.~4 of that paper indicates that for high optical depths, the
intensity is maximized along the slab normal, and monotonically declines to around 1/3 of this
maximum at the horizon.  This is very close to the behavior exhibited here in the top left panel
of Fig.~\ref{fig:B_parallel_x_polarize}, and also that for \teq{\omega /\wcyc = 10} in
Fig.~\ref{fig:B_parallel_z_intense}. The black squares exhibited in both the pertinent figure
panels are data for the \teq{\tau=10} curve illustrated in Fig.~4 of ST85, scaled by a constant
factor so that the zero zenith angle values coincide with our simulation data.  This excellent
agreement with our \teq{\tau_{\perp}=10} results when \teq{\omega\gg \wcyc} is expected since
then the magnetic differential cross section in Eq.~(\ref{eq:diff_csect}) approaches the
familiar non-magnetic one.

Turning now to the polarization comparison, Fig.~5 of ST85 demonstrates that in the absence of a
magnetic field, when \teq{\tau=10} the \underline{linear} polarization degree is zero along the
slab normal, and rises monotonically to a value of around 11-12\% when viewing almost along the
surface. While this \teq{\Pi_{\rm max}} maximum is somewhat higher than the
\teq{\tau_{\perp}=10} magnetic equatorial result at \teq{\theta_z=90^{\circ}} in
Fig.~\ref{fig:B_parallel_x_polarize}, it is somewhat lower than the polar \teq{\Pi_{\rm max}}
value along the horizon in Fig.~\ref{fig:B_parallel_z_polarize}, where \teq{V=0} and the signal
is linearly polarized; the average of our polar and equatorial evaluations is within around 9\%
of the ST85 value at \teq{\theta_z=90^{\circ}}. Moreover, for intermediate zenith angles,
\teq{50 < \theta_z < 90^{\circ}} the linear polarization degrees at \teq{\tau =10} from the
\cite{ST85} figure consistently lie between our polar and equatorial simulation values for
\teq{\vert Q/I\vert}, all exhibiting the same monotonic increase with \teq{\theta_z} towards the
slab horizon.

A material difference between our results and the non-magnetic ones of \cite{ST85} is that here
there is significant emergent circular polarization \teq{V/I} when \teq{\Bvec} is not aligned
parallel to the slab surface.  A consequence of resonant circulation of the scattering electron,
this non-zero \teq{V/I} bolsters the net polarization degree signal and introduces departures
from monotonic trends of \teq{\Pi} with \teq{\theta_z}: see
Fig.~\ref{fig:B_parallel_z_polarize}.  Inspection of the two directional opacity measures in
Eq.~(\ref{eq:tau_par_per_def}) indicates departures from the Thomson value \teq{\taut} by
1.5--3\% at \teq{\omega = 10\wcyc}, so that magnetic influences are still present. A more
precise comparison between the magnetic Comptonization polarization results here and the
non-magnetic analog in ST85 is thus best performed at higher frequencies.  Accordingly, we ran
simulations in both polar and equatorial cases for \teq{\omega /\wcyc = 10^2}, and found
excellent agreement between our angular distributions for intensity and \teq{\Pi_{\rm lin}
\equiv \vert Q/I\vert} at \teq{\tau_{\perp}=10} and the \teq{\tau = 10} ones in Figs.~4 and~5 of
ST85. We also observed that in the polar case, \teq{\vert V/I\vert < 0.04} at this much higher
frequency, a consequence of the two circular polarization cross sections coalescing when
\teq{\omega \gg\wcyc}: see Eq.~(\ref{eq:sig_pm}) in Appendix B.

For the more magnetic equatorial cases with \teq{\omega /\wcyc  \leq 2}, the intensity profiles
at \teq{\tau_{\perp} = 10} largely resemble those at super-cyclotronic frequencies in the polar
examples of Fig.~\ref{fig:B_parallel_z_intense}.  This again is due to the higher zenith
(horizon) emergences requiring a last scattering very near the slab surface, an improbable
occurrence.  The zenith angle distributions for Stokes \teq{Q} for the horizon field orientation
look very different from those of the polar case. Their relative behavior is ascribed to the
fact that the polar and equatorial cases preferentially sample complementary perspectives
relative to the field direction at any chosen zenith angle.  The \teq{Q/I} values at \teq{\omega
/\wcyc = 2, 0.5, 0.25} in Figs.~\ref{fig:B_parallel_z_polarize}
and~\ref{fig:B_parallel_x_polarize} are of opposite sign, and so if one sums them, relatively
small net \teq{Q/I} results, albeit not exactly zero.  Intuitively one expects zero net linear
polarization from an isotropic superposition of magnetic field orientations, and the combination
of the polar and equatorial configurations is the first stage of constructing such a
superposition. The exception to this ``cancellation'' is provided by the resonant case
\teq{\omega /\wcyc = 0.99}, where the scattering angle summations conspire to give mostly
negative \teq{Q/I} at \teq{\theta_z < 80^{\circ}} for both polar and equatorial field
orientations, and positive \teq{Q/I} for near-horizon viewing.

A nuance that needs brief attention concerns the co-adding of the polarization information from
all slab exit azimuths \teq{\phi_z} for the illustrations of this Section. For the polar case in
Fig.~\ref{fig:B_parallel_z_polarize}, results from all \teq{\phi_z} are statistically identical
due to the azimuthal symmetry, so the addition just improves the simulation data statistics.  In
contrast, for non-zenith field orientations such as for the equatorial depictions in
Fig.~\ref{fig:B_parallel_x_polarize}, the Stokes \teq{I, Q, V} vary with azimuth \teq{\phi_z}
and so the summations represent a mixing of azimuthal information.  Note that for a particular
observer detecting photons from a particular point on the stellar surface, individual values of
{\it both} \teq{\theta_z} and \teq{\phi_z} are selected for a particular ray propagating in
curved spacetime to infinity.

A comparison of our equatorial slab results with those from the same field orientation in 
\cite{Whitney-ApJ91} is not as straightforward as for the polar field case, since she presented
distributions for particular azimuths \teq{\phi_z}.  In doing so, we focus on \teq{\omega /\wcyc
=0.5, 0.25} values, for which Whitney presented results in Figs.~5 and~6 of her paper,
respectively. Therein, she selected azimuthal angles \teq{\phi_z = 0^{\circ}, 90^{\circ},
180^{\circ}} to present her results.  It was not clear what ranges of \teq{\phi_z} these
constituted. Nor is it apparent what injection distribution of angles and polarizations of
photons were assumed in \cite{Whitney-ApJ91}.  Since the results presented therein were for a
low opacity, \teq{\tau_{\perp}=3}, the Stokes parameter signals are generally sensitive to the
injection information; this was evident in our various exploratory simulation runs.  It was also
unclear what was the statistical quality of the results presented in Figs.~5 and~6 of 
\cite{Whitney-ApJ91}, no doubt diminished by the sub-selection of particular azimuths, and
muting details through coarse \teq{\theta_z} binning.

To best approximate her cases, we collected subsets of the data presented for the \teq{{\cal
N}=10^9} runs in Fig.~\ref{fig:B_parallel_x_polarize} with unpolarized and isotropic injection,
binned in azimuthal ranges of \teq{10^{\circ}} that were centered on \teq{\phi_z=0^{\circ}} and
\teq{\phi_z=90^{\circ}}; note that \teq{\phi_z=180^{\circ}} provides redundant information in
relation to the \teq{\phi_z=0^{\circ}} case. Comparing our \teq{\omega /\wcyc =0.5}
distributions with Fig.~5 of \cite{Whitney-ApJ91}, we find good general agreement for both
\teq{\phi_z=0^{\circ}} and \teq{\phi_z=90^{\circ}} between her and our distributions of
intensity, \teq{V/I} and effective linear polarization degree \teq{\Pi_{\rm lin} \approx \vert
Q/I\vert} throughout the zenith angle range \teq{0 < \theta_z < 90^{\circ}}. Note again that
\teq{U/I\approx 0} in this field configuration.  Comparing our \teq{\omega /\wcyc =0.25}
distributions with Whitney's Fig.~6, the agreement was just as good for both \teq{V/I} and
\teq{\Pi_{\rm lin}}, for both azimuths \teq{\phi_z = 0^{\circ},90^{\circ}}, and also for the
intensity for \teq{\phi_z = 90^{\circ}}.  However, significant deviations at the 10--30\% level
arose for zenith angles \teq{50^{\circ} < \theta_z < 80^{\circ}} for the  \teq{\phi_z =
0^{\circ}} intensity. In the absence of sufficient detail concerning the photon injection and
data binning in  \cite{Whitney-ApJ91}, it is not possible to isolate possible causes for the
origin of these differences.

\section{Polarization at High Opacities}
 \label{sec:high_opacity_pol}

In order to move beyond the testing and validation phase, in preparation for the implementation
of the radiation transport code in more sophisticated models of neutron star atmospheres, it is
necessary to explore the polarization characteristics of magnetic Thomson transport in high
opacity domains.  This is suitably done be shedding the simulations of the slab geometry and
just recording angular distributions of photon number and polarization measures subsequent to
large numbers of scatterings.  Accordingly, the ensuing results will not at first be applied
specifically to any particular neutron star locale, but will be broadly representative of
characteristics deep in atmospheric slabs.   These results are just functions of the angle
\teq{\theta} between the ultimate photon momentum \teq{\kvec} and magnetic field \teq{\Bvec}
vectors, and the scaled photon frequency \teq{\omega /\wcyc}; it will emerge that a set of quite
simple empirical approximations can encapsulate the Stokes parameter dependences on \teq{\theta}
and frequency.  These will then be employed to simulate high opacity slab results for polar
locales in Section~\ref{sec:high_tau_slab}, an extension that enables modeling of thicker regions 
than was possible in \cite{Whitney-ApJS91,Whitney-ApJ91}.  This paves the way 
for precision simulation of optically thick photospheres, whether they belong to neutron star/magnetar 
surface layers, accretion columns or magnetar burst regions.

\subsection{High Opacity Radiation Transfer Simulations}
 \label{sec:high_tau_simulate}

The simulation uses an isotropic injection at a single point, but omits a flux weighting
analogous to that in Eq.~(\ref{eq:isotropic_flux}).  The injection is also unpolarized,
employing an equal mix of  linearly-polarized \teq{\parallel} or \teq{\perp} photons in the
choice of the complex electric field vectors; see Eq.~(\ref{eq:perp_par_polarvec}). Specific
angular and polarization information at injection is irrelevant to the simulation output in this
Markovian setup. Photon frequencies are selected from a uniform distribution in \teq{\log_{10}
(\omega/\wcyc )}. To generate excellent statistics, \teq{{\cal N}_i=10^9} photons were
distributed in the frequency, angle and polarization variates.  Their paths were followed for
exactly \teq{n=100} magnetic Thomson scatterings, and then their final direction and
polarization information were recorded, regardless of their final location relative to the
injection point.   The choice of 100 scatterings was sufficient to generate a Markovian
scattering sequence and sample the differential cross section with sufficient density in polar
(\teq{\theta}) and azimuthal (\teq{\phi}) angles. We tested this by performing simulations with
\teq{10^3} and \teq{10^4} scatterings, but with somewhat fewer injected photons, and observed no
material differences within statistics (e.g. see Fig.~2.6 of \citealt{Barchas17}).

At the outset, a photon is assigned to one of 160 logarithmic frequency bins, uniformly sized in
\teq{\chi \equiv  \log_{10} (\omega/\wcyc )}. After a scattering sequence is complete, the
photon direction is recorded in one of 120 angle cosine bins, uniformly sized in \teq{\mu =
\cos\theta}. Since the differential cross section is dependent only on the difference
\teq{\phi_{fi}} in azimuths (e.g. see Eq.~(\ref{eq:dsig_perp_para}) for linear polarizations),
the resultant distributions are independent of azimuth \teq{\phi}, and so are co-added. Photon
number count and Stokes \teq{\hat{Q}} and Stokes \teq{\hat{V}} thereby generate three data
arrays, along with a derivative fourth array specifying \teq{\Pi}.  The number count
generates a photon number angular distribution \teq{A_{\omega}(\mu)} (a scaled intensity) that
will be termed the {\bf re-distribution anisotropy} as it specifies the terminal anisotropy at
high opacity when re-distributing angles and polarizations through Thomson scattering; it is
normalized to one injected photon:
\begin{equation}
   \int_{-1}^1 A_{\omega}(\mu )\, d\mu \; =\; 1\quad .
 \label{eq:Amu_norm}
\end{equation}
If one uniformly distributes a large number of injection locales in space, the cumulative 
output count distribution in \teq{\mu} would remain statistically the same.  Therefore, 
\teq{A_{\omega}(\mu ) \propto I(\mu) \;\propto n_{\gamma} (\mu )} represents a scaling of the 
intensity \teq{I(\mu)} in the direction of {\bf k} or the total density \teq{n_{\gamma} (\mu )} contributed by the 
different polarizations.  The proportionality coefficient \teq{A_{\omega}(\mu )/ I(\mu)}
is actually a function of \teq{\mu}, as will be detailed with the 
interpretation of \teq{A_{\omega}(\mu)} in Section~\ref{sec:redistribution}.
Since this scaling applies to all Stokes parameters, it drops out when 
forming the ratios \teq{\hat{Q} = Q/I} and \teq{\hat{V} = V/I}, a convenient 
circumstance in the ensuing exposition of results. 

\begin{figure*}
 \begin{minipage}{17.5cm}
\vspace*{0pt}
\centerline{\hskip 3pt\includegraphics[width=\textwidth]{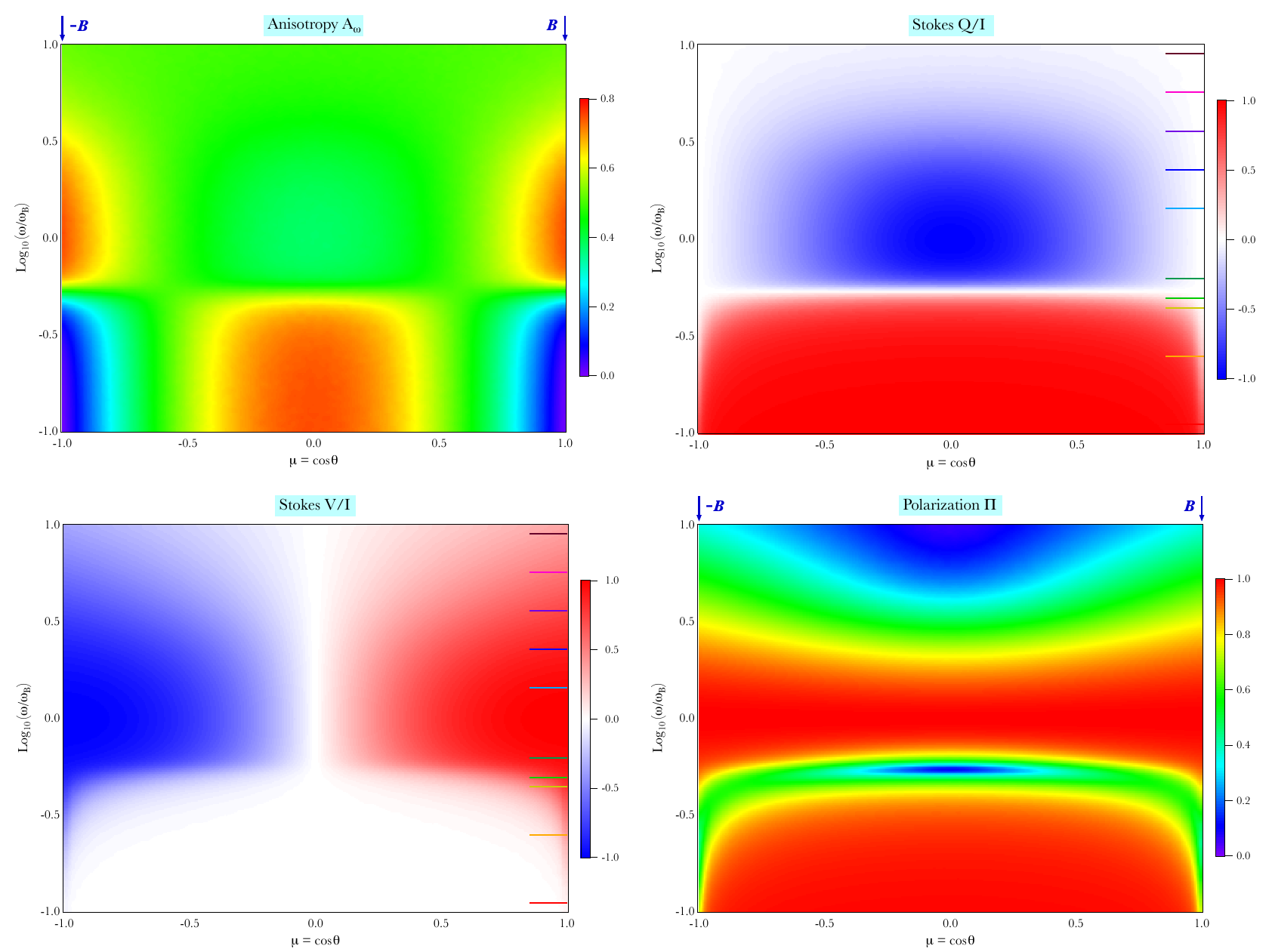}}
\vspace*{-5pt}
\caption{Maps of the angular distributions for the re-distribution anisotropy \teq{A_{\omega}}, 
Stokes parameters  \teq{Q/I},  \teq{V/I} and the {\it total} polarization degree \teq{\Pi} as 
functions of photon frequency (log scale on the ordinate). 
The angle \teq{\theta} is measured relative to the magnetic field direction, which is indicated 
in the \teq{A} and \teq{\Pi} panels, with the abscissa expressing the cosine \teq{\mu =\cos\theta}. 
The Stokes parameters were recorded after exactly 100 scattering 
events for each photon, with the injected photons sampling isotropic and unpolarized 
distributions, for a total of \teq{{\cal N}_i=10^9} photons.  The injection was also uniformly 
distributed in  \teq{\chi = \log_{10}(\omega /\wcyc )} among 160 frequency bins.
Short horizontal lines in the Q and V maps display the frequencies selected to compare 
with empirical approximations in Fig.~\ref{fig:high_tau_ang_dist}. 
 \label{fig:freq_ang_map}}
\end{minipage}
\end{figure*}

To explore the connection between the data output and photon diffusion, experiments were
performed where the position \teq{(x_n,\, y_n, \, z_n)} of a photon after the final
(\teq{n^{th}}) scattering was recorded, and collected to form statistical averages of the
diffusion process.  Here the \teq{z}-direction is aligned with \teq{\Bvec}, and \teq{(x,y)} are
Cartesian coordinates in the orthogonal plane.  The diffusion ``step'' is nominally the mean
free path \teq{\lambda = 1/[n_e\sigma (\mu )]} for scattering, and is anisotropic and
polarization dependent.  Units of \teq{n_e=1=\sigt} were chosen for expediency. The total number
of scatterings was increased to \teq{n=10^3}, and once \teq{n} exceeded around 10, the expected
correlations \teq{\langle x_n^2 + y_n^2 \rangle \propto n} and \teq{\langle z_n^2 \rangle
\propto n} were demonstrated to impressive statistical precision.  The ratio of these two
diffusion coefficients, i.e., \teq{\langle x_n^2 + y_n^2 \rangle /\langle z_n^2 \rangle} was
generally between 1 and 2, marking the anisotropy of magnetic Thomson diffusion.  In general,
this ratio was frequency-dependent, and for \teq{\omega/\wcyc=30} its numerical value was
$2.01$, signalling isotropic diffusion in a non-magnetic domain.

The information from these high opacity runs is presented as {\it frequency-angle maps} in
Fig.~\ref{fig:freq_ang_map}, with the individual legends defining the color scales for
anisotropy \teq{A_{\omega}}, \teq{Q/I}, \teq{V/I} and total polarization \teq{\Pi}.   With the
binning described just above, the raw data that these panels represent would appear more
pixellated than is displayed.  To make for a smoother depiction, we employed the Mathematica
built-in {\tt Interpolation} function with its {\tt Spline} option (the default setting of the
order is 3, i.e. cubic) to obtain the photon number and Stokes parameters as more continuous
functions of \teq{\mu} and \teq{\chi}.  The smoothing corresponds to 360 \teq{\mu} and 320
\teq{\chi} bins, both spanning the interval \teq{[-1,\, 1]}, so that the map pixellation is six
times denser than the original data.  This protocol does not change the information conveyed,
and we tested this with runs employing refined binning that were subject to poorer statistics,
and found no conflicts. The photon number density is normalized so that for each frequency the
integral of photon number over \teq{\mu} equals unity.

A synopsis of the general character of the frequency-angle maps is as follows. The angular
distribution \teq{A_{\omega}(\mu )} is fairly isotropic at high frequencies in the
quasi-non-magnetic domain, and then becomes peaked along and anti-parallel to the field as the
frequency drops and transitions through the cyclotron frequency.  This alignment of beaming with
the field direction is driven by the general preponderance of scatterings that generate small
\teq{\theta_f}, as is evident in Eq.~(\ref{eq:dsig_perp_para}). When \teq{\omega /\wcyc \sim
1/\sqrt{3}}, corresponding to approximate equality of cross sections for the two linear
polarizations, there is an abrupt transition to a domain where the density is much greater
perpendicular to the field.  Then, most of the scatterings are non-resonant
\teq{\parallel\to\parallel} events, biasing the population to a predominance of \teq{\pi /4
\lesssim \theta_f \leq \pi /2} final angles, with occasional \teq{\parallel \to \perp}
conversions; again see Eq.~(\ref{eq:dsig_perp_para}). Since we used 160 frequency bins, the
number of photons for each frequency was \teq{6.25 \times 10^6}. Thus while the runs that
generated Fig.~\ref{fig:freq_ang_map} persisted only for 100 scatterings per photon, so that
\teq{\parallel\to\perp} conversions might occur with a probability of 0.1--1 for each injected
photon, there were enough conversions in the ensemble to realize reasonable statistics in the
various distributions.  Inverse conversions \teq{\perp\to\parallel} possessed similar occurrence
probabilities; see Eq.~(\ref{eq:sig_perp_para}).

The Stokes \teq{Q/I} is generally of a small value at \teq{\omega /\wcyc \sim 10}, and becomes
more negative when transitioning through the resonance.  Again, an abrupt change in character
arises at \teq{\omega /\wcyc \sim 1/\sqrt{3}}, below which \teq{Q/I} is positive.  This
bifurcation is a direct consequence of the relative values of the cross sections in
Eq.~(\ref{eq:sig_perp_para}). Above \teq{\omega /\wcyc =1/\sqrt{3}}, since small \teq{\theta_i}
are favored, polarization conversions \teq{\parallel\to\perp} are more prevalent than
\teq{\perp\to\parallel} ones, so the \teq{\perp} state emerges  as the dominant linear
polarization, yielding \teq{Q < 0}.  The balance is inverted below \teq{\omega /\wcyc
=1/\sqrt{3}} and then the \teq{\parallel} mode is more prevalent and \teq{Q > 0}. If one inserts
the linear polarization modes in Eq.~(\ref{eq:perp_par_polarvec}) into the Stokes vector
definitions in Eq.~(\ref{eq:Stokes_polar_def}) one quickly infers that for \teq{\phi=0}
(i.e., the observer plane), that \teq{\hat{Q}_{\perp}=-1} and \teq{\hat{Q}_{\parallel} = 1}. 
This establishes the simple interpretation that \teq{Q<0} signals a preponderance of \teq{\perp}
polarization, while \teq{Q>0} marks a dominance of \teq{\parallel} modes. The angular dependence
of \teq{Q/I} is generally significant, and will be highlighted shortly.

In terms of the circular polarization, \teq{V/I} is small below the ``equipartition frequency''
\teq{\omega  =\wcyc /\sqrt{3}} as the circularity \teq{\Delta_{\hbox{\sixrm B}}(\omega )} in
Eq.~(\ref{eq:DeltaB_def}) becomes small in this domain.  At resonant and higher frequencies,
\teq{V/I} is significant up to \teq{\omega /\wcyc = 10} because of the strong circularity
imposed by the electron gyration in the scatterings. By inspection of the circular polarization
``mode-switching'' differential cross section in Eq.~(\ref{eq:mode_convert_circular}) in
Appendix B, when \teq{0 < \theta_f < \theta_i} in a scattering, the production of the \teq{+}
polarization is favored over the generation of the \teq{-} state, leading to generally positive
\teq{V} when \teq{0 < \theta < \pi /2}, as is observed in Fig.~\ref{fig:freq_ang_map}. Note that
\teq{V/I} is naturally odd in the angle cosine \teq{\mu}, with the prevailing helicity depending
on the direction of propagation \teq{\kvechat} relative to that gyration, and therefore
\teq{\Bvechat}.

The information in Fig.~\ref{fig:freq_ang_map} is inherently different from that in the slab
transport displays of Figures~\ref{fig:B_parallel_z_intense},~\ref{fig:B_parallel_z_polarize}
and~\ref{fig:B_parallel_x_polarize}.  It defines the approximate asymptotic solution to the
radiative transfer angular and polarization re-distribution problem for an infinite medium.  It
was derived in the absence of boundaries that define angle-dependent and polarization-dependent
escape probabilities, and thereby omits their critical influences.  For example, contributions
to the intensities near the zenith in Fig.~\ref{fig:B_parallel_z_intense} sample small solid
angles and therefore are statistically improbable.  At the same time, intensity contributions
for \teq{\theta_z\sim 90^{\circ}} near the slab horizon may sample large solid angles, yet they
require that the last scattering occur very near the slab surface, again an improbable
circumstance. These modify the emergent angular distributions of intensity relative to those in
Fig.~\ref{fig:freq_ang_map} that constitute regions deep down in the denser portions of an
atmospheric slab. Similar biases arise for the polarization Stokes parameters \teq{Q} and
\teq{V} when considering slab geometry.  We remark that while the results depicted in
Fig.~\ref{fig:freq_ang_map} were produced using a homogeneous medium, they apply regardless 
of density stratification along \teq{\Bvechat}, which just acts to generate a gradient in
scattering scale-lengths along the field.

\subsection{Empirical Approximations in \teq{\mu} and \teq{\omega}}
 \label{sec:empirical}

The symmetry with respect to the polar angle cosine \teq{\mu} of the maps in
Figure~\ref{fig:freq_ang_map} suggests a simple mathematical dependence.  This is borne out 
with empirical fits to the data at a variety of frequencies that are addressed here.  The origin of
this simplicity is in the quadratic \teq{\mu} dependence of the differential cross sections for
linear and circular polarization scatterings presented in Appendix~B, and will be discussed in
Section~\ref{sec:redistribution}. For a large number of scatterings where the configuration is
not changed by the action of scattering, one therefore anticipates that the photon
re-distribution anisotropy assumes the form \teq{A_{\omega}(\mu )\propto 1 + {\cal A}(\omega )
\, \mu^2}.  The evenness in \teq{\mu}, obvious in Figure~\ref{fig:freq_ang_map}, is driven by
the same property of the total cross section  \teq{\sigma (\mu )}.  Adhering to the
normalization in Eq.~(\ref{eq:Amu_norm}), the form
\begin{equation}
   A_{\omega}(\mu )\; =\; \dover{3}{2}\, \dover{1 + {\cal A}(\omega )\, \mu^2}{3 + {\cal A}(\omega )} \quad ,
 \label{eq:A_omega_mu}
\end{equation}
emerges.  Observe that this function is identically equal to 
\teq{1/2} for all frequencies when \teq{\mu = \pm 1/\sqrt{3}}; this is 
just a consequence of the Legendre 2-point quadrature evaluation
of the integrals in the radiation transport formalism. 

\begin{figure*}
     \begin{minipage}{17.5cm}
    \vspace*{0pt}
    \centerline{\hskip 3pt\includegraphics[width=\textwidth]{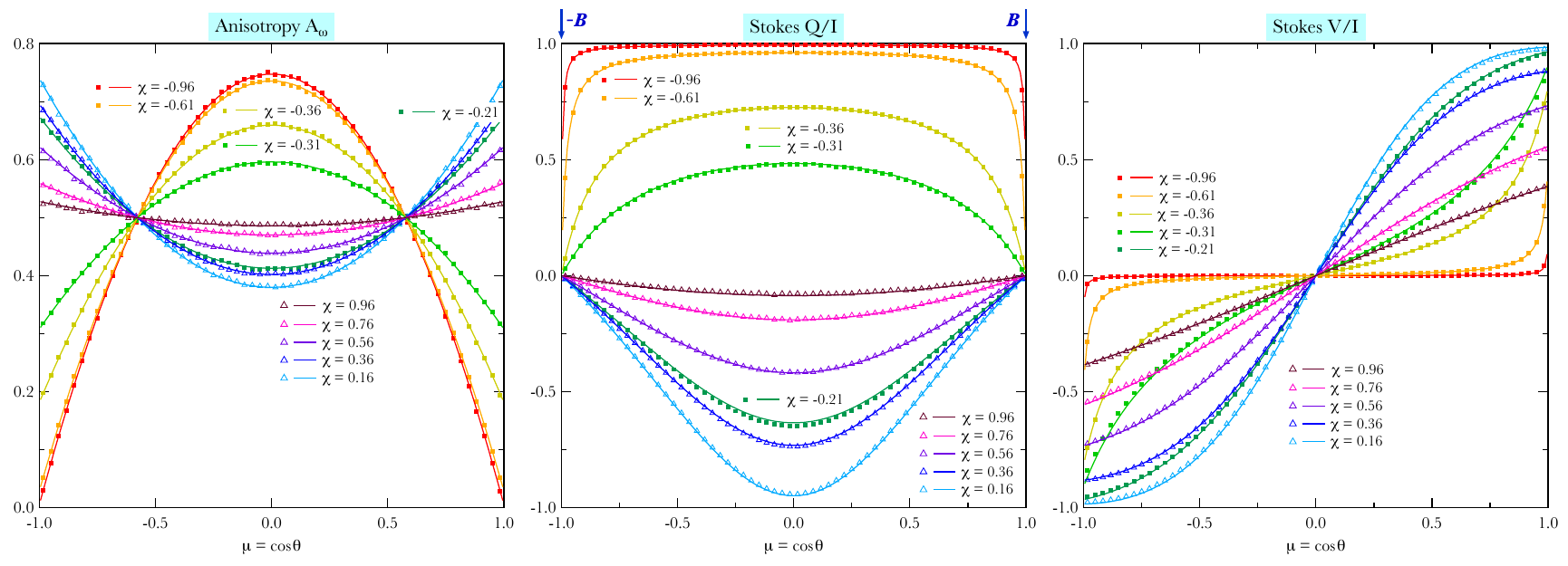}}
    \vspace*{-5pt}
    \caption{Angular distributions for \teq{A_{\omega}} (left), Stokes \teq{\hat{Q}_{\omega}} 
    (center) and Stokes \teq{\hat{V}_{\omega}} (right) for high opacity runs of 
    radiative transfer at select photon frequencies \teq{\omega} identified
    by the parameter \teq{\chi = \log_{10}(\omega /\wcyc )}, as labelled.  These correspond to 
    horizontal cuts of the respective panels in the frequency-angle maps in Fig.~\ref{fig:freq_ang_map}.  
    Again, the angle \teq{\theta} is measured relative to \teq{\Bvec}, as indicated 
    in the middle panel. The squares (for \teq{\chi < 0}) and open triangles (for \teq{\chi > 0}) 
    represent data acquired for the runs with 100 scatterings for each of the 
    \teq{{\cal N}/160=6.25 \times 10^6} photons per frequency bin.  The curves are computed 
    using the relatively compact analytic forms of the empirical approximations in 
    Eqs.~(\ref{eq:A_omega_mu}), (\ref{eq:Q_omega_mu}) and (\ref{eq:V_omega_mu}),
    demonstrating the precision of these fits at around the 2\% level or better. 
     \label{fig:high_tau_ang_dist}}
    \end{minipage}
\end{figure*}

To highlight the symmetries of the frequency-angle maps, we formed horizontal sections of 
them at select frequencies identified by the markers along the right axes of the 
\teq{Q/I} and \teq{V/I} panels in Fig.~\ref{fig:freq_ang_map}.  The data for 
these frequencies were presented as the ``curves'' with discrete points in 
Fig.~\ref{fig:high_tau_ang_dist}, employing the same color coding as the markers 
in Fig.~\ref{fig:freq_ang_map}.  In the case of \teq{A_{\omega}(\mu )} depicted
in the left panel, the coefficient \teq{{\cal A}(\omega )} was determined numerically by 
fitting the data at each of these selected frequencies plus about another dozen 
more (not displayed).  The result was a data ``curve'' in frequency space that 
was then fit by an empirical function of \teq{\omega /\wcyc}.  This 
fitting protocol was most conveniently detailed using the variables
\begin{equation}
   \psi \; =\; 16\chi^2
   \quad ,\quad
   \chi = \log_{10} (\omega /\wcyc )\quad .
 \label{eq:psi_chi_def}
\end{equation}
The resulting frequency function for the anisotropy fits was 
determined on the range \teq{-1 \leq \chi \leq 1} to be 
\begin{equation}
    {\cal A}(\omega ) \; =\;
    \begin{cases}
        \dover{2 - \psi /10 - 12\psi^2/11 +  24\psi^3/19}{1 + 4\psi^5/5} -1 \; ,\;\; \omega < \wcyc\; ,\\[15pt]
        \dover{1 - 5\psi/38 + 17\psi^2/1840 - \psi^3/4250 }{1 + \psi^2/80}  \; ,\;\; \omega \geq \wcyc\; .
    \end{cases}
    \label{eq:calA_def}
\end{equation}
This is an empirical form suitably compact for numerical applications. It possesses a signature
value of \teq{{\cal A}(\omega ) =1} at the cyclotron resonance, where \teq{A_{\omega}(\mu ) =3(
1 + \mu^2)/8}. In the magnetar surface X-ray domain where \teq{\omega /\wcyc\ll 1},  \teq{{\cal
A}(\omega )  \approx -1} and \teq{A_{\omega}(\mu )  \approx 3( 1 - \mu^2)/4}, so that photon
propagation is clearly suppressed  along or anti-parallel to \teq{\Bvec}. Also, \teq{{\cal
A}(\omega )} approaches zero when \teq{\chi = +1}  in the quasi-non-magnetic domain, so that
then \teq{A_{\omega}(\mu )\approx 1/2} and the photons are almost isotropic. Outside the
interval \teq{-1 \leq \chi \leq 1}, the \teq{\chi = \pm 1} endpoint values for \teq{{\cal A}}
can be adopted. We note that a full phase matrix analysis of the radiative transfer would not
generate this mathematical form, instead capturing the analytics of the differential cross
section, yet it may in principal be more complicated; its determination is deferred to a future
investigation.

The analytic approximation for the anisotropy formed by the combination of
Eqs.~(\ref{eq:A_omega_mu}) and~(\ref {eq:calA_def}) is represented as the solid curves in the
left panel of Fig.~\ref{fig:high_tau_ang_dist} for the selected frequencies.  The precision of
its fit to the simulation data points is better than 2\%, instilling confidence in its utility.

The linear and circular polarization data were subjected to fitting protocols similar to the
anisotropy ones. Since linear polarization is zero for propagation in either direction parallel
to the field, one anticipates that \teq{Q \propto 1-\mu^2}. In forming the \teq{Q/I} ratio, the
quadratic dependence of the anisotropy appears in the denominator, so that the fit should be a
Pad\'e approximant in \teq{\mu}.  Guided by this, we arrived at a linear polarization empirical
fit defined by
\begin{equation}
   \dover{\QP}{\IP} \;\equiv\; \hat{Q}_{\omega} (\mu )
   \; =\; \dover{{\cal A}(\omega ) \, \left[ \mu^2-1 \right] }{1 + {\cal A}(\omega )\, \mu^2} \quad .
 \label{eq:Q_omega_mu}
\end{equation}
We have now introduced \teq{\IP \propto A_{\omega}} anticipating the phase matrix 
interpretation of this anisotropy in Section~\ref{sec:redistribution},  and its Stokes 
counterparts \teq{\QP} and \teq{\VP}.  Note that the frequency dependence of the 
numerator is just that for the anisotropy, as opposed to some other function of \teq{\omega}.  
This is not a coincidence, and should emerge from a full polarization phase matrix 
analysis of the transport; this effective coupling between \teq{I} and \teq{Q} is 
apparent in the Stokes parameter form of the cross section in
Eq.~(\ref{eq:sigma_tot_Stokes}).  

The circular polarization is necessarily odd in \teq{\mu} due to this 
parity property of the cross section: see Eq.~(\ref{eq:sig_pm}).  The 
obtained fitting function was 
\begin{equation}
   \dover{\VP}{\IP} \;\equiv\; \hat{V}_{\omega} (\mu )
   \; =\; \dover{2 {\cal C}(\omega ) \,  \mu }{1 + {\cal A}(\omega )\, \mu^2} \quad ,
 \label{eq:V_omega_mu}
\end{equation}
for
\begin{equation}
    {\cal C}(\omega ) \; =\;
    \begin{cases}
        \dover{1 - 7\psi /25 + 9\psi^2/25 -  \psi^3/59}{1 + \psi^{3.8}}  \; ,\;\; \omega < \wcyc \; ,\\[15pt]
        \dover{1 - \psi/90 }{1 + 3\psi^{1.2}/25}  \; ,\;\; \omega \geq \wcyc\; .
    \end{cases}
    \label{eq:calC_def}
\end{equation}
This function, which applies to the domain \teq{-1 \leq \chi \leq 1}, 
appears because the circularity of the polarized transfer inherently 
differs from its linearity, whether proximate to the cyclotron resonance or not.
Signature values are \teq{{\cal C}(\omega ) =1} at the cyclotron resonance, \teq{\chi=0},
and it is very close to zero when \teq{\chi =-1}.   When \teq{\chi = +1} is around 0.19, 
and does approach zero for \teq{\omega \gg \wcyc}, albeit somewhat slowly.

Having assembled these empirical fits to the frequency-dependent anisotropy and 
Stokes parameters, numerical checks on their validity are in order.  The high opacity 
simulation was run again, but with Eqs.~(\ref{eq:A_omega_mu}--\ref{eq:calC_def}) 
used to provide an alternative, polarized injection (see below).  For \teq{n=100} scatterings for
each photon, the subsequent results were indistinguishable from Fig.~\ref{fig:freq_ang_map}.
Yet this provides a Markovian test that is not an incisive probe.  We therefore performed such 
a simulation for just \teq{n=1} scattering, and the results in Figs.~\ref{fig:freq_ang_map}
and~\ref{fig:high_tau_ang_dist} were reproduced with excellent precision.  This validation 
demonstrated that the empirical forms do indeed constitute the true asymptotic state 
of the polarized magnetic Thomson transfer system.

A further, independent check was to actually fold the analytic forms in 
Eqs.~(\ref{eq:A_omega_mu}), (\ref{eq:Q_omega_mu}) and~(\ref{eq:V_omega_mu})
through the phase matrix analysis formulation of \cite{Chou86} numerically.
The post-scattering Stokes parameter information was obtained via numerical 
integration over the phase matrix form in Eq~(2.24) of \cite{Barchas17}, 
adapted from \cite{Whitney-ApJS91}, over scattering 
solid angles and weighted by the pre-scattering Stokes parameters.  
The two frequency-dependent functions \teq{{\cal A}(\omega )} and
\teq{{\cal C}(\omega )} were left as free parameters, and their values 
for each \teq{\omega} were derived numerically by demanding that the 
Stokes parameters were invariant under the scattering operation.  
The solution for these two functions agreed numerically with the 
empirical determinations in Eqs.~(\ref {eq:calA_def}) and~(\ref{eq:calC_def})
to excellent precision.  Details of this validation protocol will 
be expanded in a future paper.

Note that in hydrostatic models of magnetar atmospheres, 
radiation at \teq{\omega/\omegaB \ll 1} is dominated by \teq{\perp} mode
photons (\teq{Q<0}) because they emerge from hotter regions deep in the atmosphere 
\citep[e.g.][]{Ozel-2001-ApJ,Ho03ApJ} where they are more numerous.  This property is not evident in 
the \teq{Q/I} panels of Figs.~\ref{fig:freq_ang_map} and~\ref{fig:high_tau_ang_dist} 
since temperature weighting and stratification are currently not  included in our simulation.

\subsection{Interpreting the Re-distribution Anisotropy}
 \label{sec:redistribution}

To correctly interpret the simulation results presented in 
Section~\ref{sec:high_tau_simulate}, it is necessary to identify the relationship 
between the intensity \teq{I (\mu )} and the re-distribution anisotropy 
\teq{A_{\omega}(\mu )}, and extend such to all Stokes parameters.
What is output from the high opacity simulation are the statistics 
of all the pertinent products of electric field components, averaged per scattering.
These data are not generated in the spatial or radiative transfer domain. 
As such, they represent conditional probabilities of a re-distribution mapping from one
polarization/anisotropy configuration to another, i.e., defining the probability of a 
scattering into new directions and new polarization vectors given these quantities
prior to the photon scattering.  The mapping function is a 4-tensor or matrix 
\teq{\cal R} in polarization space, and will be termed the {\bf re-distribution tensor} 
or {\bf phase matrix} \citep{Chou86} for magnetic Thomson scattering.  
For an unpolarized scattering process, i.e. one-dimensional
in polarization space (intensity only; \teq{Q=U=V=0}), this must reduce to the differential 
cross section divided by the total cross section.

Let \teq{\boldsymbol{S} = (I,Q,U,V)} be the true Stokes vector 
and \teq{\boldsymbol{P} = (\IP , \QP , \UP , \VP )} be its corresponding 
quantity (putatively dimensionless) 
in the re-distribution mapping.  We expect that \teq{\boldsymbol{S}\propto \boldsymbol{P}}, 
with a single coefficient of proportionality that can be dependent on a photon's 
direction.  The evolution of \teq{\boldsymbol{P}} with 
scattering is negligible in the high scattering limit, and this can be expressed 
via a ``polarization equilibrium'' re-distribution equation
\begin{equation}
   \boldsymbol{P} (\kvec ) \; =\; \int {\cal R} (\kvec_i\to \kvec )\, 
       \boldsymbol{P} (\kvec_i )\, d\Omega_i\quad . 
 \label{eq:re_dist_equil}
\end{equation}
The subscript \teq{i} denotes initial quantities prior to a scattering, which deflects a photon
from direction \teq{\kvechat_i} to \teq{\kvechat}. The right hand side of
Eq.~(\ref{eq:re_dist_equil}) describes the establishment of a new photon direction and a new
photon polarization given a pre-scattering polarization configuration, integrating over all
photon directions using the solid angle differential \teq{d\Omega_i}.  Setting this equal to the
left hand side expresses the asymptotic condition that the system polarization does not change
on average in a single scattering, but allows the photon direction to be altered.  This
precisely describes what is modeled by the high opacity simulations of
Section~\ref{sec:high_tau_simulate}.

If one were to specialize this re-distribution formalism to unpolarized systems, such as
effectively arises in the high opacity  ``non-magnetic domain'' \teq{\omega /\wcyc \gg 1}, then
only the \teq{\IP} portion of \teq{\boldsymbol{P}} need be retained.  The re-distribution is
then in angle only, expressed using the differential cross section, with the structure of a
conditional probability leading to a normalizing factor incorporating the total unpolarized
cross section.  Thus, the analog of Eq.~(\ref{eq:re_dist_equil}) would then be
\begin{equation}
   \IP (\kvec ) \; =\; \int  \dover{1}{\sigma (\kvec_i )} \, \dover{d\sigma (\kvec_i\to \kvec )}{d\Omega_i} 
       \, \IP (\kvec_i ) \, d\Omega_i \quad .
 \label{eq:re_dist_equil_1D}
\end{equation}
Integrating over \teq{\kvec} directions (\teq{d\Omega}) one quickly sees that this is 
properly normalized when introducing the cross section \teq{\sigma (\kvec_i )} 
in the denominator of the integrand.  This analogy enables one 
to directly connect \teq{{\cal R}} with the magnetic Thomson scattering 
phase matrix in Eq.~(21) of \cite{Chou86}, which expresses the conditional 
probability weighting via ratios of the differential and total cross section, 
yet both being dependent on the details of the polarization.   Thus, now, 
\teq{\sigma (\kvec_i )} depends on the Stokes parameters, expressed in 
Eq.~(4) of \cite{Whitney-ApJS91}, and in Eq.~(\ref{eq:sigma_tot_Stokes}) here.

Since Eq.~(\ref{eq:re_dist_equil}) can be scaled by any constant, we can normalize 
it as suits.  Furthermore, our coordinate choices have \teq{U=0}, so noting
\teq{\IP \propto A_{\omega}(\mu )}, one can write 
\begin{equation}
   \boldsymbol{P} \; =\;  \bigl( \IP, \, \QP, \, 0, \, \VP \bigr) 
   \; \equiv\; \IP \bigl( 1, \, \hat{Q}, \, 0, \, \hat{V} \bigr) 
 \label{eq:Pvec_redistribute}
\end{equation}
for the interpretation of 
the simulations in Section~\ref{sec:high_tau_simulate}.  As the phase matrix version
of \teq{\cal R} in Eq.~(21) of \cite{Chou86} possesses simple quadratic dependence on 
both \teq{\mu_i} and \teq{\mu} (final photon angle cosine relative to  \teq{\Bvechat}), the 
observed quadratic dependence of \teq{\boldsymbol{P}} on \teq{\mu} is guaranteed by 
Eq.~(\ref{eq:re_dist_equil}), and motivates our protocol for seeking its solutions
via our Monte Carlo experiment.  The numerical evaluations for \teq{\boldsymbol{P}} 
obtained in our simulations nicely satisfy a numerical evaluation of Eq.~(\ref{eq:re_dist_equil}) 
using the phase matrix of \cite{Chou86}, establishing the correspondence between the 
two approaches.

The true Stokes vector \teq{\boldsymbol{S}} describes {\sl fluxes} of quadratic forms 
of electric field components and thus obeys a radiation transport equation.  The radiative 
transfer in polarized scattering systems is expressible as an integro-differential equation
\citep[e.g.][]{Chandra60}, with the scattering transfer captured in 
integrals over the differential cross section.   This is simplest to express in the 
unpolarized case, for which only the intensity \teq{I} comes into consideration. 
The familiar form for an equilibrium situation with an angular distribution 
that does not evolve with scatterings can quickly be written down.  It is 
\begin{equation}
   \sigma (\kvec ) \, I(\kvec ) \; =\; \int  \dover{d\sigma (\kvec_i\to \kvec )}{d\Omega_i} 
       \, I(\kvec_i ) \, d\Omega_i \quad ,
 \label{eq:Stokes_equil_1D}
\end{equation}
a radiation transfer counterpart to Eq.~(\ref{eq:re_dist_equil_1D}).   Thus, one infers that
\teq{\IP (\kvec ) \propto \sigma (\kvec ) \, I(\kvec )}.  A partner equation for the 
photon number density angular distribution \teq{n_{\gamma}(\mu )} that 
satisfies the equilibrium Boltzmann equation (time domain) assumes the same form, 
noting that \teq{I (\mu) \propto n_{\gamma} (\mu )}.  Extending this to the full 
polarization configuration, we assert the correspondence 
\begin{equation}
   \boldsymbol{P} (\kvec ) \; \propto\; \sigma (\kvec ) \, \boldsymbol{S} (\kvec ) \quad .
 \label{eq:Pvec_Svec_correspond}
\end{equation}
This can be inserted into Eq.~(\ref{eq:re_dist_equil}) to develop the radiative transfer analog 
of Eq.~(\ref{eq:Stokes_equil_1D}) for the full polarization configuration.
The desired relationship between the two polarization vectors 
indicates that the coupling depends on both the direction {\it and} the 
polarization of the photons.  

Taking the intensity portion of Eq.~(\ref{eq:Pvec_Svec_correspond}) yields 
the result
\begin{equation}
   n_{\gamma}(\mu ) \;\propto\; I(\kvec) \; \propto\; \dover{A_{\omega}(\mu )}{\sigma (\kvec )}
 \label{eq:n_gamma_form}
\end{equation}
for our azimuthally-symmetric system, where \teq{\sigma (\kvec ) \to \sigma (\omega, \, \mu )}. 
This is the sought-after interpretation of the re-distribution anisotropy \teq{A_{\omega}(\mu
)\propto \IP}, defining a path for using it to obtain true intensity-related anisotropies. The
cross section \teq{\sigma (\kvec )} must be computed using the asymptotic Stokes parameters,
inserted into Eq.~(\ref{eq:sigma_tot_Stokes}). The resulting \teq{I (\mu )} or
\teq{n_{\gamma}(\mu )} can be employed in formal opacity calculations treating magnetic Thomson
diffusion. It is notable that while the scaling \teq{1/\sigma (\kvec )} impacts the intensity
component and those for the other Stokes parameters, by forming ratios \teq{Q/I} and \teq{V/I},
this scale factor cancels out, so that \teq{\hat{Q}} and \teq{\hat{V}} are invariant under the
mapping from re-distribution (\teq{\boldsymbol{P}}) space to Stokes (\teq{\boldsymbol{S}})
space.  This convenience was exploited in the presentations in
Sections~\ref{sec:high_tau_simulate} and~\ref{sec:empirical}.

While the logic leading to the \teq{ \boldsymbol{P} \propto \sigma \, \boldsymbol{S} }
correspondence is intuitive and obvious, its verity is also amenable to scrutiny via simulation.
 To this end, we performed a number of simulations modified from those addressed in
Sections~\ref{sec:high_tau_simulate} and~\ref{sec:empirical}.  Instead of the exit criterion of
a fixed total number of scatterings, the simulation termination was made when the cumulative
distance travelled by each photon exceeded a pre-defined and large value \teq{d_{\rm max}}. 
This was not a constraint on the spatial displacement from the point of injection, but on a
linear addition of the pathlengths travelled between each scattering.  Thus it incorporated
information on the mean free path \teq{\lambda \propto 1/\sigma (\kvec )} for magnetic Thomson
scatterings, and is a true modeling of radiative transfer as opposed to just angle and
polarization re-distribution probabilites. This simulation was performed for a representative
range of frequencies \teq{0.1 \leq \omega /\wcyc \leq 10}.  To excellent precision, the emergent
polarization information generated intensity anisotropies \teq{I (\mu )\propto
A_{\omega}(\mu)/\sigma} and the same \teq{\hat{Q}(\mu )} and \teq{\hat{V}(\mu )} distributions
produced by the fixed number of scattering simulations of Section~\ref{sec:high_tau_simulate}. 
Results for this comparison will be presented in another paper.  These experiments thus
numerically validate the \teq{ \boldsymbol{S} \propto \boldsymbol{P}/\sigma } relationship.

\begin{figure}
\vspace*{-0pt}
\centerline{\includegraphics[width=8.5cm]{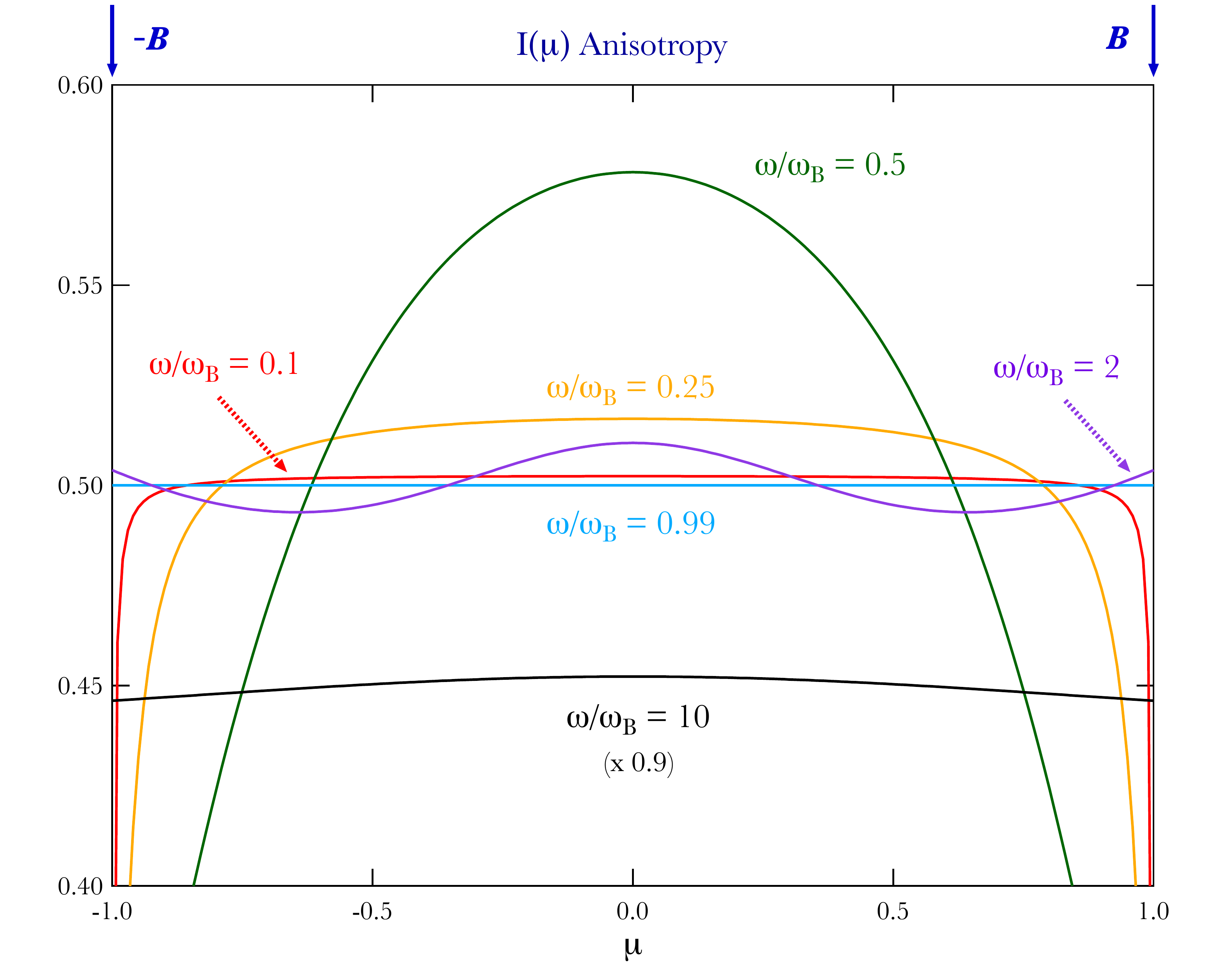}}
\vspace*{-8pt}
\caption{Intensity \teq{I (\mu )} as a  
function of \teq{\mu=\cos{\theta}}, for our six familiar frequencies
\teq{\omega/\wcyc = 0.1, 0.25, 0.5, 0.99, 2, 10}.  These depictions, germane 
to high opacity domains, are obtained by dividing the re-distribution anisotropy 
\teq{A_{\omega}(\mu )} in Eq.~(\ref{eq:A_omega_mu})  by the 
polarized cross section in Eq.~(\ref{eq:sigma_tot_Stokes}): see Eq.~(\ref{eq:intensity_anisotropy}).
All polarization information was generated using Eqs.~(\ref{eq:Q_omega_mu}) 
and~(\ref{eq:V_omega_mu}), and the empirical approximations in 
Eqs. ~(\ref{eq:calA_def}) and~(\ref{eq:calC_def}).
Distributions were normalized to unity on \teq{[-1,1]}; the 
\teq{\omega/\wcyc = 10} curve was multiplied by \teq{0.9} to aid visual clarity.
\label{fig:ngamma_mu}}
\end{figure}

To complement the \teq{A_{\omega}(\mu )} information presented in
Fig.~\ref{fig:high_tau_ang_dist}, we plot in Fig.~\ref{fig:ngamma_mu} the resultant intensity
anisotropy \teq{ I(\mu )\propto n_{\gamma} (\mu )} described in Eq.~(\ref{eq:n_gamma_form}). 
The curves are mostly normalized to unit area, and also represent the intensity anisotropy. 
Thus, hereafter, we posit the form 
\begin{equation}
   I(\mu ) \; =\; \dover{A_{\omega}(\mu )}{\Lambda_{\omega} \,\sigma (\omega, \, \mu )}
   \;\; ,\quad
   \Lambda_{\omega} \; =\; \int_{-1}^1 \dover{A_{\omega}(\mu )\, d\mu}{\sigma (\omega, \, \mu )} \quad .
 \label{eq:intensity_anisotropy}
\end{equation}
Again, \teq{\sigma (\omega, \, \mu )} is the cross section form 
in Eq.~(\ref{eq:sigma_tot_Stokes}) with the full polarization information 
embodied in \teq{\IP \to A_{\omega}(\mu)}, \teq{\hat{Q}} and \teq{\hat{V}}
from the empirical approximations of Section~\ref{sec:empirical} incorporated.  
This \teq{ I(\mu )} form will be employed in Section~\ref{sec:high_tau_slab}.
It is evident that the overall anisotropy is much smaller than that evinced 
by \teq{A_{\omega}(\mu )}.  Yet anisotropy is still significant, 
particularly around \teq{\omega/\wcyc \sim 1/\sqrt{3}} where the 
interplay between linear and circular polarizations is complex, 
and also around \teq{\mu \approx \pm 1} at lower frequencies.
The distributions are approximately isotropic at both the cyclotron 
resonance and in the non-magnetic regime \teq{\omega\gg \wcyc}.

\vspace{-8pt}
\subsection{High Opacity Slab Simulations}
 \label{sec:high_tau_slab}

The real utility of developing the connection between re-distribution anisotropy 
\teq{A_{\omega}(\mu )} and the true intensity one, and the three empirical approximations
for \teq{A_{\omega}(\mu )}, \teq{\hat{Q}(\mu )} and \teq{\hat{V}(\mu )}, is that they
circumvent the need to compute atmospheric slabs with large thicknesses 
and high opacities.  Regions deep in the atmospheres that sample 
\teq{\tau_{\parallel, \perp}\gg 10} are just zones of high opacity that do not 
intimately connect to the escape through the upper atmospheric boundary.
Accordingly, it is expedient to use the empirical approximations to define 
suitable polarized and anisotropic injections at the base of the slab, 
capturing the key information of scattering diffusion deeper in the atmosphere.
This can be done simply using an accept/reject procedure.  This is performed 
here for the magnetic polar case to take advantage of its azimuthal symmetry, 
though in principle the following injection protocol can be modified routinely 
through rotations to address any field orientation in the slab.  

First, the direction of the injected photon is specified exactly as before via 
Eq.~(\ref{eq:isotropic_flux}) in using two variates \teq{\xi_{\theta}} and \teq{\xi_{\phi}} selected 
randomly on the interval \teq{[0,\, 1]}.  These establish the angles \teq{\theta_0} and \teq{\phi_0},
respectively, and the azimuthal symmetry guarantees that the value of \teq{\phi_0} 
can always be accepted.  In contrast, as the polar angle injection distribution 
is not isotropic in general, we introduce a new random variate \teq{\xi_I} on 
the interval \teq{[0,\, I_{\rm max}]} for the intensity anisotropy.  Here, 
\teq{I_{\rm max}} is the maximum value of \teq{I(\mu_0 )} on the 
interval \teq{-1 \leq \mu_0 \leq 1 } (see Fig.~\ref{fig:ngamma_mu}).
Then, if \teq{\xi_I \leq n_{\gamma} (\mu_0)} for \teq{\mu_0 \equiv \cos \theta_0}, 
the polar angle is accepted, otherwise new \teq{\xi_{\theta}} and \teq{\xi_I} 
variates are chosen and the process repeated until an acceptance occurs.
The injection is then further flux-weighted by a factor of \teq{\mu_0} as 
with all prior slab runs.

The polarization injection protocol is similarly expedient, employing the electric
field vector forms in Eq.~(\ref{eq:Stokes_polar_def}).  Each photon has 
a scaled intensity of \teq{\hat{I}_0 =1}, and the polarization degree \teq{\Pi} 
for the injection is given by the combination of information from 
Eqs.~(\ref{eq:Q_omega_mu}) and~(\ref{eq:V_omega_mu}):
\begin{equation}
   \Pi \; =\; \sqrt{  \bigl( \hat{Q}_0 \bigr)^2 + \bigl( \hat{V}_0 \bigr)^2 }  \quad ,
 \label{eq:Pi_inject_high_tau}
\end{equation}
for 
\begin{equation}
   \hat{Q}_0 \; =\; \hat{Q}_{\omega} (\mu_0 ) 
   \quad ,\quad
   \hat{V}_0 \; =\; \hat{V}_{\omega} (\mu_0 ) \quad .
 \label{eq:Q0_V0_def}
\end{equation}
A new random variable \teq{\xi_{\Pi}} is sampled on \teq{[0, \, 1]} to 
determine the polarization state of the photon.  If \teq{\xi_{\Pi} \geq \Pi}, 
then the injection is deemed unpolarized, and it is sufficient to 
randomly select linear modes of either the \teq{\parallel} state
\teq{(\hat{\cal E}_{\theta} =1,\, \hat{\cal E}_{\phi} = 0)} or the 
\teq{\perp} one \teq{(\hat{\cal E}_{\theta} =0,\, \hat{\cal E}_{\phi} = 1)}.
With large photon count statistics, this generates unpolarized information,
and a \teq{+/-} circular polarization choice could also be implemented.
If, on the other hand, \teq{\xi_{\Pi} < \Pi}, the injected photon is polarized, 
and one can choose \teq{{\cal E}_{\theta}} to be real without loss of 
generality, which implies that \teq{{\cal E}_{\phi}} is purely imaginary.  
The inversion of Eq.~(\ref{eq:Stokes_polar_def}) then yields
\begin{equation}
  \hat{\cal E}_{\theta} \; =\; \sqrt{ \dover{\Pi + \hat{Q}_0}{2\Pi \vphantom{\bigl(}} }
  \; ,\quad
  \hat{\cal E}_{\phi} \; =\; \dover{i \hat{V}_0}{2\Pi \, \hat{\cal E}_{\theta} }
  \; =\; i\, \hbox{sgn}(\hat{V}_0) \, \sqrt{ \dover{\Pi - \hat{Q}_0}{2\Pi \vphantom{\bigl(}} } \; .
 \label{eq:Ehat_theta_phi_inj}
\end{equation}
These define an elliptically-polarized photon.  In the limits
where \teq{\hat{Q}_0 \to \pm \Pi}, the circular polarization \teq{\hat{V}_0} is zero, 
and these then constitute the \teq{\perp} and \teq{\parallel} modes
identified just above.  For large numbers of photons, this protocol 
establishes a polarization injection statistically commensurate with the 
high opacity distributions for \teq{\hat{Q}} and \teq{\hat{V}}.

\begin{figure}
\vspace*{-0pt}
\centerline{\includegraphics[width=8.3cm]{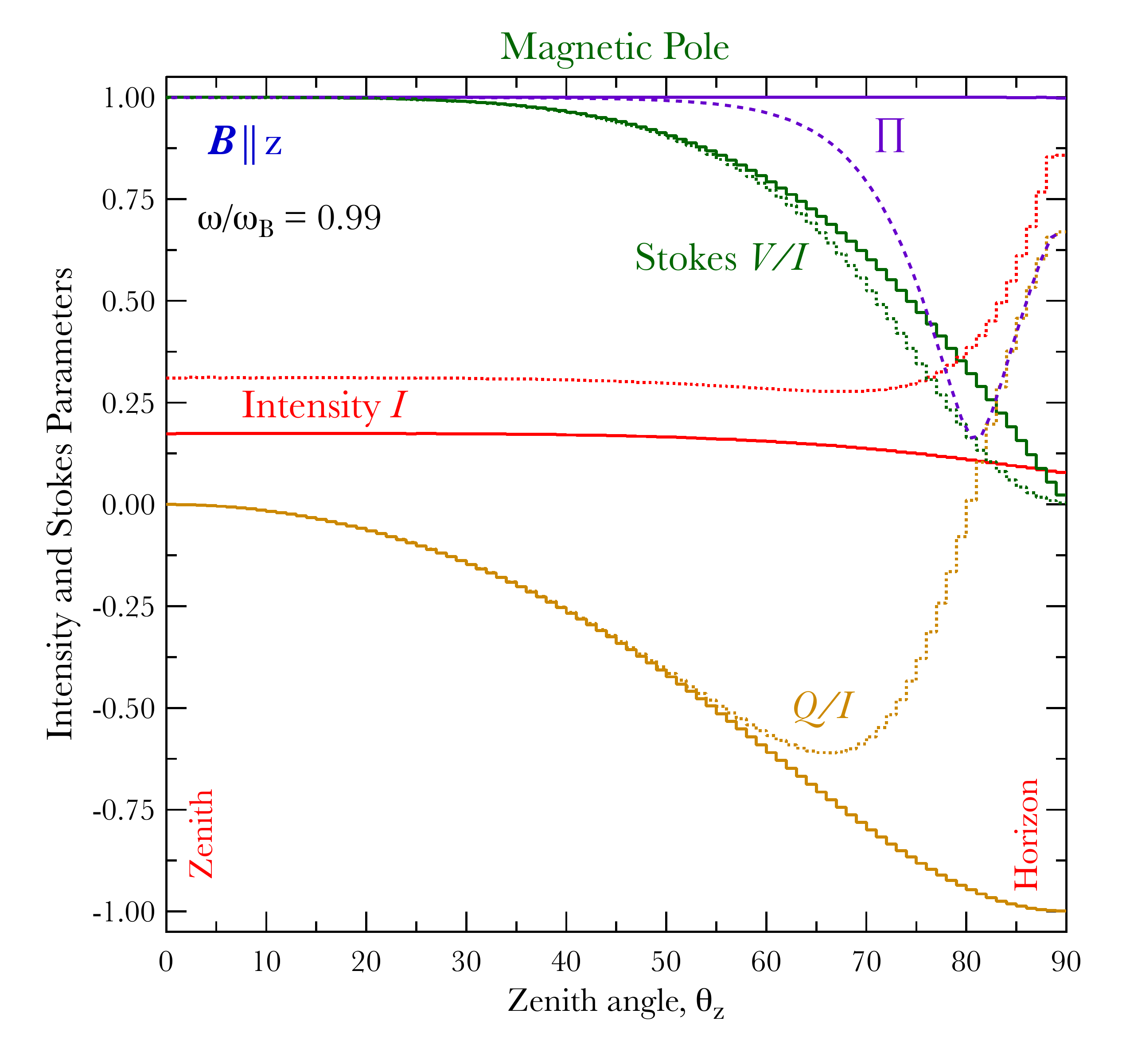}}
\vspace*{-10pt}
\caption{Angular distributions of intensity (red) and Stokes parameters 
\teq{Q/I} (tan) and \teq{V/I} (green) as functions 
of the zenith angle \teq{\theta_z}, for \teq{\omega/\wcyc=0.99} and the polar case 
where \teq{\Bvec} is parallel to the slab normal. The purple curves constitute 
the resultant polarization degree.  The solid histograms display the 
results of \teq{\tau_\parallel=10} with polarized injection at the slab base using 
the empirical forms as described in the text for the high opacity simulations. 
The injection was anisotropic, effectively comprising mixtures of linearly
and circularly-polarized photons so that the ensemble satisfies 
Eqs.~(\ref{eq:A_omega_mu}), (\ref{eq:Q_omega_mu}) and (\ref{eq:V_omega_mu}).
The dotted histograms/curves are from the \teq{\omega/\wcyc=0.99} panels of 
Figs.~\ref{fig:B_parallel_z_intense} and~\ref{fig:B_parallel_z_polarize},
displaying the results for isotropic injection for a slab depth of \teq{\tau_\parallel=10}.
\label{fig:slab_pole_anis_inj}}
\vspace{-20pt}
\end{figure}

Using this polarized, anisotropic injection protocol, simulations were performed for the
magnetic polar cases that were the focus of Figs.~\ref{fig:B_parallel_z_intense}
and~\ref{fig:B_parallel_z_polarize}, to discern how the introduction of anisotropy and
polarization to injection influenced the emergent Stokes parameters at the top of the slab. For
the most part, the various distributions were fairly similar to the isotropic, unpolarized
injection case once the optical depth \teq{\tau_{\parallel}} exceed around 7 or so, with either
modest or small  differences.  This is not surprising given that numerous scatterings obscure
the injection information. The exception was for the resonant \teq{\omega /\wcyc =0.99} example,
and results for this frequency are presented in Fig.~\ref{fig:slab_pole_anis_inj}. Therein, a
comparison between \teq{\tau_{\parallel}=10} results for the updated polarized injection and the
Figs.~\ref{fig:B_parallel_z_intense} and~\ref{fig:B_parallel_z_polarize} one is forged. 
Differences are small for zenith angles \teq{\theta_z \lesssim 45^{\circ}}, but become
significant or large for viewing perspectives somewhat near the slab horizon, i.e. perpendicular
to the field.  The intensity excess near the horizon observed for the isotropic injection case
is eliminated because the scattering depopulates the distribution orthogonal to \teq{\Bvec}.  At
the same time, circular polarization becomes somewhat more influential in the resonant
transport, and the linear polarization signatures correspond to a preponderance of \teq{\perp}
photons as \teq{Q/I} approaches \teq{-1}.  

This cameo comparison illustrates the importance of developing an understanding of the Stokes
parameter distributions in high opacity domains germane to the deeper portions of neutron star
atmospheres.  Our illustrative protocol can be applied to
any field orientation, thereby capturing any location on the neutron star surface.

\section{Discussion and Conclusion}
 \label{sec:discussion} 

The polarizations and anisotropies determined here in the high opacity domain directly impact the
effective Eddington limiting X-ray luminosity for a
magnetized compact object, above which radiation pressure-driven winds arise.  The familiar
non-magnetic value, \teq{\LEdd = 4\pi G\mns cm_p/\sigt} for a stellar mass \teq{\mns}
\citep[e.g.,][]{Rybicki79}, is predicated upon isotropic scattering with the Thomson cross
section \teq{\sigt}. This is strongly modified by the magnetic field, particularly well below
the cyclotron frequency.  \cite{Paczynski1992AcA}
employed an angle-averaged Rosseland mean opacity in adapting the Eddington limit to
magnetar conditions; various problems with employing such a Rosseland mean 
were discussed in \cite{vanPutten-2013-MNRAS}. The polarization-dependent
anisotropies and cross sections developed here in
Sections~\ref{sec:high_tau_simulate}-\ref{sec:redistribution} can be used to render estimates of
the effective Eddington luminosity more precise, leveraging the combination of polarization
information embedded in the \teq{A_{\omega}(\mu )} and \teq{\sigma (\omega ,\, \mu )}, 
and capturing the important interplay between circular and linear polarization in the vicinity 
of the cyclotron resonance.

Another potential application of this atmospheric transport simulation is to millisecond pulsars
(MSPs).   These serve as a prime science focus of NASA's {\sl NICER} X-ray mission on the
International Space Station, with the goal of measuring mass-to-radius ratios that can constrain
the neutron star equation of state.  In particular, {\sl NICER} has been able to infer a ``hot
spot'' emission geometry for the soft X rays from the surface of MSP J0030+0451 (Riley et al.
2019; Miller et al. 2019) that is far from the antipodal one commonly associated with a pure
dipolar configuration for isolated neutron stars. The surface temperature of J0030+0451 (spin
period \teq{P=4.87}ms) is in the range of \teq{T\sim 1.5 - 3 \times 10^6}K \citep[\teq{0.13 -
0.26}keV;][]{Bogdanov2009ApJ} and its surface polar field is \teq{4.5 \times 10^8}Gauss,
establishing a characteristic \teq{\omega /\wcyc} scale of \teq{3kT/\hbar \wcyc \sim 150}. This
corresponds to the non-magnetic domain of our study that is fairly well modeled using the
\teq{\omega /\wcyc = 10} examples in the various figures.   The local surface anisotropy, an
important quantity for determining contributions to the pulse profile that serves as a principal
diagnostic for {\sl NICER}, can be informed by our simulation, for example by
\teq{\tau_{\parallel}=10} runs like those depicted in the upper left panels of
Figures~\ref{fig:B_parallel_z_intense} and~\ref{fig:B_parallel_x_polarize}.  The anisotropy in
this \teq{\omega /\wcyc \gg 1} regime is very close to that of \cite{ST85}, and is approximately
independent of the magnetic field orientation in the slab.

Our anisotropy results exhibit general character similar to the non-magnetic, hydrostatic
atmosphere results of \cite{Zavlin1996AandA}. Yet there are differences between their
anisotropies and those here, dictated by the significant contributions of free-free and
bound-free opacities at low field strengths. Figure~1 of the
hydrostatic atmosphere model of \cite{HoLai-2001-MNRAS} demonstrates that free-free
absorption dominates scattering opacity at energies below 1--2 keV and deeper 
(\teq{\rho \gtrsim 1}g cm$^{-3}$) in a stratified, dense atmosphere of a magnetar.
Future enhancement of the {\sl MAGTHOMSCATT} simulation will
include such free-free opacity, simply implemented in the Monte Carlo technique. The radiation
transfer equation solution of \cite{HoLai-2001-MNRAS}, which is developed for magnetic fields
along the slab zenith (polar case), is extrapolated to arbitrary field orientations using a
simplistic diffusion approximation.  While this is likely suitable for MSP studies, our code's
facility in treating arbitrary field orientations will afford a profound improvement for
magnetars and accreting X-ray pulsars, for which departures from radiation isotropy are
significant and the diffusion approximation is no longer valid.

While the simulation was 
constructed using magnetic Thomson scattering for cold electrons in atmospheres, it is routinely
generalizable to incorporate the Doppler boosting/broadening and aberration effects associated
with warm plasma.  It thus has good potential for application to the more tenuous environs 
of magnetar magnetospheres, for example in modeling magnetar burst emission in hard X rays.
This is a problem that has been explored by \cite{Taverna17} in the magnetic Thomson domain 
via solution of the radiative transfer equation.  Such an extension of our Monte Carlo approach
will require a suitable reframing of QED scattering cross sections for a domain
where the classical formalism focused upon here is no longer appropriate.

In conclusion,  this paper presents the details of a versatile Monte Carlo simulation that has
been developed to model polarized radiative transfer in neutron star surface layers.  The code
employs an electric field vector formalism that enables a breadth of utility in terms of the
relationship between linear, circular and elliptical polarizations.  It is therefore adaptable
to address dispersive photon transport in plasmas and the magnetized quantum vacuum.   The 
{\sl MAGTHOMSCATT} code was validated for both intensity and Stokes parameter measures in 
a variety of ways.  The determination of analytic approximations at high optical depths for the 
Stokes parameters and anisotropy relative to the field direction helps define injection conditions 
deep in the simulation slab geometries, expediting simulations for full atmospheres, and 
potentially applicable to other high opacity neutron star magnetospheric environments such 
as accretion columns and magnetar burst regions.

Our frequency-dependent results identify 
informative polarization signatures that will be exploited by NASA's upcoming IXPE X-ray
polarimetry mission, presently scheduled for launch in 2021.  The next steps in our program will
be to integrate results over large surface regions and imbue the code with modules for modeling
hydrostatic structure, thereby introducing vertical stratification of temperature, pressure and
density into the atmospheres.  This will enable exploration of the interplay between photon
frequencies and polarization-dependent, photospheric optical depths that is central to
signatures of magnetar soft X-ray emission.

\section*{Acknowledgments}

\noindent
The authors thank the referee for comments helpful to improving the communicability 
of the paper.
M.~G.~B. acknowledges the generous support of the National Science Foundation
through grants AST-1517550 and AST-1813649.  M.~G.~B. also thanks the 
Netherlands Organisation for Scientific Research NWO for support for a 
visit to the Anton Pannekoek Institute at the University of Amsterdam, where
part of the research for this paper was performed.

\section*{Data Availability}

\noindent
The data underlying this article is housed on Rice University servers, 
and will be shared on reasonable request to the corresponding author
(M.~G.~B.).

\newpage

 \onecolumn

\begin{flushleft}
{\bf Appendix A: Developments for the Cross Section}
\end{flushleft}

As the vector formalism for magnetic Thomson scattering has only received very limited
treatment in the literature before, we outline our developments here at length for future reference.
The starting point for the derivations pertaining to the differential and total cross sections 
is the complex polarization for the scattered photon, 
specified using Eqs.~(\ref{eq:calEf_eval}) and~(\ref{eq:alpha_vec_def}):
\begin{equation}
   \calEvechat_f \; =\; \dover{ \kvechat_f \times \bigl( \kvechat_f \times \alphavec \bigr)}{\omega^2 - \wcyc^2} 
   \quad ,\quad
    \alphavec \; =\; \omega ^2 \calEvechat_i
      -i \omega \wcyc \,\calEvechat_i \times \Bvechat
      - \wcyc^2 ( \calEvechat_i \cdot \Bvechat ) \Bvechat \quad .
 \label{eq:calEhatf_append}
\end{equation}
This can be inserted into the dipole scattering formula to generate the differential cross section 
in Eq.~(\ref{eq:dsig_magThom}).  The squaring of the polarization vector is expedited using 
standard vector identities:
\begin{equation}
   \Bigl[ \kvechat_f \times \bigl( \kvechat_f \times \alphavec \bigr) \Bigr] \cdot
    \Bigl[ \kvechat_f \times \bigl( \kvechat_f \times \alphavec^{\ast} \bigr) \Bigr] 
    \; =\; \bigl( \kvechat_f \times \alphavec \bigr) \cdot  \bigl( \kvechat_f \times \alphavec^{\ast} \bigr)
    \; =\; \alphavec \cdot  \alphavec^{\ast} 
            - \bigl( \kvechat_f \cdot \alphavec \bigr) . \bigl( \kvechat_f \cdot \alphavec^{\ast} \bigr) \quad .
 \label{eq:vector_ident_app1}
\end{equation}
This defines the numerator for the differential cross section in Eq.~(\ref{eq:dsig_magThom}).
Expressing \teq{\alphavec} in terms of its real and imaginary vector components, 
\teq{\alphavec = \alphavecre + i \alphavecim}, one can establish the following identities 
expressed in terms of real and positive-definite quantities:
\begin{equation}
   \alphavec \cdot  \alphavec^{\ast} \; =\; \vert \alphavecre \vert^2 +  \vert \alphavecim \vert^2
   \quad ,\quad
   \alphavec \cdot  \alphavec^{\ast} 
            - \bigl( \kvechat_f \cdot \alphavec \bigr) . \bigl( \kvechat_f \cdot \alphavec^{\ast} \bigr)
   \; =\; \vert \alphavecre \vert^2 +  \vert \alphavecim \vert^2 
            -  \bigl( \kvechat_f \cdot \alphavecre \bigr)^2 - \bigl( \kvechat_f \cdot \alphavecim \bigr)^2 \quad .
 \label{eq:alphavec_ident1}
\end{equation}
From the last identity herein, the following inequality then immediately follows:
\begin{equation}
   0 \;\leq\; \bigl( \kvechat_f \times \alphavec \bigr) \cdot  \bigl( \kvechat_f \times \alphavec^{\ast} \bigr)
   \;\leq\; \alphavec \cdot  \alphavec^{\ast} \quad ,
 \label{eq:kvec_alphavec_bound}
\end{equation}
a result that is of considerable use in posing the accept-reject protocol for 
choosing the direction of the scattered photon and its associated 
polarization: see Eq.~(\ref{eq:dsig_mag_acc_rej}).   By inspection of the form for 
\teq{\alphavec} in Eq.~(\ref{eq:alpha_vec_def}), it is apparent that there a 
domains where \teq{ \vert \alphavecre \vert \gg  \vert \alphavecim \vert} if \teq{\calEvec_i} is real, for which 
\teq{ \alphavec \cdot  \alphavec^{\ast} \approx \vert \alphavecre \vert^2 -  \bigl( \kvechat_f \cdot \alphavecre \bigr)^2}
and the differential cross section is maximized when \teq{\kvechat_f} is approximately orthogonal 
to \teq{\alphavecre}.  This situation is intuitively expected given the 
\teq{\calEvechat_f \propto  \kvechat_f \times \bigl( \kvechat_f \times \alphavec \bigr)} form.

The determination of the total cross section requires the integration of the forms in 
Eq.~(\ref{eq:alphavec_ident1}) over all solid angles, \teq{d\Omega_f}.  The 
polar coordinates for such an integration can be aligned relative to any vector 
\teq{\betavec}.   From this one deduces that 
\begin{equation}
   \int \vert \boldsymbol{\beta} \vert^2 \, d\Omega_f 
   \; =\; 4\pi \, \vert \boldsymbol{\beta} \vert^2
   \quad \hbox{and}\quad
   \int \bigl( \kvechat_f \cdot \boldsymbol{\beta} \bigr)^2 \, d\Omega_f 
   \; =\; \dover{4\pi}{3} \, \vert \boldsymbol{\beta} \vert^2\quad ,
   \quad \hbox{for}\quad 
   \boldsymbol{\beta} \; =\; \alphavecre ,\; \alphavecim \quad .
 \label{eq:solid_angle_integs}
\end{equation}
It follows that 
\begin{equation}
   \int \Bigl[ \bigl( \kvechat_f \times \alphavec \bigr) \cdot  \bigl( \kvechat_f \times \alphavec^{\ast} \bigr) \Bigr] \, d\Omega_f 
   \; =\; \dover{8\pi}{3} \, \Bigl\{ \vert \alphavecre \vert^2 +  \vert \alphavecim \vert^2 \Bigr\}
   \; =\; \dover{8\pi}{3} \, \alphavec \cdot  \alphavec^{\ast} \quad .
 \label{eq:sigma_tot_integ}
\end{equation}
This identity establishes the form for the total cross section in Eq.~(\ref{eq:sig_mag_Thom_form}).
An alternative path to its derivation is directly via the last identity in Eq.~(\ref{eq:vector_ident_app1}),
using direct solid angle integration.  This employs the result
\begin{equation}
   \int \bigl( \kvechat_f \cdot \boldsymbol{\beta} \bigr) \, \bigl( \kvechat_f \cdot \boldsymbol{\gamma} \bigr)\, d\Omega_f 
   \; =\; \boldsymbol{\gamma} \cdot \betavec \,  \int \bigl( \kvechat_f \cdot \betavechat \bigr)^2 \, d\Omega_f 
   + \int \bigl( \kvechat_f \cdot \boldsymbol{\beta} \bigr) \, \Bigl( \kvechat_f \cdot
   \bigl\lbrack \boldsymbol{\gamma} - \bigl( \boldsymbol{\gamma} \cdot \betavechat \bigr) \betavechat \bigr\rbrack \Bigr) \, d\Omega_f
   \; =\; \dover{4\pi}{3} \,  \boldsymbol{\gamma} \cdot \betavec   \quad .
 \label{eq:solid_angle_integ2}
\end{equation}
The manipulation here is to resolve the vector \teq{\boldsymbol{\gamma}} into components 
parallel to [\teq{\bigl( \boldsymbol{\gamma} \cdot \betavec \bigr) \, \betavechat }]
and perpendicular to \teq{\betavec}, yielding respectively the first and second integrals in the 
middle expression.  The second integral is simply demonstrated to be zero.

It is straightforward to express the differential cross section in terms of the incoming 
polarization vector \teq{\calEvechat} and \teq{\Bvechat} using Eq.~(\ref{eq:alpha_vec_def}).
When squaring, the last  form in Eq.~(\ref{eq:vector_ident_app1}) is preferred, and 
the first portion of this is routinely generated:
\begin{equation}
   \alphavec \cdot  \alphavec^{\ast} \; =\; 
   \omega^4 + \wcyc^2 (\wcyc^2 - 2\omega^2) \Bigl\vert \calEvechat_i \cdot \Bvechat \Bigr\vert^2 
      + \omega^2 \wcyc^2 \Bigl\vert \calEvechat_i \times \Bvechat \Bigr\vert^2
         + 2i\, \omega^3 \wcyc \, \Bvechat \cdot \bigl( \calEvechat_i \times \calEvechat_i^* \bigr) \quad .
 \label{eq:alpha_dot_alphastar}
\end{equation}
The only notes for deriving this identity are that the \teq{i \omega \wcyc^3} term is identically zero
because of orthogonality of \teq{\calEvechat \times \Bvechat} and \teq{\Bvechat}, and 
the cyclic permutation of the triple scalar product is employed to distill the \teq{i \omega^3 \wcyc} term
into a convenient form.  With this identity, the expanded form for the total cross section 
in Eq.~(\ref{eq:sig_magThom}) follows. The remaining term on the right of Eq.~(\ref{eq:vector_ident_app1}) is
routinely evaluated to generate the full differential cross section:
\begin{eqnarray}
   \dover{ (\omega^2 - \wcyc^2)^2}{r_0^2}\, \dover{d\sigma}{d\Omega_f} 
   & = &  \omega^4  \, \Bigl\{ 1 -  \bigl( \kvechat_f \cdot \calEvechat_i \bigr)^2 \Bigr\} 
    + \wcyc^4 \, \Bigl\vert \calEvechat_i \cdot \Bvechat \Bigr\vert^2 \, \Bigl\{ 1 -  \bigl( \kvechat_f \cdot \Bvechat \bigr)^2 \Bigr\} 
    +\, \omega^2\wcyc^2 \left\{ \left\vert \calEvechat_i \times \Bvechat \right\vert^2
         - 2\; \bigl\vert \calEvechat_i \cdot \Bvechat \bigr\vert^2 \right\}  \nonumber\\[-6.5pt] 
 \label{eq:diff_csect}\\[-6.5pt]
   && +\, \omega^2\wcyc^2 \left\{  -  \left\vert \kvechat_f \cdot \bigl( \calEvechat_i \times \Bvechat \bigr) \right\vert^2 
      +  (\kvechat_f \cdot \Bvechat ) \;  \Bigl[ ( \kvechat_f \cdot \calEvechat_i ) (\calEvechat_i^\ast \cdot \Bvechat ) + 
        (\kvechat_f \cdot \calEvechat_i^\ast ) (\calEvechat_i \cdot \Bvechat ) \Bigr]  \right\} \nonumber\\[1.0pt]
  && +\; i \, \omega^3 \wcyc\,  \bigl( \calEvechat_i \times \calEvechat_i^* \bigr) \cdot
            \Bigl\lbrack \kvechat_f \times ( \kvechat_f \times \Bvechat ) \Bigr\rbrack
        + i \, \omega \wcyc^3\, ( \kvechat_f \cdot \Bvechat  ) 
     \bigl( \calEvechat_i \times \calEvechat_i^* \bigr) \cdot \Bigl[ \Bvechat \times \bigl( \Bvechat \times \kvechat_f \bigr) \Bigr] \quad . \nonumber
\end{eqnarray}
This expression will be employed in the algorithm for radiative transfer in 
the atmosphere, as discussed in Section~\ref{sec:MC_technique}.  Using the integral identities 
in Eqs.~(\ref{eq:solid_angle_integs}) and~(\ref{eq:solid_angle_integ2}) it can be routinely integrated 
over solid angles to yield the total cross section in Eq.~(\ref{eq:sig_magThom}),
observing that the \teq{i \omega\wcyc^3} term integrates to zero.
It is technically applicable for non-dispersive light propagation where 
\teq{\omega = \vert \kvec_i \vert = \vert \kvec_f \vert} and the eigenmodes are purely transverse;
this circumstance is modified 
when dispersion in a magnetized plasma is treated \citep{Canuto71PRD}.
When \teq{\omega\gg \wcyc} the influence of the magnetic field is negligible,
and only the leading term on the RHS of  Eq.~(\ref{eq:diff_csect}) 
is retained, yielding the familiar non-magnetic scattering result
\teq{d\sigma /d\Omega_f = r_0^2 \{ 1 -  \bigl( \kvechat_f \cdot \calEvechat_i \bigr)^2 \} }.

\vspace{10pt}
%
\begin{flushleft}
{\bf Appendix B:  Cross Section -- Special Cases}
\end{flushleft}

\begin{figure}
\vspace*{-0pt}
\centerline{\includegraphics[width=9.5cm]{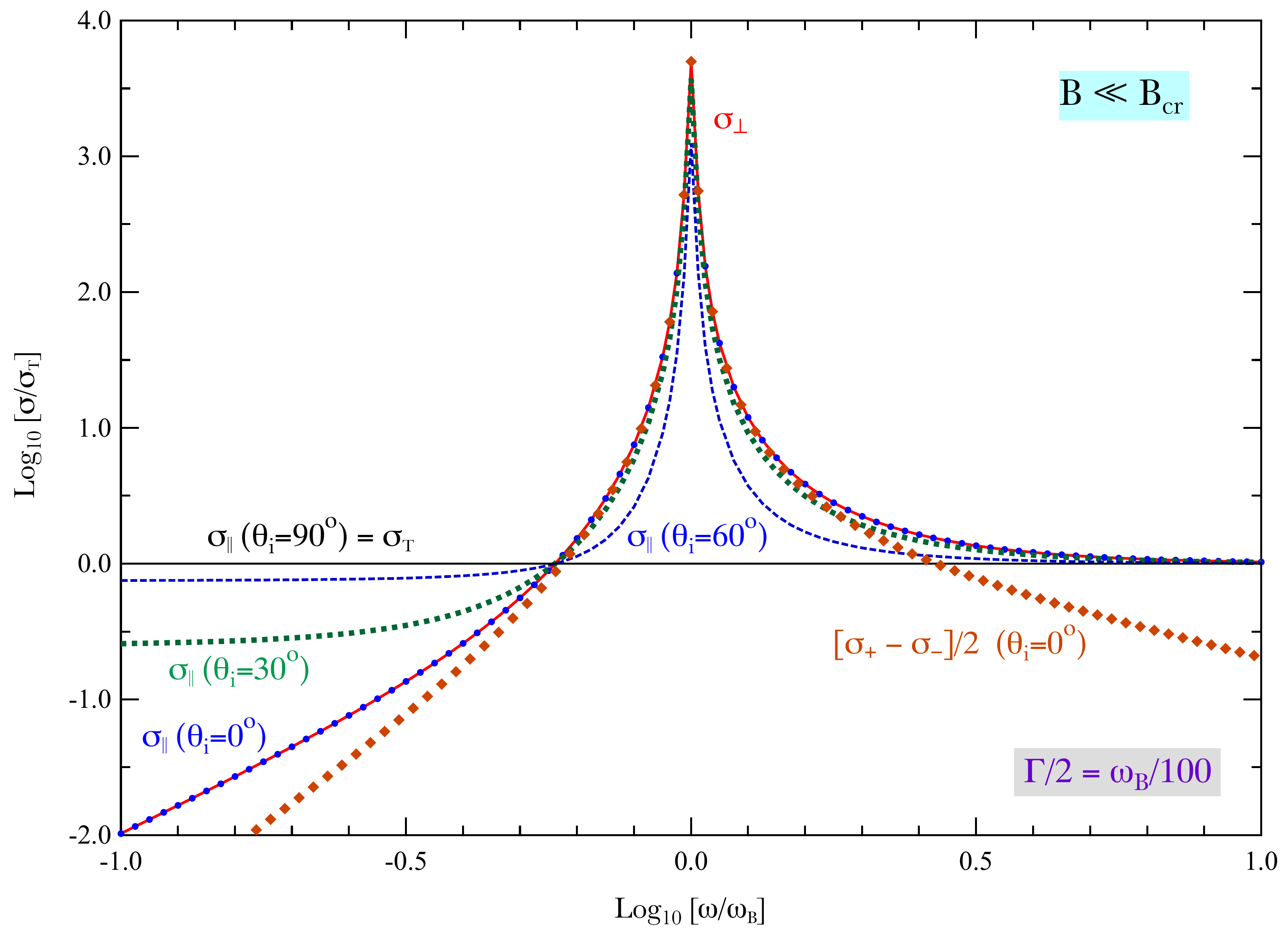}}
\vspace*{-5pt}
\caption{The total cross section \teq{\sigma_{\perp, \parallel}} for magnetic 
Thomson scattering (in units of \teq{\sigt}) for the two standard linear polarization 
states \teq{\perp} (extraordinary mode) and \teq{\parallel} (ordinary mode), 
computed using Eq.~(\ref{eq:sig_perp_para_sum}) in Appendix B.  The cross sections 
are functions of the frequency ratio \teq{\omega /\wcyc} (log scale) and sum 
over final polarization states.  For the \teq{\parallel} mode, four different incidence 
angles \teq{\theta_i} to the field direction are depicted, with \teq{\theta_i=0} being 
the blue dots.  For the \teq{\perp} mode (red curve), 
the cross section is independent of \teq{\theta_i}, and coincides with
the \teq{\theta_i=0^{\circ}} result for the \teq{\parallel} mode.  Linear polarization cross sections 
are resonant at the cyclotron frequency \teq{\wcyc} when \teq{\theta_i < 90^{\circ}}, 
truncated using a small width \teq{\Gamma = \wcyc/50}, and are identically equal to the 
Thomson value \teq{\sigt} when \teq{\omega = \wcyc/\sqrt{3}}.  Also depicted 
is the difference factor \teq{[\sigma_+ - \sigma_-]/2} for  \teq{\theta_i=0^{\circ}}
(brown diamonds) of the cross sections \teq{\sigma_{\pm}} for the two circular 
polarization modes, i.e., the \teq{\DeltaB (\omega )} circularity function expressed in 
Eq.~(\ref{eq:DeltaB_def}).
 \label{fig:csect_mag}}
\end{figure}

To aid insight into the radiative transfer simulation elements, some results pertaining to 
familiar polarization states are presented.  Without loss of generality, the magnetic field 
will be chosen to be in the \teq{z}-direction so that \teq{\Bvechat = (0,\, 0,\, 1)}.  The
initial (i) and final (f) electromagnetic wave unit vectors in a scattering event will be specified by
spherical polar coordinates
\begin{equation}
   \kvechat_j \; =\; \bigl( \sin\theta_j \cos\phi_j ,\, \sin\theta_j \sin\phi_j ,\, \cos\theta_j \bigr)
   \quad \hbox{for} \quad  j \; =\; i,\, f\quad .
 \label{eq:kvechat_j}
\end{equation}
 Considering first two orthogonal linear polarization states, it is customary to identify these as being 
 parallel (\teq{\parallel}) to the plane defined by \teq{\kvechat_j} and \teq{\Bvechat}
 (ordinary mode), with the second state (denoted \teq{\perp}) being perpendicular to this plane
 (extraordinary mode).  Thus, the unit electric field vectors of these two orthogonal polarization modes are given by
\begin{equation}
   \calEvechat_{\perp j} \; \equiv\; e^{i\varphi_j} \dover{\kvechat_j \times \Bvechat}{\vert \kvechat_j \times \Bvechat\vert }
   \; =\; e^{i\varphi_j} \bigl( \sin\phi_j ,\, -  \cos\phi_j ,\,  0 \bigr) 
   \quad ,\quad
   \calEvechat_{\parallel j} \; \equiv \; \kvechat_j \times \calEvechat_{\perp j} 
   \; =\; e^{i\varphi_j} \bigl( \cos\theta_j \cos\phi_j ,\, \cos\theta_j \sin\phi_j ,\, - \sin\theta_j \bigr) \quad ,
 \label{eq:perp_par_polarvec}
\end{equation}
%
%
%
%
%
observing that \teq{\vert  \kvechat_j \times \calEvechat_{\perp j} \vert = 1} is automatically guaranteed 
for the orthogonal vector triad constituted by \teq{\kvechat_j}, \teq{\calEvechat_{\perp j}} and 
\teq{\calEvechat_{\parallel j}}.  Note also that the complex phase \teq{\varphi_j} is explicitly introduced 
to encompass the possibility of complex electric field vectors \teq{\calEvec_j}.  If these forms are inserted 
into Eq.~(\ref{eq:Stokes_polar_def}), one quickly determines that \teq{V=0} for both linear polarizations.
For the initial wave 
prior to scattering, one can set \teq{\varphi_i=0} without loss of generality.  Yet, given the forms in 
Eqs.~(\ref{eq:calEf_eval}) and~(\ref{eq:alpha_vec_def}), it is apparent that for the scattered wave, 
in general \teq{\varphi_f} is non-zero. The linear polarization states in Eq.~(\ref{eq:perp_par_polarvec}) 
simply yield  \teq{ \calEvechat_{\perp i} \times \calEvechat_{\perp i}^{\ast} =\mathbf{0}
= \calEvechat_{\parallel i} \times \calEvechat_{\parallel i}^{\ast}},
expediting the algebraic development of the differential and total cross sections.  In addition, 
the normalization results
\begin{equation}
  \Bigl\vert \calEvechat_{\perp i} \cdot \Bvechat \Bigr\vert \; =\; 0  
  \quad ,\quad \Bigl\vert \calEvechat_{\perp i} \times \Bvechat \Bigr\vert \; =\; 1
  \qquad \hbox{and}\qquad
  \Bigl\vert \calEvechat_{\parallel i} \cdot \Bvechat \Bigr\vert \; =\; \sin\theta_i 
  \quad ,\quad \Bigl\vert \calEvechat_{\parallel i} \times \Bvechat \Bigr\vert \; =\; \cos\theta_i
 \label{eq:calE_dot_times_calE_perp_para}
\end{equation}
render the evaluation of Eq.~(\ref{eq:alpha_dot_alphastar}) simple, since it only 
depends on quantities for the incident photon.
Thus, the total scattering cross section for \teq{\perp} and \teq{\parallel} states 
and the resultant unpolarized (up) cross section 
\teq{\sigma_{\rm up} = (\sigma_{\perp}+\sigma_{\parallel})/2} can 
be expressed in units of the Thomson cross section \teq{\sigt = 8\pi r_0^2/3}:
\begin{equation}
   \sigma_{\perp}\;=\;\sigt\, \SigmaB (\omega) 
   \quad ,\quad
   \sigma_{\parallel} \; =\; \sigt\,\sin^2\theta_i + \sigt\, \SigmaB (\omega ) \,\cos^2\theta_i
   \quad \Rightarrow\quad
   \sigma_{\rm up} \; =\;  \dover{\sigt}{2} \Bigl\{ \sin^2\theta_i  + \SigmaB (\omega)
       \left[1+\cos^2\theta_i\right] \Bigr\}\quad ,
 \label{eq:sig_perp_para_sum}
\end{equation}
with
\begin{equation}
    \SigmaB (\omega ) \; =\; \dover{\omega^2(\omega^2+\wcyc^2)}{ (\omega^2 - \wcyc^2)^2}
    \; =\; \dover{1}{2} \biggl\{ \dover{\omega^2}{(\omega - \wcyc)^2}  
         + \dover{\omega^2}{(\omega + \wcyc)^2} \biggr\} \quad .
 \label{eq:SigmaB_def}
\end{equation}
Using the second form for \teq{\SigmaB (\omega)}, which isolates the helicity contributions
to the scattering, it is quickly seen that the polarized forms in Eq.~(\ref{eq:sig_perp_para_sum}) 
are identical to those in Eq.~(16) of \cite{Herold79}, who took the non-relativistic limit of a quantum mechanical treatment.
An alternative path to the same result is to instead evaluate \teq{\alphavec} in Eq.~(\ref{eq:alpha_vec_def}) directly:
\begin{equation}
  \alphavec_{\perp} \; =\; \omega ^2 \calEvechat_{\perp i}
  -i \omega \wcyc \,\calEvechat_{\perp i} \times \Bvechat \quad ,\quad 
  \alphavec_{\parallel} 
  \; =\; \omega ^2 \calEvechat_{\parallel i}
  -i \omega \wcyc \cos\theta_i \,\calEvechat_{\perp i}
  + \wcyc^2 e^{i\varphi_i}\sin\theta_i \,\Bvechat \quad ,
 \label{eq:perp_para_alpha}
\end{equation}
observing that the second term for \teq{\alphavec_{\parallel}} is the 
\teq{\calEvechat_{\parallel i} \times \Bvechat} contribution.  These forms
quickly yield
\begin{equation}
   \alphavec_{\perp} \cdot  \alphavec^{\ast}_{\perp} 
   \; =\; \omega^2(\omega^2+\wcyc^2)\quad,
   \quad \alphavec_{\parallel} \cdot  \alphavec^{\ast}_{\parallel} 
   \; =\; (\omega^2-\wcyc^2)^2\sin^2\theta_i+\omega^2(\omega^2+\wcyc^2)\cos^2\theta_i\quad ,
 \label{eq:perp_para_alpha_dot_alphastar}
\end{equation}
and inserting these into Eq.~(\ref{eq:sig_mag_Thom_form}) again yields 
Eq.~(\ref{eq:sig_perp_para_sum}).  A depiction of the linearly-polarized 
cross section is given in Fig.~\ref{fig:csect_mag}.
The Figure is labelled with \teq{B\ll B_{\rm cr}}, the domain 
for which its classical derivation is formally valid.  Yet, at soft X-ray frequencies it can be applied to supercritical 
fields such as are encountered in magnetars, since the quantum derivation \citep[e.g.][]{Herold79} reduces to 
the classical one as long as \teq{\hbar \omega \ll m_ec^2}. 
Fig.~\ref{fig:csect_mag} illustrates the identity of 
\teq{\sigma_{\perp}} and \teq{\sigma_{\parallel}} when \teq{\theta_i=0}, i.e. for 
photons propagating exactly along \teq{\boldsymbol{B}}, evident in Eq.~(\ref{eq:sig_perp_para_sum}).
In this special case of aligned incidence, the scaled acceleration vector \teq{\alphavec}
in Eq.~(\ref{eq:alpha_vec_def}) is always perpendicular to \teq{\boldsymbol{B}} and 
of fixed magnitude, so the wave excites a circular motion of the target electron.

In order to obtain the differential and total cross sections for specific incident {\it and} scattered polarization states, 
it is not possible to directly work from Eq.~(\ref{eq:diff_csect}) as the coupling between \teq{\kvec_f} and 
polarization state of this scattered wave needs to be isolated.
For \teq{p,q=\perp,\parallel} denoting the polarization states of incident \teq{(p)} and scattered \teq{(q)} waves,
we write Eq.~(\ref{eq:dsig_magThom}) as
\begin{equation}
   \dover{d\sigma_{p\rightarrow q}}{d\Omega_f} 
   \; =\; r_0^2 \, \dover{ \calEvec_{qf}\cdot \calEvec_{qf}^*}{\vert \calEvec_{pi} \vert^2} 
   \;=\; r_0^2\, \dover{ \bigl\vert\bigl[\kvechat_f\times(\kvechat_f\times\alphavec_p)\bigr] 
         \cdot \calEvechat_{qf}^* \bigl\vert^2}{(\omega^2 - \wcyc^2)^2} 
   \;=\; r_0^2\, \dover{ \bigl\vert \alphavec_p \cdot \calEvechat_{qf}^* \bigr\vert^2}{(\omega^2 - \wcyc^2)^2} \quad .
 \label{eq:dsig_magThom_p_q}
\end{equation}
Herein,  \teq{\calEvec_{qf} = (\calEvec_f \cdot \calEvechat_{qf}^*) \calEvechat_{qf}} is the projection 
of the polarization vector \teq{\calEvec_{f}} onto the unit vector \teq{\calEvechat_{qf}}. 
The polarization vector \teq{\calEvec_{qf}} can be expressed in terms of \teq{\kvechat_f} 
and \teq{\alphavec_p} through Eq.~(\ref{eq:calEf_eval}), and for these linear polarization 
considerations it assumes one of the forms in Eq.~(\ref{eq:perp_par_polarvec}).
In prescribing the last part of Eq.~(\ref{eq:dsig_magThom_p_q}) the vector identity 
\teq{\kvechat_f\times(\kvechat_f\times\alphavec_p)=(\kvechat_f\cdot\alphavec_p)\kvechat_f-\alphavec_p} 
was employed, together with the transversality condition \teq{\kvechat_f\cdot\calEvechat^{\ast}_{qf}=0}
appropriate for our zero dispersion presumption. 
Inserting Eq.~(\ref{eq:perp_par_polarvec}) and Eq.~(\ref{eq:perp_para_alpha}) into the 
final form in Eq.~(\ref{eq:dsig_magThom_p_q}), and defining  \teq{\phi_{fi}=\phi_f-\phi_i} as the 
change in azimuthal angle in a scattering, one quickly obtains forms for the polarization-dependent 
differential cross sections:
\newpage
\begin{eqnarray}
  \qquad\qquad\qquad\qquad\dover{d\sigma_{\perp\rightarrow\perp}}{d\Omega_f} 
   & = & \dover{r_0^2}{ (\omega^2 - \wcyc^2)^2} \Bigl( \omega^4\cos^2\phi_{fi}+ \omega^2\wcyc^2\sin^2\phi_{fi}\Bigr) \quad , \nonumber\\[0pt]
             \dover{d\sigma_{\perp\rightarrow\parallel}}{d\Omega_f} 
   & = & \dover{r_0^2}{ (\omega^2 - \wcyc^2)^2} \cos^2\theta_f\Bigl( \omega^4\sin^2\phi_{fi}+ \omega^2\wcyc^2\cos^2\phi_{fi}\Bigr) \quad , \nonumber\\[0pt]
 \label{eq:dsig_perp_para}
             \dover{d\sigma_{\parallel\rightarrow\perp}}{d\Omega_f} 
   & = & \dover{r_0^2}{ (\omega^2 - \wcyc^2)^2} \cos^2\theta_i\Bigl( \omega^4\sin^2\phi_{fi}+ \omega^2\wcyc^2\cos^2\phi_{fi}\Bigr) \quad , \\[0pt]
             \dover{d\sigma_{\parallel\rightarrow\parallel}}{d\Omega_f} 
   & = & r_0^2\sin^{2}\theta_i\sin^{2}\theta_f+\dover{r_0^2}{ (\omega^2 - \wcyc^2)^2} \biggl[\cos^2\theta_i\cos^2\theta_f
             \Bigl( \omega^4\cos^2\phi_{fi}+ \omega^2\wcyc^2\sin^2\phi_{fi}\Bigr) \Big. \nonumber \\[-3pt]
   && \hspace{145pt}  
         +2\omega^2 \bigl(\omega^2-\wcyc^2\bigr) \sin\theta_i\cos\theta_i\sin\theta_f\cos\theta_f\cos\phi_{fi}\biggr] \quad . \nonumber
\end{eqnarray}
All results in Eq.~(\ref{eq:dsig_perp_para}) are identical to those in Eq.~(1) of \cite{BS76_MNRAS}, which were 
derived from the classical formalism in \cite{Canuto71PRD}.  Integrating these results over the 
phase difference \teq{\phi_{fi}}, and forming the difference between 
the \teq{\perp\to\parallel} and \teq{\parallel\to\perp} results, which captures a 
key portion of the mode switching information, yields
\begin{equation}
    \dover{d\sigma_{\perp\rightarrow\parallel}}{d\cos\theta_f} -  \dover{d\sigma_{\parallel\rightarrow\perp}}{d\cos\theta_f}
    \; =\; \dover{3}{8}\sigt\, \SigmaB (\omega) \,
    \Bigl\{ \cos^2\theta_f - \cos^2\theta_i \Bigr\} \quad .
 \label{eq:mode_convert_linear}
\end{equation}
This essentially defines the detailed balance between \teq{\perp} and \teq{\parallel} polarizations in the radiative transfer problem,
with production of \teq{\perp} being favored in directions more oblique and pependicular to the field \teq{\Bvec}.  
Integrating Eq.~(\ref{eq:dsig_perp_para}) over solid angle yields the total cross section for the four possible linear polarization configurations 
in units of the Thomson cross section \teq{\sigt}:  
\begin{eqnarray}
   \hspace{120pt} \sigma_{\perp\rightarrow\perp}
   \;=\;\dover{3}{4}\sigt\, \SigmaB (\omega)  \quad &,& \quad
   \sigma_{\perp\rightarrow\parallel} 
   \; =\; \dover{1}{4}\sigt\, \SigmaB (\omega)   \quad,\nonumber \\[-7pt]
\label{eq:sig_perp_para}\\[-7pt]
   \sigma_{\parallel\rightarrow\perp} 
   \;=\;\dover{3}{4}\sigt\, \SigmaB (\omega) \, \cos^2\theta_i \quad &,& \quad
   \sigma_{\parallel\rightarrow\parallel} 
   \; =\; \sigt\,\sin^2\theta_i + \dover{1}{4}\sigt\, \SigmaB (\omega) \, \cos^2\theta_i \quad . \nonumber
\end{eqnarray}
These trivially sum over the final polarizations to produce Eq.~(\ref{eq:sig_perp_para_sum}).

The total cross sections for circular polarized initial states can be derived in a similar way. 
The unit electric field vectors for the incident photons of positive (subscript $+$) and negative (subscript $-$)
helicity are given by
\begin{equation}
   \calEvechat_{\pm i} \; \equiv\; \dover{1}{\sqrt{2}} \bigl( \calEvechat_{\perp i}\pm i \calEvechat_{\parallel i} \bigr)
   \; =\; \dover{1}{\sqrt{2}}e^{i\varphi_i} \Bigl( \sin\phi_i \pm i\cos\theta_i\cos\phi_i,
              \, -  \cos\phi_i \pm i\cos\theta_i\sin\phi_i,\,  \mp i\sin\theta_i \Bigr)  \quad .
 \label{eq:plus_minus_polarvec}
\end{equation}
%
%
%
%
If these definitions are inserted into the forms for the Stokes parameters in Eq.~(\ref{eq:Stokes_polar_def}), 
one quickly determines that \teq{V_{\pm} = \pm \cos\theta_i}, i.e.  when \teq{\theta_i=0} and propagation is 
parallel to \teq{\Bvechat}, \teq{V=1} for the $+$ polarization and \teq{V=-1} for the $-$ polarization.
From these complex field vectors, one can routinely form \teq{\alphavec_{\pm} = (\alphavec_{\perp} \pm i\alphavec_{\parallel}) /\sqrt{2}}, 
and then employ 
\begin{equation}
  \Bigl\vert \calEvechat_{\pm i} \cdot \Bvechat \Bigr\vert^2 \; =\; \dover{1}{2} \sin^2\theta_i  
  \quad , \quad 
  \Bigl\vert \calEvechat_{\pm i} \times \Bvechat \Bigr\vert^2 \; =\; \dover{1}{2}(1+\cos^2\theta_i) 
  \quad  , \quad
  \Bvechat \cdot \bigl( \calEvechat_{\pm i} \times \calEvechat_{\pm i}^* \bigr)\;=\; \mp i\cos\theta_i \quad  .
   \label{eq:calE_B_dot_times_calE_pm}
\end{equation}
%
%
as identities for the circularly-polarized incident radiation, leveraging Eq.~(\ref{eq:alpha_dot_alphastar}) to yield
\begin{equation}
   \alphavec_{\pm} \cdot  \alphavec^{\ast}_{\pm} \; =\; 
   \omega^4 + \dover{1}{2}\wcyc^2 \bigl( \wcyc^2 - 2\omega^2 \bigr) \sin^2\theta_i 
   + \dover{1}{2}\omega^2 \wcyc^2 (1+\cos^2\theta_i)
      \pm 2\, \omega^3 \wcyc \cos\theta_i \quad .
 \label{eq:alpha_dot_alphastar_pm}
\end{equation}
The \teq{\omega^3\wcyc} term is not zero because \teq{\calEvechat_{\pm i}} 
and \teq{\calEvechat_{\pm i}^*} are not parallel vectors.  The 
total cross sections can then be expressed as
\begin{equation}
   \sigma_{\pm}  \; = \; \dover{1}{2}\sigt\,\sin^2\theta_i + \dover{\sigt}{(\omega^2-\wcyc^2)^2}
   \biggl[\dover{1}{2}\omega^2(\omega^2+\wcyc^2) \bigl(1+\cos^2{\theta_i} \bigr)
            \pm 2\omega^3\wcyc\cos\theta_i \biggr] 
    \; = \; \dover{1}{2} \bigl(\sigma_{\perp}+\sigma_{\parallel} \bigr) 
              \pm \sigt \, \Delta_{\hbox{\sixrm B}} (\omega ) \, \cos\theta_i \;\; . 
  \label{eq:sig_pm}
\end{equation}
with the second form isolating a circularity function
\begin{equation}
   \DeltaB (\omega ) \;=\; \dover{2\omega^3\wcyc}{(\omega^2 - \wcyc^2)^2} 
   \;=\; \dover{1}{2} \biggl\{ \dover{\omega^2}{(\omega - \wcyc)^2} 
          - \dover{\omega^2}{(\omega + \wcyc)^2} \biggr\}
 \label{eq:DeltaB_def}
\end{equation}
that is depicted in Fig.~\ref{fig:csect_mag}.
Eq.~(\ref{eq:sig_pm}) is in agreement with Eq.~(17) of \cite{Herold79}. 
The second form highlights the relationship between the circular polarization results and the linear ones.
The circularity contribution is 
an odd function of \teq{\cos\theta_i}, character that maps into an anti-symmetry of 
Stokes parameter \teq{V} in the polar angle relative to 
\teq{\Bvec} that is ubiquitous in the results presented in Section~\ref{sec:high_opacity_pol}.
The circularity is zero 
when photons propagate perpendicular to the field, \teq{\theta_i= \pi /2}.
When \teq{\theta_i=0}, and the incoming wave eigenstates are those of circular polarization,
the circularity is at a maximum and one quickly deduces from 
Eq.~(\ref{eq:sig_pm}) that \teq{\sigma_+ = \sigt\, \omega^2/(\omega - \wcyc )^2} 
is resonant at the cyclotron frequency, 
but that \teq{\sigma_- = \sigt\, \omega^2/(\omega + \wcyc )^2} is not.
The positive helicity state drives the scattered electron in the sense it 
naturally gyrates in the field, precipitating a resonance at the 
cyclotron frequency.  In contrast, the negative helicity state ``opposes'' the gyration and 
does not lead to resonant scatterings.

Without detailing lengthy forms for the polarization-dependent differential cross sections, 
it suffices to posit a circular polarization analog to Eq.~(\ref{eq:mode_convert_linear}) that 
encapsulates the mode switching information:
\begin{equation}
    \dover{d\sigma_{-\rightarrow +}}{d\cos\theta_f} -  \dover{d\sigma_{+ \rightarrow -}}{d\cos\theta_f}
    \; =\; \dover{3}{8}\sigt\, \DeltaB (\omega) \,
    \bigl( \cos\theta_f - \cos\theta_i \bigr) \, \bigl( 1 - \cos\theta_f \cos\theta_i \bigr)\quad .
 \label{eq:mode_convert_circular}
\end{equation}
Thus, when \teq{\theta_f} is small, and less than the typical \teq{\theta_i} involved in a 
scattering, the production of the \teq{+} polarization is favored over the generation of the \teq{-}
state, leading to generally positive \teq{V}, as is observed in the magnetic polar slab simulation 
results in Fig.~\ref{fig:B_parallel_z_polarize}, and the \teq{0 < \theta < \pi /2} values for 
\teq{V/I} when \teq{\omega > \wcyc/\sqrt{3}} in the high opacity data displayed in 
Fig.~\ref{fig:freq_ang_map}.

\end{document}